\RequirePackage{fix-cm}
\documentclass[smallextended]{svjour3}       
\smartqed  
\usepackage{graphicx}
\usepackage{latexsym}
\usepackage{natbib}
\usepackage{lineno}
\usepackage{txfonts}
\usepackage{changepage}
\usepackage{lipsum}
\usepackage{amstext}
\usepackage[usenames]{xcolor}
\newcommand{\rd}{\bf}

\newcommand{\orange}{}

\def\note #1]{{\bf #1]}}
\def\noterd #1]{{\bf\rd #1]}}
\def\dd{{\rm d}}
\def\bolddelta{\delta\kern-0.45em\delta\kern-0.45em\delta}
\def\boldr{\mbox{\boldmath$r$}}
\def\bolda{\mbox{\boldmath$a$}}
\def\boldF{\mbox{\boldmath$F$}}
\def\bolddelr{\bolddelta \boldr}
\def\eye{{\rm i}}
\def\ee{{\rm e}}
\def\muHz{\,\mu{\rm Hz}}
\def\div{{\rm div}\,}

\def\modsg{{\rm M_{SG}}}
\def\modrg{{\rm M_{RG}}}

\def\Msun{\,{\rm M}_\odot}
\def\Rsun{\,{\rm R}_\odot}
\def\CD{{\cal D}}
\def\CF{{\cal F}}
\def\CI{{\cal I}}
\def\CK{{\cal K}}
\def\CL{{\cal L}}
\def\CN{{\cal N}}
\def\CO{{\cal O}}
\def\CP{{\cal P}}
\def\CR{{\cal R}}

\begin{document}

\title{Giant star seismology
}


\author{S. Hekker        \and
        J. Christensen-Dalsgaard 
}


\institute{S. Hekker \at
             Max-Planck-Institut f\"ur Sonnensystemforschung, Justus-von-Liebig-Weg 3, 37077 G\"ottingen, Germany \\
              Stellar Astrophysics Centre, Department of Physics and Astronomy, Aarhus University, Ny Munkegade 120, 8000 Aarhus C, Denmark\\
              Tel.: +49 (0)551/384 979-265\\
              \email{Hekker@mps.mpg.de}           
           \and
           J. Christensen-Dalsgaard \at
              Stellar Astrophysics Centre, Department of Physics and Astronomy, Aarhus University, Ny Munkegade 120, 8000 Aarhus C, Denmark\\
}

\date{Received: date / Accepted: date}

\maketitle

\begin{abstract}
The internal properties of stars in the red-giant phase undergo significant changes on relatively short timescales. Long near-{\orange un}interrupted high-precision photometric timeseries observations from dedicated space missions such as CoRoT and {\it Kepler} have provided seismic inferences of the global and internal properties of a large number of evolved stars, including red giants. These inferences are confronted with predictions from theoretical models to improve our understanding of stellar structure and evolution. Our knowledge and understanding of red giants have indeed increased tremendously using these seismic inferences, and we anticipate that more information is still hidden in the data. Unraveling this will further improve our understanding of stellar evolution. This will also have significant impact on our knowledge of the Milky Way Galaxy as well as on exo-planet host stars. The latter is important for our understanding of the formation and structure of planetary systems. 
\end{abstract}

\def\note #1]{{\bf #1]}}
\section{Introduction}

Stars are bodies formed by baryonic mass. They are an important source of electromagnetic radiation in the universe allowing for studies of many phenomena, from distant galaxies to the interstellar medium and extra-solar planets. However, due to their opacity it is not trivial to study the internal structure of stars \citep{eddington1926}. The only way to probe and study the internal stellar structure directly is through global stellar oscillations.  Many stars across the Hertzsprung-Russell (HR) diagram oscillate (driven by different mechanisms), which allows for in-depth studies of stars and stellar evolution.

\subsection{History}
\textit{Asteroseismology} -- the study of global properties of stars and their internal structure through their global intrinsic oscillations -- is already more than a century old. An early remarkable result was obtained for Cepheids revealing the period-luminosity relation \citep{leavitt1912}, which played an important role in measuring distances of galaxies and star clusters and ultimately the expansion of the universe. Observations of pulsators which display large (photometric) variations have been ongoing ever since. These stars include among others, Cepheids, high amplitude $\delta$ Scuti stars, RR Lyrae stars, white dwarfs and {\orange Miras}. An early result for which asteroseismic observations and stellar modelling were inconsistent was presented by \citet{petersen1973} on double-mode Cepheids. He showed that stellar masses inferred from the periods and period ratios were inconsistent with the location of the stars in the HR diagram as predicted by stellar evolution models. This problem motivated a revision of stellar opacity tables which led to very good agreement between the models and the observations \citep[e.g.,][]{moskalik1992}. Despite the long-term research on such bright large amplitude oscillators there are still open questions concerning their oscillations \citep[e.g., Blazkho effect,][]{blazhko1907} and internal properties. For recent reviews on classical oscillators see e.g. \citet{balona2010,handler2013,szabo2015} and references therein.

Lower-amplitude oscillations were not detected for several more decades. 
In the 1960's oscillations were  first  discovered in the Sun \citep{leighton1962}. Subsequent observations provided details of the solar interior and constraints for general stellar modelling beyond that previously possible  \citep{jcd1996,jcd2002}. Seismology was pivotal in the solar neutrino problem \citep{bahcall1972,trimble1973} and confirmed that the solution lies within particle physics \citep{elsworth1990sn}.  Through sensitive neutrino detections, \citet{mcdonald2001} later showed that  low energy neutrinos do indeed change flavour. The Nobel Prize 2015 for physics was awarded for this discovery.  Unfortunately, our understanding of the Sun remains incomplete.  The  `new solar abundances' \citep{asplund2005,asplund2009} result in a solar structure that deviates much further from that obtained from the oscillations \citep{bahcall2005,guzik2008,basu2008,basu2013} as compared with the `old solar abundances' \citep{grevesse1993,grevesse1998}. Additionally, the 11-yr solar activity cycle discovered by \citet{schwabe1843} and visible in the solar oscillations \citep{elsworth1990sc,howe2002,hathaway2015} is not fully understood. This includes dynamo processes involving interaction between rotation and convection, restructuring the magnetic field between the toroidal and poloidal components, in a manner that is yet to be resolved. This also gives rise to surface manifestation such as sunspots when flux tubes break through the solar surface \citep[e.g.,][]{charbonneau2014,cameron2015}. Additionally, the reasons for the long solar minimum between cycle 23 and 24 are still subject of discussion \citep{basu2012,basu2013solmin, jiang2015}. Furthermore, oscillations sensitive to the activity suggest the presence of a 2-yr cycle in addition to the 11-yr cycle \citep{broomhall2012} which is yet to be understood.

The road to the detection of low-amplitude oscillations in stars other than the Sun was paved by radial-velocity measurements -- much like  the variations used to discover pulsations in the Sun.  Ground-based spectroscopic surveys first identified excess  oscillation power in Arcturus \citep{smith1987} and Procyon \citep{brown1991} as well as the first evidence for individual modes in $\eta$ Bootis \citep{kjeldsen1995b,kjeldsen2003}, $\alpha$ Centauri A \citep{bouchy2001}, $\xi$ Hydrae \citep{frandsen2002}, and $\mu$ Her \citep{bonanno2008}. These seminal discoveries relied on  single-site spectroscopic programs and inspired  longer multi-site spectroscopic campaigns on nearby bright stars, such as $\alpha$ Cen A \citep{bedding2004,butler2004} and B \citep{carrier2003,kjeldsen2005}, $\beta$ Hydrae \citep{bedding2007}, $\nu$ Indi \citep{carrier2007}, $\eta$ Serpentis \citep{deridder2006} and Procyon \citep{hekker2008,arentoft2008,bedding2010procyon}.  Complementary photometric multi-site campaigns were also pursued  \citep{stello2006M67I,stello2007} and although both strategies revealed detailed {\orange oscillation patterns}, they were yet unable to fully constrain the internal structure of the stars \citep{miglio2005,huber2011procyon}. 

Space-based instruments contributed to the progression of the field, with the star-tracker of the WIRE (Wide field InfraRed Explorer) mission \citep{buzasi2002,stello2008} and
the Hubble Space Telescope \citep{gilliland2008} 
both detecting low-amplitude oscillations in other stars.  Soon after, the  era of dedicated photometric space-based missions heralded a revolution for asteroseismology.   Observations with the MOST  mission \citep[Microvariability and Oscillations of STars;][]{matthews2000} have contributed significantly to the revolution of red-giant asteroseismology \citep{barban2007,kallinger2008,kallinger2008HD20884} and our understanding of classical pulsators such as $\delta$ Scuti stars \citep{casey2013}, Slowly Pulsating B stars \citep[SPB;][]{aerts2006,jerzykiewicz2013}  and rapidly oscillating Ap stars \citep{huber2008,gruberbauer2011}, as well as other pulsators such as pre-main sequence stars \citep{zwintz2013} and Wolf-Rayet stars \citep{david-uraz2012}. CoRoT \citep[Convection, Rotation and planetary Transits;][]{baglin2006} and \textit{Kepler} \citep{borucki2008} have contributed greatly to the detection of pulsations in, and understanding of, many different kinds of stars such as; massive stars \citep{belkacem2009,degroote2010,kurtz2015}, RR Lyrae stars \citep{kolenberg2010} and sub-dwarf B stars \citep{Ostensen2014}. Additionally, many break-through results have been reported on low-amplitude oscillations in Sun-like stars, subgiants and red-giant stars (see recent reviews by \citet{chaplin2013,hekker2013rev,mosser2016} and references therein).

\subsection{Observations of stellar oscillations}
\label{sec:introobs}
\begin{figure}
\centering
\begin{minipage}{\linewidth}
\centering
\includegraphics[width=\linewidth]{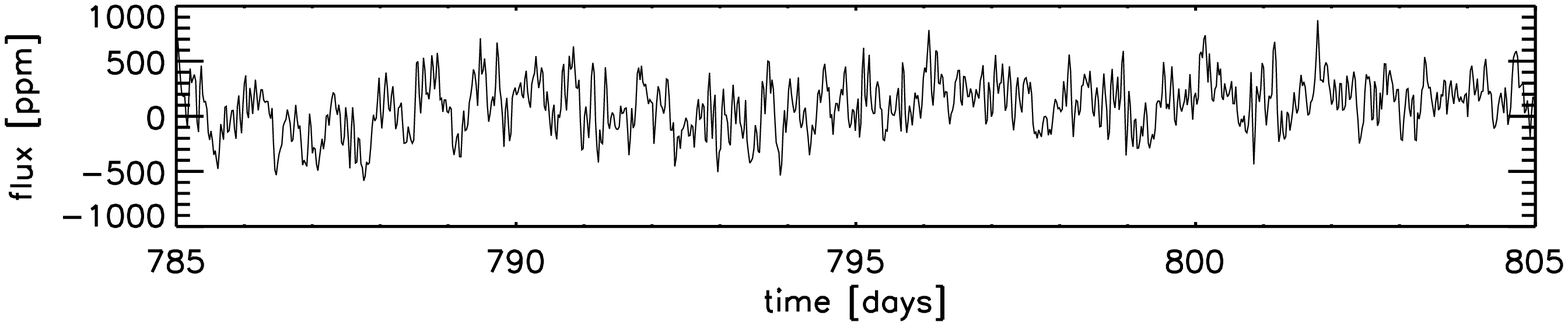}
\end{minipage}
\begin{minipage}{\linewidth}
\centering
\includegraphics[width=\linewidth]{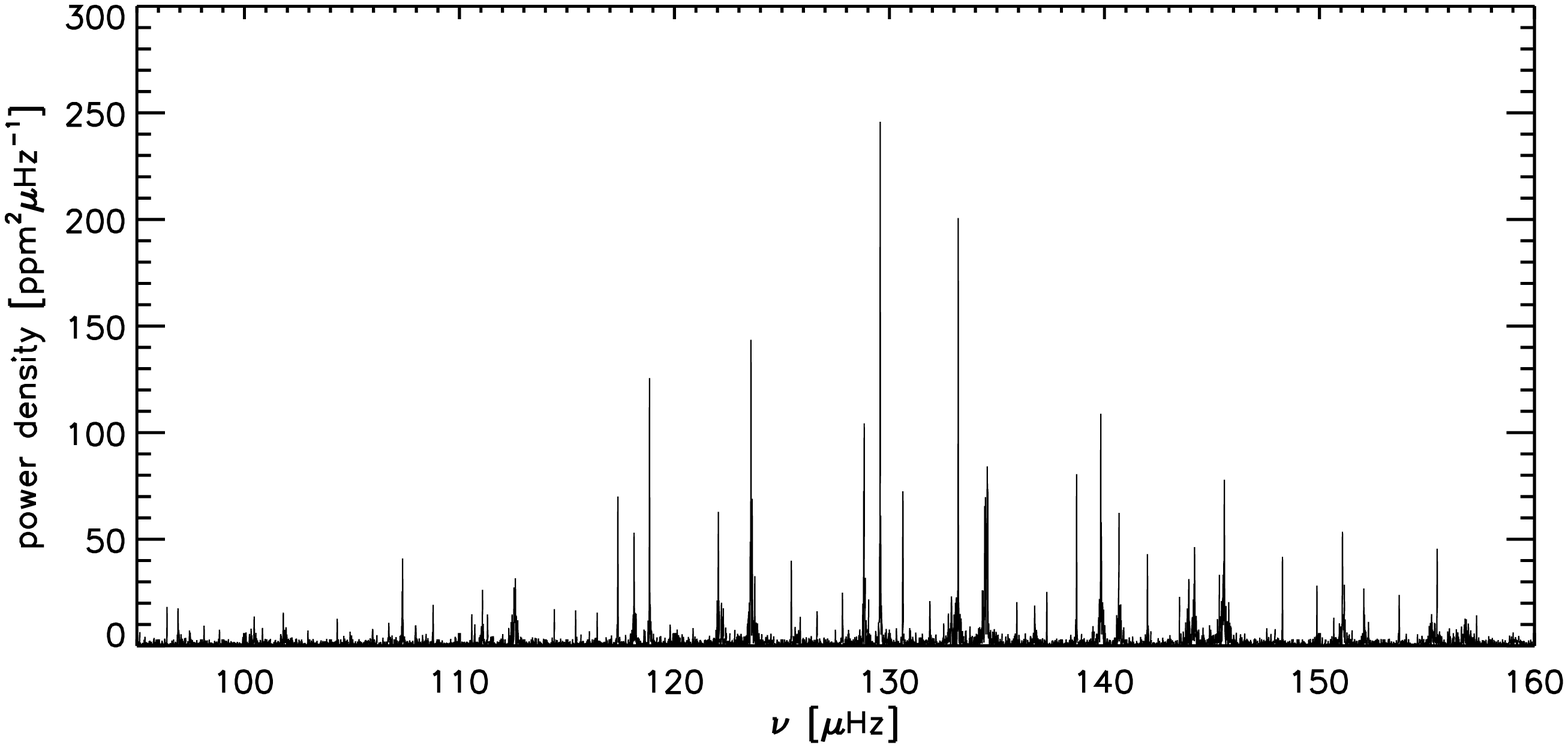}
\end{minipage}
\caption{A 20 day long subset of a \textit{Kepler} photometric timeseries (or light curve) of a red-giant star (KIC 9145955, top panel) and the Fourier power spectrum of the same star using a 1060-days long dataset (bottom panel). Note that for the Fourier power spectrum only the frequency range in which oscillations occur is shown here.}
\label{fig:ts+ps}
\end{figure}

Oscillations can be determined from timeseries data of either intensity variations or radial-velocity (RV) variations. Intensity variations reflect the brightness variations of a star induced by stellar oscillations. RV variations reveal the outward and inward movement of the stellar surface due to stellar oscillations through a Doppler shift of the spectrum. Part of a timeseries of intensity variations of a red-giant star observed with the \textit{Kepler} space telescope is shown in the top panel of Fig.~\ref{fig:ts+ps}. 

There are significant differences between intensity and RV variation measurements in the sense of their sensitivity to other intrinsic stellar features. For example, granulation (the visible effect of convection at a star's surface) has higher amplitude in intensity variations than in RV variations, relative to the oscillations. Additionally, stars other than the Sun can be observed only in integrated light, causing cancellation effects that differ between intensity and RV observations. Since solar-like oscillations are mainly in the radial direction, RV observations with the projection onto the line of sight have a reduced sensitivity to the oscillations near the limb, increasing the response to modes of slightly higher degree and hence the diagnostic potential, compared with intensity observations.

To extract oscillation frequencies the timeseries data are most commonly transformed to frequency space by a Fourier transform. The resulting Fourier power spectrum reveals the oscillation frequencies as sharp peaks.  
An example of a Fourier power spectrum of a red-giant star observed with the \textit{Kepler} space telescope is shown in the bottom panel of Fig.~\ref{fig:ts+ps}. \newline
\newline
\indent In this review we will discuss the internal structure and structure changes of subgiants and red giants including AGB stars in more detail (Sect. 2). These stars oscillate with intrinsically damped oscillations stochastically excited by convection in the outer layers of the stars, i.e., some of the convective energy is converted into energy of eigenmodes of the star. Consequently, these oscillations allow for observational investigations of internal stellar structure. As such oscillations are present in the Sun, they are referred to as \textit{solar-like oscillations}. Solar-like oscillations are expected to be present in all stars with turbulent outer layers. 
In Section 3 we discuss the diagnostics that can be obtained from timeseries data and Fourier power spectra. An overview of  stellar pulsation theory is presented in Section 4. We highlight ground-breaking results from the past years in Section 5. Finally, we discuss some promising prospects of asteroseismology and stellar structure of giants in Section 6.

\def\note #1]{{\bf #1]}}
\section{Giant star evolution}
\begin{figure}
\centering
\begin{minipage}{\linewidth}
\centering
\includegraphics[width=\linewidth]{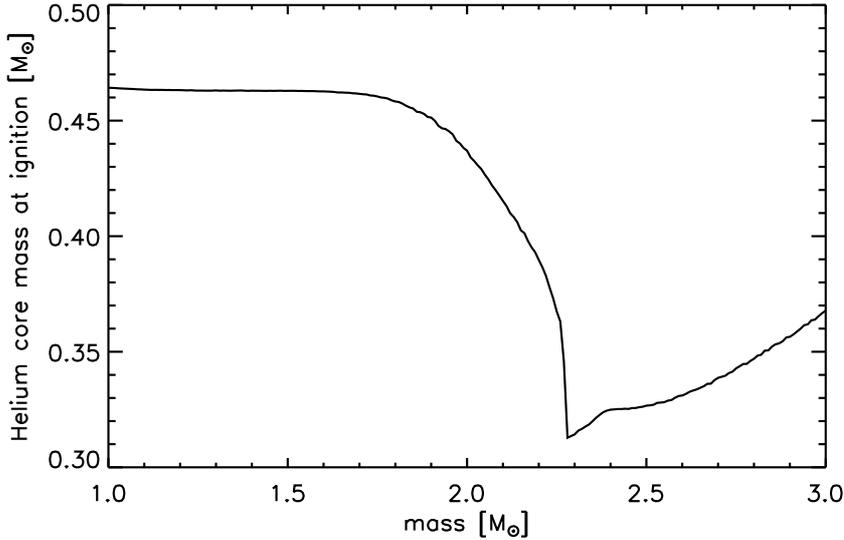}
\end{minipage}
\caption{Helium core mass at ignition vs. stellar mass for stellar models of solar metallicity computed with the MESA stellar evolution code \citep{paxton2011}.}
\label{fig:Hecoremass}
\end{figure}

Here we provide an overview of the internal structures of low- to intermediate-mass stars (roughly 0.8 to 10 M$_{\odot}$) in their respective evolutionary stages. We discuss the low-mass stars and intermediate-mass stars separately. \textit{Low-mass stars} are those stars that ignite helium in the core under degenerate conditions. This occurs in stars with masses between $\sim$0.48~M$_{\odot}$ and $\sim$2~M$_{\odot}$. The lower limit is defined by the lower limit of the critical mass needed to ignite helium, while the upper limit depends on the chemical composition of the stars. If mass loss is small enough, and time long enough, all low-mass stars will go through the giant phases discussed below. At the current age of the universe ($\sim$14 Gyr) only stars with $M>0.8\,{\rm M}_\odot$ will have reached these late stages of evolution, unless substantial mass transfer in a binary system has taken place. \textit{Intermediate-mass stars} are stars that do not develop a degenerate core and have a more gentle onset of core-helium burning. These stars range in mass between $\sim$2~M$_{\odot}$ and 8-10~M$_{\odot}$ depending on metallicity \citep{kippenhahn2012}. For stars with degenerate cores the helium-core mass at ignition is the same, while this decreases for more massive stars without degenerate cores (see Fig.~\ref{fig:Hecoremass}). To illustrate the internal structure changes of low- and intermediate-mass stars we show the paths of a 1~M$_{\odot}$ and a 3~M$_{\odot}$ model in the Herzsprung-Russel diagram (Fig.~\ref{fig:HRD}) and so-called Kippenhahn diagrams of these models in Figs~\ref{fig:kip1} and \ref{fig:kip3}.   

The description provided here is aimed to provide an insight in the many internal structure changes that a giant star undergoes. For a more complete picture and details of stellar evolution we refer the reader to \citet{kippenhahn2012}.

\subsection{Low-mass stars ($M \lesssim 2$~M$_{\odot}$)}

\subsubsection{End of Hydrogen core burning phase: end of main sequence}
For stars with masses below 1-1.5~M$_{\odot}$ (depending on chemical composition) 
hydrogen (H) in the central regions fuse to helium via proton-proton (pp) chains under conditions of radiative energy transport. Conversely, stars with masses above this range develop a convective core on the main-sequence, with the CNO (Carbon-Nitrogen-Oxygen) catalytic reactions predominately responsible for the conversion of H into $^4$He. CNO burning takes place at higher temperatures ($T$) than proton-proton interactions. Additionally, the energy production ($\epsilon$) of CNO burning is more temperature-dependent than that of proton-proton chains \citep[$\epsilon_{\rm CNO} \sim T^{17}$ vs. $\epsilon_{\rm pp} \sim T^4$;][]{iliadis2007}. The respective conditions in the core significantly impact the structure and  evolution at the end of the main sequence and therefore the two regimes, i.e. stars with \textit{radiative cores} and \textit{convective cores} on the main sequence, are discussed separately.

\paragraph{Stars with radiative cores}
For stars with a radiative core the fusion of H to $^4$He is predominantly enacted via the pp chains. 
The increase in the mean molecular weight ($\mu$) resulting from hydrogen fusion affects hydrostatic equilibrium. To sustain pressure support to the overlying layers of the star the core must contract. This increases the temperature in the inner parts of the star, leading to an increase in the efficacy of energy transport and hence in the luminosity ($\sim$0.7\% increase in brightness every 100~Myr in a 1~M$_{\odot}$ star). This is matched by a corresponding increase in the energy generation. 

The gradual depletion of H from the radiative core generates a smooth transition to
a chemically inhomogeneous structure with hydrogen burning in an extended shell
around a growing inert degenerate helium core. Complete depletion of hydrogen in the centre
marks the end of the main-sequence and the transition to the hydrogen shell burning
phase.

\begin{figure}
\centering
\begin{minipage}{\linewidth}
\centering
\includegraphics[width=\linewidth]{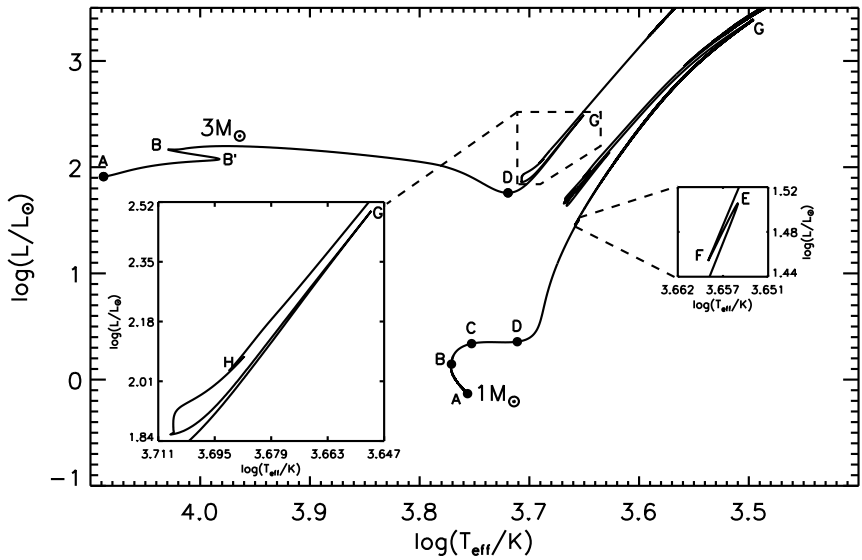}
\end{minipage}
\caption{Herzsprung-Russell diagram of a 1~M$_{\odot}$ and a 3~M$_{\odot}$ model. The insets show the "bump" of the 1~M$_{\odot}$ model (see Section~\ref{par:bump}) and the helium core burning phase for the 3~M$_{\odot}$ model (see Section~\ref{sect:imHeburn}). The models are computed using the MESA stellar evolution code \citep{paxton2011} with solar metallicity. The letters indicate different phases of evolution: A = zero-age main-sequence; B$'$ = core hydrogen mass fraction $\approx 0.05$, B = start of thick shell burning; C = maximum extent of thick shell (in mass); D = start of thin shell burning; E = maximum bump luminosity; F = minimum bump luminosity; G = tip of the red-giant branch; H = end of helium-core burning, and correspond to the letters indicated in Figs~\ref{fig:kip1} and \ref{fig:kip3}.}
\label{fig:HRD}
\end{figure}

\paragraph{Stars with convective cores}
\label{sect:cc}
There is a physical limit to which energy can be transported by radiation. The Schwarzschild criterion \citep{schwarzschild1906} states that if the temperature gradient inside a star is too steep, convection will take over as the primary means of energy transport.  In addition to efficient energy transport, the associated bulk mass motions of convection ensure that the composition of any convective region is well mixed. In a more formal sense convection is activated once the radiative temperature gradient exceeds the  adiabatic temperature gradient \citep{schwarzschild1906}.  

The Schwarzschild criterion defines the likely regimes in stellar interiors in which convection will develop. The first regime is where there is a large energy flux, the second where the stellar material is opaque to photons {\orange (such as in ionization regions of abundant elements)} and energy transfer by radiation is therefore inefficient. In stars with $M \gtrsim 1.1$~M$_{\odot}$ the central conditions are sufficient for the activation of the CNO cycle -- reactions that are highly temperature dependent. A consequence for these stars is that the burning region becomes ever more centrally concentrated, and  the large energy flux and steep temperature gradient drives a convective core.  

An alternative to the Schwarzschild criterion is the condition derived by \citet{ledoux1947} which, in addition to the temperature gradients, takes into account the spatial variation of the mean molecular weight. In some stars, particularly those with $M \gtrsim 2.25$~M$_{\odot}$, a composition gradient may develop outside the shrinking (Ledoux) convective core. Application of the Schwarzschild criterion would render this region convectively unstable, however according to Ledoux, the presence of a molecular weight gradient can have a stabilizing effect. The true behaviour of material in the stellar interior under these circumstances remains unclear \citep{gabriel2014}. However, in stellar evolution codes some form of slow (slow compared to convection) mixing is applied in these regions in order to match various observational constraints \citep[see e.g.][]{lattanzio1983, langer1985}. This slow mixing process can also be referred to as \textit{semi-convection}.

In stars below $\sim 2 \Msun$ with a convective core, the core grows for most of the main-sequence lifetime. This is caused by the increasing contribution to the energy production from the highly temperature-sensitive CNO cycle, as oxygen is gradually converted to nitrogen in parts of the cycle. A further extension of the mixed core can be caused by material that approaches the boundary of stability with momentum and overshoots into the radiative layer. This process, \textit{convective-core overshoot}, extends the burning region and brings in fresh H-rich fuel prolonging the main-sequence evolution. As with a radiative core the decrease in hydrogen abundance with nuclear burning and the resulting increase in $\mu$ causes an increase in luminosity. 
However, with a convective core this depletion takes place uniformly in an extended region.
When hydrogen is nearly depleted in this region ($X_{\rm core} \approx 0.05$, B$'$ in Fig.~\ref{fig:HRD}) the star contracts to maintain the energy production. This contraction leads to an increase in effective temperature ($T_{\rm eff}$) and luminosity ($\cal L$) until hydrogen is completely depleted in the centre (the \textit{``hook"} in the Hertzsprung-Russell diagram: B$'$-B in the 3M$_{\odot}$ track in Fig.~\ref{fig:HRD}).  At this point the central burning and convection cease abruptly. This is the end of the main-sequence phase.

\subsubsection{Hydrogen shell burning phase: subgiants and red-giant branch stars}
\label{sect:Hburn}

\paragraph{Subgiants} Without the central nuclear reactions the star must find an alternative way to generate energy to compensate for the energy loss from the core. The star turns to another available source of energy in core contraction and the corresponding release of gravitational potential energy. For stars with $M \lesssim 1.1$~M$_{\odot}$ the central density is large enough that electron degeneracy dominates and provides significant pressure support. Therefore, low-mass stars can remain in thermal and hydrostatic equilibrium with a degenerate, isothermal core as they smoothly transition to hydrogen-shell burning. As a consequence the contraction phase and transition to giant is gradual and the timescale much longer compared with the higher-mass counterparts (see Section~\ref{sect:Hg}). Shell burning is initially in the form of an extended burning region outside the core. Slowly, the core mass increases as the ashes of hydrogen-shell burning are deposited on the He core. As the core is degenerate this corresponds to a reduction in radius, accompanied by an expansion of the envelope. The boundary where contraction changes to expansion is located near the hydrogen-burning shell. This behaviour is typical of a more general exchange in evolving stellar models between contraction and expansion at shell-burning sources,  referred to as the \textit{mirror principle}.

The shell-burning region, dominated by the CNO cycle, is confined to increasingly narrow mass. The envelope expands and cools while the star evolves from the main sequence towards the \textit{Hayashi line}. The Hayashi line is the locus in the Herzsprung Russell diagram of fully convective stars, where a star cannot decrease its temperature further (otherwise it cannot maintain hydrostatic equilibrium). Thus, upon approaching the Hayashi line further increase of the radius causes an increase in luminosity. A large convective region develops in the envelope due to the increased photospheric opacity at lower temperatures (contribution from H$^-$ ions). The star is now on the red-giant branch, just on the hot side of the Hayashi line with a large convective envelope on top of a small radiative core.

\begin{figure}
\centering
\begin{minipage}{\linewidth}
\centering
\includegraphics[width=\linewidth]{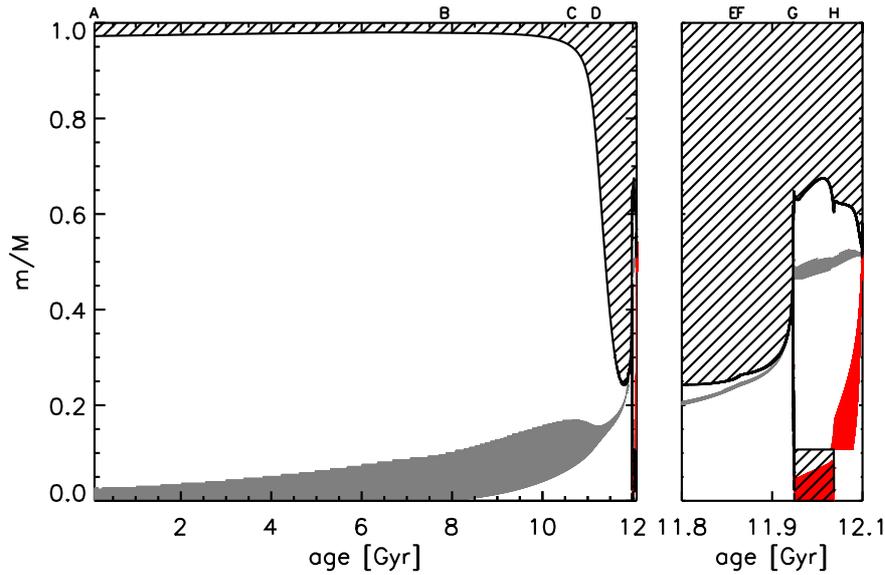}
\end{minipage}
\caption{Kippenhahn diagram of a 1~M$_{\odot}$ model shown in Fig.~\ref{fig:HRD}. Regions where convection takes place are hatched. Regions where nuclear burning produces more than $10 \,{\rm erg\,g^{-1}\,s^{-1}}$ are shown in grey for hydrogen burning and in red in case of helium burning. The right panel shows a zoom of the late stages of evolution. The letters at the top indicate different stages of evolution corresponding to the phases indicated in Fig.~\ref{fig:HRD}.}
\label{fig:kip1}
\end{figure}

\paragraph{Red-giant branch stars} On the red-giant branch, H-burning ashes are deposited on the degenerate core as the shell burns and moves outwards. Due to the degenerate conditions an increase in mass results in contraction and heating of the inert helium core. This also heats the hydrogen-burning shell, which reacts to the higher temperatures by compressing the burning region and increasing the energy generation. The density contrast becomes so large that the envelope and core are essentially decoupled. Therefore, the hydrogen burning in the shell is completely determined by the properties of the helium core and not by the envelope. Hence, the luminosity of the star is now related to the mass of the inert degenerate helium core and does no longer depend on the total mass of the star. Therefore, stars with the same core mass, but a spread in total mass, follow the same path in the Hertzsprung-Russell diagram.

\paragraph{First dredge-up}
\label{par:1dredge}
The convective envelope penetrates deep into the star to the regions where the chemical composition has been altered by nuclear processes that took place during the main sequence. The processed material is then subsequently transported to the surface. This is the \textit{first "dredge-up"} phase, i.e. chemical elements from deeper layers are dredged up towards the surface of the stars: for example the $^{12}$C/$^{13}$C ratios are lowered. The convective region reaches a maximum depth in mass and recedes because of the advance of the hydrogen-burning shell, leaving behind a chemical (mean molecular weight) discontinuity.

\paragraph{Bump}
\label{par:bump}
The hydrogen shell, in which burning takes place, moves gradually outwards (in mass) while the helium-core mass steadily increases. In a simplified picture the luminosity decreases when the hydrogen-burning shell reaches the chemical discontinuity left behind by the deepest extent of the convective envelope due to the decrease in the mean molecular weight at the chemical discontinuity,
causing the luminosity to decrease again following $\mathcal{L}\propto\mu^7M_{\rm core}^7$ \citep{refsdal1970}. After this the core mass keeps increasing at constant $\mu$ causing a resumption of the increase in the luminosity.

In fact, the situation is more complex with the luminosity beginning to decrease prior to the shell burning through the discontinuity.
As discussed by \citet{jcd2015} the reason is that the decrease in $\mu$
above the discontinuity starts affecting the hydrostatic structure,
and hence the temperature, within and above the hydrogen-burning shell
before it reaches the discontinuity. This causes the decrease in the luminosity.
This zig-zag in the evolution path is the so-called \textit{RGB-bump} (see right inset in Fig.~\ref{fig:HRD}). The bump is visible for stars up to about 2.2~M$_{\odot}$ as an over-density of stars in stellar clusters at the bump luminosity. For more massive stars helium-core burning starts before the hydrogen-burning shell approaches the composition discontinuity left behind by the first dredge-up. In these cases a bump-like structure is not present on the red-giant branch.

\paragraph{High-luminosity red-giant branch stars}The process on the RGB continues till the core reaches a temperature of $\sim$$10^8$~K (at a core mass of $\sim$0.45~M$_{\odot}$) at which helium is ignited in a thermal run-away process. This is the so-called \textit{helium flash}. 
We note here that additional mixing processes, such as thermohaline mixing \citep[e.g.][]{eggleton2006,eggleton2008,charbonnel2007,charbonnel2010,angelou2011,angelou2012}, are necessary to include in models of stars ascending the red-giant branch to match the observed chemical compositions of these stars.

\subsubsection{Onset of Helium burning: He-flash}
\label{sect:He-flash}
At a temperature of $\sim$$10^8$~K in the inert helium core, helium fusion can be ignited in a triple-alpha process.\footnote{Interestingly, \citet{hoyle1954} predicted that this could only occur if carbon possessed a resonant state, i.e. a state with a very particular energy, which we now know is true.}  In the highly degenerate core the pressure does not depend on the temperature and hence there is no thermostatic control to expand and cool the core. The onset of (unstable) burning in these degenerate conditions results in a thermal runaway process creating for a very short time (of order a few hours!) an enormous overproduction of nuclear energy. This energy is absorbed by the expansion of non-degenerate layers outside the degenerate core and does not reach the stellar surface.  

The onset of helium fusion takes place at the location of maximum temperature. The temperature is generally at its maximum in a concentric shell around the centre of the degenerate core due to gravo-thermal energy and neutrino losses. Stellar models predict that the first main helium flash is followed by a series of sub-flashes. Each subsequent sub-flash is located closer to the centre such that eventually the degeneracy in the centre is completely lifted and the star is back in equilibrium with helium burning in a convective core. As neither the energy of the flash, nor the energy of the subflashes reaches the stellar surface, the existence of sub-flashes in real stars is not observationally confirmed, i.e., they could be artefacts of stellar models.

\subsubsection{Helium core burning phase: red clump}
\label{sect:Hecoreburn}
The star has now two sources of energy generation: hydrogen burning in a shell around the core producing helium, while in the core helium is consumed to produce carbon and oxygen. Due to the expansion and accompanying decrease in density and temperature of the hydrogen-burning shell after the helium-flash it generates less energy (this is however still the main source of energy generation in the star). Therefore, the luminosity decreases while the core expands and the envelope contracts (mirror principle, see Section~\ref{sect:Hburn}). The star is back in equilibrium and settles in the red clump. 

All stars that have gone through a helium flash have very similar core masses and hence very similar luminosities on the horizontal branch. Therefore, all stars that go through a He-flash end up at a very similar spot in the Herzsprung-Russell diagram, with only some dependence on $T_{\rm eff}$ owing to their total masses (with lower masses being slightly hotter) and composition (stars with higher contents of heavier elements are cooler).

\paragraph{Properties of the convective core} The helium core burning stars have a central convective core that  becomes enriched in carbon and oxygen during helium burning. The opacity in the temperature-density regime present in the core is dominated by free-free transitions and increases with increasing carbon and oxygen abundance. This causes an increase in the radiative temperature gradient in the core leading to a discontinuity in the radiative temperature gradient at the boundary between the convective core and the radiative envelope. In models with no mixing beyond the convective core the nuclear burning gives rise to an increasing composition discontinuity at the edge of the core. Convective overshoot and/or semiconvection increases the size of the core and, depending on the implementation, may lead to a smooth composition profile or further discontinuities in composition \citep[see][for an overview of these processes]{constantino2015}. As discussed in Sections \ref{sect:glitchestheory} and \ref{sec:redclump} the detailed properties of the composition profile can have a strong effect on the behaviour of the oscillations of the star. Towards the end of the central helium burning sudden mixing in the models between the carbon-oxygen-rich core and the overlying helium-rich layers may occur at the edge of the core. This leads to an abrupt increase in the helium content of the core and a loop in the Herzsprung-Russell diagram. These are so-called \textit{breathing pulses}. The occurrence and appearance of these breathing pulses in stellar models depend on the criterion used for convection and may only be an artefact of the way convection is included in models and may not be present in real stars. 

After some time helium is exhausted in the convective core and the star will undergo some rapid evolution towards a shell-burning phase with burning taking place in a helium shell and hydrogen shell surrounding the core. This phase of evolution is the asymptotic giant branch (AGB).

\subsubsection{Helium and Hydrogen shell burning phase: Asymptotic giant branch}
\label{sect:AGB}
In the low-mass regime, the asymptotic giant branch is characterized by an inert carbon-oxygen core surrounded by two burning shells of which the helium shell is thermally unstable. In this phase a star is again moving in the Herzsprung-Russell diagram towards the Hayashi line and at the same time increasing its luminosity and radius. In the early AGB (E-AGB) phase hydrogen is burning outwards and the temperature in this shell drops. Consequently, the hydrogen-burning shell supplies only a small fraction of the energy for some time. However, as the temperature in the hydrogen shell increases again in between thermal pulses (see below) burning is recovered and dominates the energy production.

\paragraph{Thermal pulses}
\label{sect:TP}
As the star ascends the asymptotic giant branch, the helium-burning shell narrows while providing most of the energy to the stellar surface. Eventually the helium-burning shell advances in mass towards the hydrogen-burning shell and their separation, the inter-shell region, becomes too narrow. This, along with the high temperature dependence of the helium-burning reactions, results in a thermal runaway and the onset of the \textit{thermally pulsing (TP) AGB phase}. 

The thermally pulsing AGB phase is characterized by long periods of quiescent hydrogen-shell burning, followed by instabilities of the helium-burning shell \citep{schwarzschild1965, weigert1966}. Each instability or \textit{`helium-shell flash'} grows in amplitude for the first  5-10 pulses before approaching a maximum in helium luminosity. As per \citet{iben1981} we describe the TP-AGB cycle by four distinct phases:
\begin{itemize}
\item On phase: The sudden deposition of energy from the shell flash drives an inter-shell convection/burning zone cycling the products of the triple-alpha process into the region below the hydrogen shell. 

\item Power down phase:  Although the carbon-oxygen core is highly electron degenerate the shell-flash instability occurs in a non-degenerate region of the stellar interior. The energy generated by the flash helps drive expansion of the star and thereby extinguishing the hydrogen-burning shell. The flash is able to generate luminosities of the order 10$^7$ or 10$^8$ solar luminosities. However, due to the expansion this increase in luminosity is not manifested at the surface.  

\item Third dredge-up\footnote{Note that no second dredge-up {\orange takes} place in low-mass stars. The second dredge-up for intermediate-mass stars is described in Section~\ref{sect:IMAGB}.}: As the star expands and cools, convection is able to penetrate beyond the hydrogen-burning shell into regions homogenised by the inter-shell convection zone. Hence the products of He burning are mixed into the envelope where they can be observed at the stellar surface. 

\item The inter-pulse phase: Eventually the helium-burning luminosity drops below the surface luminosity and the outer regions can once more contract. The hydrogen shell can reignite where it provides most of the luminosity until once again the interior conditions arise for a successive helium-shell instability. The inter-pulse phase lasts for $\approx$$10^4$ years.
\end{itemize}

\paragraph{Third dredge-up}
Third dredge-up plays an important role in the chemical enrichment of the galaxy. 
The significant amount of carbon produced in these stars and brought to the stellar surface is in some cases able to raise the C/O  abundance ratio to become larger than unity. In addition  TP-AGB stars have been identified as a site of the s-process nucleosynthesis\footnote{slow-neutron-capture-process: a nucleosynthesis process that occurs at relatively low neutron density and intermediate temperature conditions}, which is responsible for the production of half of the elements beyond iron. Through efficient mass-loss processes these elements are expelled into the interstellar medium. 

\paragraph{Post-thermally-pulsing AGB phase}At some point the envelope mass is insufficient to allow TP to continue.  When the envelope mass drops below about a few ($\sim$5) per cent of the total mass, the envelope contracts and shell burning extinguishes: the star becomes a white dwarf.
This is the post-AGB phase. The number of thermal pulses the star experiences and
the final white dwarf mass depend on the competition between mass loss and core growth in the
AGB phase.
For extensive reviews of the AGB stars we refer to \citet{iben1983}, \citet{herwig2005} and \citet{karakas2014}.

\subsection{Intermediate-mass stars ($M\sim$ 2-10~M$_{\odot}$)}

\begin{figure}
\centering
\begin{minipage}{\linewidth}
\centering
\includegraphics[width=\linewidth]{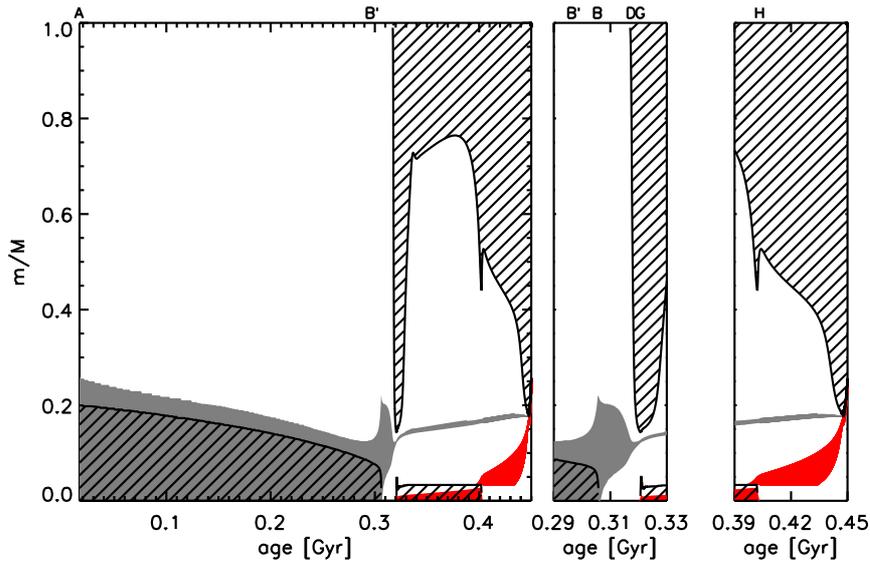}
\end{minipage}
\caption{Kippenhahn diagram of a 3~M$_{\odot}$ model as shown in Fig.~\ref{fig:HRD}. Regions where convection takes place are hatched. Regions where nuclear burning produces more than $10 \,{\rm erg\,g^{-1}\,s^{-1}}$ shown in grey for hydrogen burning and in red in case of helium burning. The central panel shows a zoom of the subgiant and red-giant branch. The right panel shows a zoom of the late stages of evolution.The letters at the top indicate different stages of evolution corresponding to the phases indicated in Fig.~\ref{fig:HRD}.}
 \label{fig:kip3}
\end{figure}

\subsubsection{End of Hydrogen core burning phase: end of main sequence}
Intermediate-mass stars have a convective core on the main sequence and show the same behaviour as low-mass stars with a convective core as described in section~\ref{sect:cc}.

\subsubsection{Hydrogen shell-burning phase: subgiants and red-giant branch stars}
After a short resettling at the end of the main sequence, hydrogen-shell burning intensifies in the region around the core. As for the low-mass star this shell burning steepens the hydrogen profile at the edge of the He-core leading to a narrowing of the burning shell when the lower hydrogen tail has been consumed. The core contraction and envelope expansion below and above the shell burning layer respectively (mirror principle) increase the radius of the star. 

\paragraph{Hertzsprung gap}
\label{sect:Hg}
In intermediate-mass stars the core is non-degenerate following central hydrogen exhaustion. \citet{schoenberg1942} demonstrated that there is a maximum relative core mass that an isothermal, non-degenerate core can have whilst maintaining hydrostatic and thermal equilibrium and deriving all its energy from a nuclear burning shell source. Without electron degeneracy to supply additional pressure support, contraction on a Kelvin-Helmholtz timescale will develop when the core mass exceeds this Sch\"onberg-Chandrasekhar limit, so that the star can maintain equilibrium between the energy it generates in the interior and that which it loses at the surface.  Whether stars in the intermediate-mass range reach the Sch\"onberg-Chandrasekhar limit depends on their hydrogen-exhausted core mass. This mass depends on the amount of overshoot the convective core experienced on the main sequence. 
Upon reaching the Sch\"onberg-Chandrasekhar limit stars cross the Hertzsprung-Russell diagram rapidly and move onto the red-giant branch. The fast timescales involved in this phase of evolution leads to a dearth of intermediate-mass stars observed in this region of the HR diagram:  the so-called \textit{Hertzsprung gap}.

At the bottom of the red-giant branch, intermediate-mass stars develop a deep outer convective region to transport energy more efficiently and prevent the star from cooling beyond the Hayashi line. This deepening of the convective envelope causes a change in the surface chemical composition due to the first dredge-up (Section \ref{par:1dredge}). At the same time the core continues to contract and heat. When the core temperature has increased to about $10^8$~K, helium is ignited. We note that the bump (Section \ref{par:bump}) is only present on the red-giant branch for stars with masses roughly below 2.2~M$_{\odot}$. More massive stars ignite helium before the hydrogen-burning shell reaches the chemical discontinuity left behind by the first dredge-up. 

\subsubsection{Onset of Helium burning: non-degenerate ignition}
In the non-degenerate regime, the luminosity at which helium ignites is a monotonically increasing function of the core mass.  Because the pressure and temperature are related, the thermostatic feedback allows intermediate mass stars to gently ignite helium in their core. 

\subsubsection{Helium core burning phase: secondary clump}
\label{sect:imHeburn}
Stars in the secondary clump have two sources of burning. Firstly, helium burning that produces carbon and oxygen is present in the core. Like the CNO reactions, energy production via the triple-alpha process is highly temperature dependent. The reactions are concentrated towards the centre and give rise to a convective core. Burning proceeds quiescently with the core growing as a function of time. Secondly, hydrogen burning is taking place in a shell around the core. The latter provides most of the total energy output as a rather small release of nuclear energy is sufficient in the core to compensate for the energy loss from the core and prevent the core from contracting. As the core masses of these stars can be different when they ignite helium, these stars do not ``clump" as the red-clump stars. Instead they form the secondary clump at lower luminosities and effective temperatures.  Additionally these stars loop through the red-giant region during central helium burning (see left inset in the HR Diagram (Fig.~\ref{fig:HRD}) where the stars leave the Hayashi line to become hotter and subsequently move back towards the Hayashi line). The temperature range that the loops cover increases with increasing stellar mass. The cause of these loops lies in the chemical composition profile in the central regions of stars that had a convective core on the main-sequence. 

\subsubsection{Helium and Hydrogen shell burning phase: Asymptotic giant branch}
\label{sect:IMAGB}
When helium burning terminates in the core the burning continues in two shells
around the inert carbon-oxygen core.  For stars with masses of 4-8~M$_{\odot}$ the hydrogen shell is at best barely active. This allows the convective envelope to penetrate down reaching layers through which the hydrogen shell has burned (similar as for low-mass stars on the red-giant branch). This so-called \textit{second dredge-up} brings processed material that is generated by helium and hydrogen burning such as carbon, oxygen, nitrogen and helium to the surface.  For all stars that do not ignite carbon in their core, the AGB proceeds in roughly the same fashion as for low-mass stars (see also Section~\ref{sect:AGB}). They experience episodic thermal pulses and efficient mass loss until they lose most of their envelope and become a white dwarf. However, their s-process nucleosythesis will differ greatly depending on stellar mass.

\subsection{Why do stars become giants?}
The description written above represents  our current understanding of stellar evolution based on stellar models and observations. However, the reason why stars become red giants is actually not understood. The mirror phenomenon mentioned before seems to play an essential role in stars to become a red giant. However, it is not understood what physical mechanism(s) drive the mirror nor what other physical mechanisms are essential in a star to become a giant. 

A number of studies have investigated the question `Why do stars become giants?' proposing reasons related to central gravitational field \citep{hoeppner1973,weiss1983}, the effective equation of state \citep{eggleton1991,eggleton1998}, gravothermal instability in the core \citep{iben1993}, thermal instabilities in the stellar envelope \citep{renzini1992} and mean molecular weight gradient \citep{stancliffe2009}. Some of these studies have been met with fierce opposition whilst others have devised conditions that were later shown to be necessary but not sufficient in all stars. Currently, it is clear that a strong gravitational field and a mean molecular weight gradient play an important role \citep{stancliffe2009} in stars for them to become a giant. For an extensive overview regarding the literature addressing `Why do stars become giants?' we refer the reader to \citet{sugimoto2000}, while \citet{faulkner2005} provided  a detailed analytical investigation inspired by the scientific legacy of Fred Hoyle.

\subsection{Rotation}
In the description of the internal structure of stars provided here, rotation has not been taken into account. However, it is plausible that all stars rotate as the clouds of gas and dust that they are formed from contain angular momentum. For slowly rotating stars it is generally assumed that second-order effects can be neglected and that the hydrostatic structure of the star is not affected. For low-mass stars on the main sequence and more evolved stars this will generally be the case. Intermediate-mass stars on and shortly after the main sequence may however be faster rotators before they slow down under the influence of magnetic braking. The rotation can have a strong impact on the thermal structure and radiative transfer, possibly inducing meridional flows and instabilities affecting mixing processes. Additionally, the shape of the star may become aspherical. Work on including rotation in stellar structure models has been performed by for example \citet{palacios2006,eggenberger2010,eggenberger2012}, based on earlier work by \citet{zahn1992}. A critical issue is the treatment of the evolution of the internal angular velocity, including transport of angular momentum, which currently fails to reproduce the seismically inferred internal rotation (see Section~\ref{sect:resrot}). 
One of the main conclusions for red-giant stars so far is that rotationally induced mixing and meridional circulation do not provide enough mixing of chemicals to explain the abundance anomalies observed around the bump luminosity of globular clusters (see also Section~\ref{sect:future}).
For a detailed analysis of the evolution of rotating stars, see \citet{maeder2009}.

\def\note #1]{{\bf #1]}}
\section{Seismic diagnostics}

Seismic diagnostics are by definition obtained from a signal that varies over time. In the context of oscillating stars, timeseries data (see top panel of Fig.~\ref{fig:ts+ps}) are most often taken from photometric fluxes or radial velocities (see Section 1). Some stellar parameters can be directly determined from the timeseries data. However, most seismic diagnostics are obtained from a Fourier transform (Figs~\ref{fig:ts+ps}, \ref{fig:ps_bg} and \ref{fig:ps_numax}) of the timeseries data. 

Important characteristics of the timeseries data are the total length or timespan of the data ($T$) and the typical time sampling ($\delta t$). These translate in Fourier space in the \textit{frequency resolution} $\delta\nu=1/T$, and the highest frequency at which one can reliably obtain results, i.e. the \textit{Nyquist frequency} $\nu_{\rm Nyq} = 1/(2\delta t)$. The Nyquist frequency is a hard limit for evenly sampled timeseries. However, astrophysical datasets are usually not exactly evenly sampled, which allows for measurements with higher frequencies, so-called super-Nyquist determinations \citep{murphy2013}. It was shown by \citet{eyer1999} that in cases of serious oversampling or undersampling the Nyquist frequency can be derived as $\nu_{\rm Nyq} = 1/(2p)$ with $p$ being the greatest common divisor of all differences between consecutive observation times. In practise a realistic estimate of $\nu_{\rm Nyq}$ in the case of unevenly sampled data is to use the inverse of twice the median of all time differences between two consecutive measurements in the entire timeseries \citep{aerts2010}.

The power in Fourier transforms is commonly normalized using \textit{Parseval's theorem}. This theorem states that the integral of the square of a function is equal to the integral of the square of its transform, i.e. the total power in the Fourier transform is equal to the total of the squared flux variations (for intensity) in the timeseries. Alternatively, Fourier transforms can be normalized using \textit{`peak-scaling'} in which the Fourier transform is normalized to recover the full sine-amplitude of an injected signal. The power can either be computed per bin, i.e. the frequency resolution, or per frequency unit. In the latter case it is the \textit{power density} that is shown which has the advantage that its value does not depend on the frequency resolution.

The fact that integration times of the observations are not infinitely short causes \textit{apodization} $\eta_a$. This affects the power at all frequencies with the largest impact close to the Nyquist frequency. The apodization can be accounted for by multiplying the power by $\eta_a^2$, where $\eta_a^2$ is defined as:\footnote{sinc$(x) \equiv 1 \mbox{ for } x=0$ and sinc$(x) \equiv \sin(x)/x \mbox{ otherwise}$ \label{fn:sinc}}
\begin{equation}
\eta_a^2= {\rm sinc}^2 \left [\pi/2\left(\frac{\nu}{\nu_{\rm Nyq}}\right)\right].
\label{apodization}
\end{equation}

Gaps in the timeseries data impact on features in the power spectrum. This \textit{window function}, i.e. the pattern of observations and gaps, causes alias frequencies (or sidelobes) to occur at $n(1/T_{\rm gap})$, with $n$ an integer and $T_{\rm gap}$ the typical time between gaps (for instance one day for ground-based single-site observations).  

When the data quality between the observations varies significantly, such as can happen with ground-based spectroscopic (multi-site) campaigns, one can opt to compute a weighted Fourier transform. This weighting can be performed to optimize the noise level, but also to optimize the window function and reduce the sidelobes caused by gaps \citep{arentoft2009}.

For a crash course on data analysis in asteroseismology including the statistics and uncertainties of timeseries data we refer the reader to \citet{appourchaux2014}.

\subsection{Variance and typical timescale in timeseries data}
\label{sect:rms}
For stars with a convective outer layer, such as low-mass dwarfs, subgiants and red giants, the variations in the flux are dominated by granulation and oscillations. As both granulation, i.e. the visible pattern of convection, and oscillations depend on surface gravity it has been possible to calibrate relations between the variance in the flux and surface gravity \citep{hekker2012,bastien2013} as well as between the typical timescales present in the timeseries data and surface gravity \citep{kallinger2016}.

\subsection{Background signal in Fourier spectrum}
\label{sect:bg}
For stars with a convective outer layer a \textit{frequency-dependent background signal} is present in the Fourier spectrum. This background consists of stellar intrinsic phenomena including activity features, such as spots and flares that are observable features of magnetic fields, rotation and granulation. In addition to intrinsic stellar background signal an observed power spectrum also includes white noise and instrumental effects. These instrumental effects can incorporate for instance degrading of CCDs but also incidental cosmic ray hits or telescope jitter. 

\begin{figure}
\centering
\includegraphics[width=\linewidth]{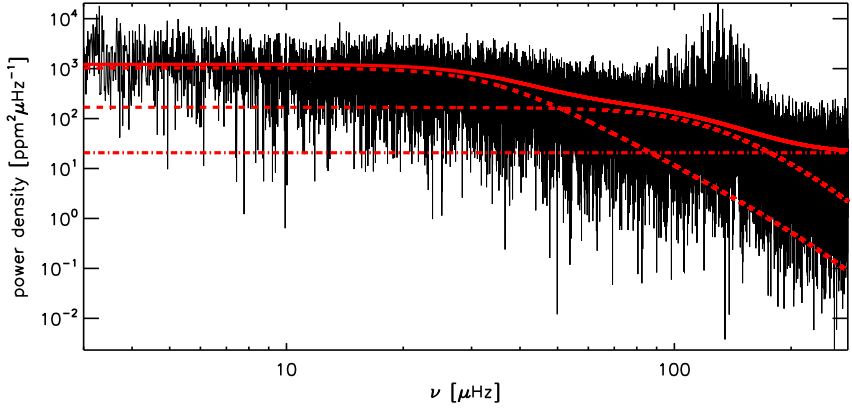}
\caption{Fourier power density spectrum of a red giant (KIC 9145955) in log-log space. The solid red line shows the background fit with the different background components shown with dashed lines. The white-noise level is indicated with a dotted-dashed line.}
\label{fig:ps_bg}
\end{figure}

All these features together form a background on top of which the oscillation modes are visible as relatively narrow peaks. This background can be fitted with the following function \citep[e.g.][]{harvey1985,kallinger2014}:
\begin{equation}
P(\nu)=P'_n+\eta_a(\nu)^2\sum_i \frac{ a_i^2/b_i}{1+(\nu/b_i)^{c_i}},
\label{background}
\end{equation}
where $P'_n$ is the white noise; $a_i$, $c_i$ and $b_{i}$ are for the $i$-th background component the rms amplitude, exponent, and the frequency at which the power of the component is equal to half its value at zero frequency (the characteristic frequency), respectively. The factor $\eta_a^2$ is the apodization defined in Eq.~(\ref{apodization}). The exponent $c_{i}$ provides a measure of the temporal correlation of the signal and determines the slope of the decay of the background in Fourier space \citep{mathur2011}. The number of functions ($i$) needed depends on the presence of activity, (super)granulation, or faculae (bright spots) on the star. Based on state-of the-art data, a two-component fit is in most cases necessary and sufficient \citep{karoff2013,kallinger2014}. An example of a background fit is shown in Fig.~\ref{fig:ps_bg}.

\subsection{Oscillation signal}
\label{sect:osc}
Oscillations establish themselves as a series of relatively narrow peaks on top of the background described in the previous subsection. The oscillations are confined to a limited range in frequency. Within this frequency range the individual modes of stochastic oscillations have a Lorentzian shape with a width that represents the lifetime of the mode and a height determined by the intrinsic amplitude of the mode and geometrical effects. In the limit of infinite lifetime, i.e. an oscillation that appears coherent over the timespan of the observations, the mode is unresolved and takes the form of a sinc$^2$ function (see footnote~\ref{fn:sinc}) in power. Each oscillation mode is characterized by its quantum numbers: \textit{radial order} $n$, related to the number of nodes in the radial direction (cf. Section \ref{sec:modeorder}); \textit{degree} $l$, the number of nodal lines on the surface and \textit{azimuthal order} $m$, the number of nodal lines crossing the stellar equator. The frequencies, width and amplitudes of the individual modes as well as the overall shape of, and patterns in, the oscillation power excess have valuable diagnostic power. Here, we first discuss the global features of the oscillation power excess, i.e. single measures that provide a diagnostic. We subsequently provide more details regarding the individual frequencies and the diagnostics that can be extracted from them. 

In red giants all non-radial modes have a mixed character, with a gravity-mode behaviour (buoyancy is the restoring force) in the core and an acoustic behaviour (pressure is the restoring force) in the envelope. Observationally, the acoustic behaviour is most prominent, and we first discuss the related observed properties. Afterwards we consider the more profound aspects of the star that are revealed by the mixed nature of the modes.

\begin{figure}
\centering
\includegraphics[width=\linewidth]{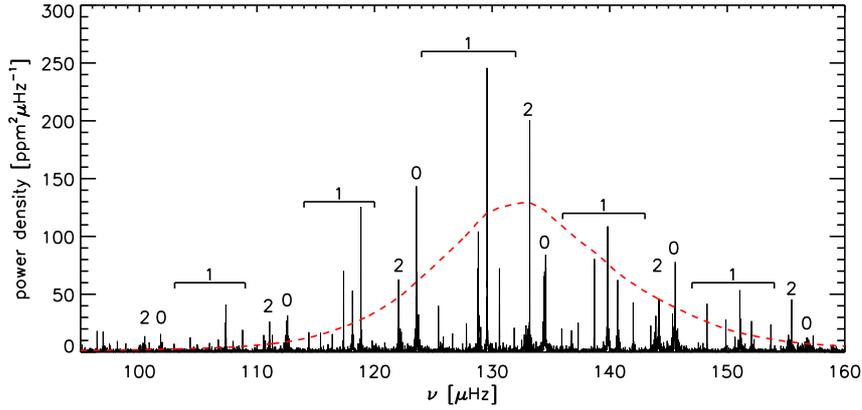}
\caption{Oscillations in the background-corrected power density spectrum of KIC~9145955. The numbers indicate the degree ($l$) of the modes (see Section~\ref{sect:indf}). The red dashed curve shows a heavily (triangular) smoothed power spectrum indicating the power excess envelope of the oscillations. The amplitude of the smoothed power spectrum is enhanced for visual purposes. }
\label{fig:ps_numax}
\end{figure}

\subsubsection{Frequency of maximum oscillation power ($\nu_{\rm max}$)}
\label{sect:numax}
All solar-like oscillations in a star form together a bell-shaped power excess above the granulation and background signal centred around a specific frequency (see Fig.~\ref{fig:ps_numax}, where the amplitude of the power excess envelope is enhanced for visual purposes). This specific frequency is often referred to as \textit{frequency of maximum oscillation power} or $\nu_{\rm max}$. This frequency has been linked empirically to the acoustic cut-off frequency  
\begin{equation}
\nu_{\rm ac}=\frac{c}{4\pi H_p}
\label{eq:nu_ac}
\end{equation}
\citep[][using the approximation for an isothermal atmosphere]{lamb1932}; here $c$ is the adiabatic sound speed and $H_p$ is the pressure scale height. (See Eq.~(\ref{eq:fullac}) and subsequent text for a theoretical explanation of the acoustic cut-off frequency.)
It can be shown that $\nu_{\rm max}$ provides a direct measure of the surface gravity ($g$) when the effective temperature ($T_{\rm eff}$) is known \citep[e.g.][]{brown1991,kjeldsen1995}:
\begin{equation}
\nu_{\rm max} \propto \frac{g}{\sqrt{T_{\rm eff}}} \propto \frac{M}{R^2\sqrt{T_{\rm eff}}}
\label{eq:numax}
\end{equation}
with $M$ and $R$ the stellar mass and radius, respectively. A theoretical basis for this relation has been investigated by \citet{belkacem2011} and is discussed further in Section~\ref{sec:excitation}.

The value of $\nu_{\rm max}$ can be estimated as the centroid of a Gaussian fit to the oscillation power excess \citep{kallinger2012}. Alternatively, one can use the peak of the oscillation power excess in the smoothed power spectrum \citep{huber2009} or the first moment of the area under the smoothed power envelope \citep{hekker2010method}. All these methods use slightly different, but equally valid, definitions of $\nu_{\rm max}$ and therefore can provide different values. Comparisons between values obtained with different methods show that this difference is generally within a few percent  \citep{hekker2011comp,hekker2012,verner2011}.

As stated above, for high-precision data such as the \textit{Kepler} timeseries the main sources of the signal in the timeseries data are the granulation and the oscillations.  The amplitudes of these signals are correlated with $\nu_{\rm max}$, and therefore the frequency of maximum oscillation power can also be directly estimated from the variance in the timeseries \citep{hekker2012}.

\begin{figure}
\centering
\begin{minipage}{\linewidth}
\centering
\includegraphics[width=\linewidth]{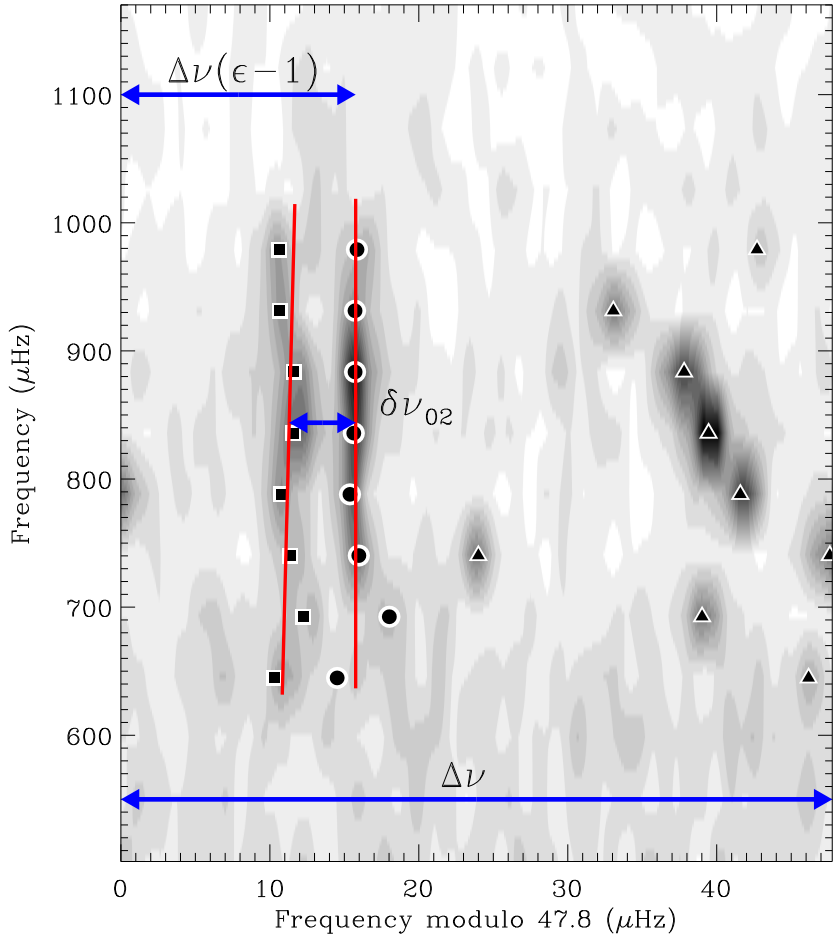}
\end{minipage}
\caption{\'Echelle diagram of KIC 11395018 showing the frequencies (black points) as determined by \citet{mathur2011FF}: modes with $l = 0$, $l = 1$, and $l = 2$ are indicated with circles, triangles and squares, respectively. For reference, a grey-scale map showing the power spectrum (smoothed to 1 $\mu$Hz resolution) is included in the background. The fits made to the $l = 0$ and $l = 2$ modes are shown by red lines. The values of $\Delta\nu$ and $\delta\nu_{02}$ (at $\nu_{\rm max}$) and the relationship $\Delta\nu(\epsilon-1)$, i.e. the absolute position of the $l = 0$ ridge, are indicated by the blue arrows, as labeled. {\orange Image reproduced with permission from \citet{white2011}, copyright by AAS.}}
\label{fig:echelle}
\end{figure}

\subsubsection{Frequency pattern}
\label{sect:pattern}
Following asymptotic theory 
\citep[][and Section~\ref{sec:astheory}]{tassoul1980}, acoustic oscillation modes (pressure as restoring force) of solar-like oscillators follow a distinct pattern:
\begin{equation}
	\nu_{n\,l} \simeq \Delta\nu \left( n+\frac{l}{2}+\epsilon \right)
	-d_{n\,l} \; ,
\label{eq:tassoul}
\end{equation}
with $\nu$ cyclic oscillation frequency, $\Delta\nu$ large frequency separation (Section~\ref{sect:dnu}), $\epsilon$ a phase term (Section~\ref{sect:epsilon}) and $d_{nl}$ a small correction to the leading order asymptotics, which is zero for $l = 0$. 

Based on the asymptotic expression, \citet{mosser2011} developed the \textit{universal pattern} for red-giant spectra, according to which all parameters are assumed to be a function of $\Delta\nu$.
This is equivalent to assuming that the underlying physics of the parameters varies as a function of the global stellar parameters. The universal pattern has the following form to describe pure acoustic modes \citep{mosser2011,mosser2012}:
\begin{equation}
\nu_{n_p\,l}=\left(n_p +\frac{l}{2}+\epsilon(\Delta\nu)-\hat d_{l}(\Delta\nu)+\frac{\alpha(\Delta\nu)}{2}[n_p-n_{\rm max}]^2\right)\Delta\nu,
\label{eq:UP}
\end{equation}
where $n_{\rm max}=\nu_{\rm max}/\Delta\nu$. The phase term $\epsilon$ and non-radial correction $\hat d_{l}$, with $\hat d_0 = 0$, are described by scaling laws on the form $A+B\log \Delta \nu$ \citep{mosser2010}. The second-order term (or curvature) in the asymptotic expression is represented by the parameter $\alpha = 0.015 \Delta\nu^{-0.32}$ \citep{mosser2012}. 

The reason for the universal behaviour of red giants is not yet fully understood, as the asymptotic approximation is fundamentally related to the behaviour of the solution of the oscillation equations near the singularity at $r=0$ \citep[][and Section~\ref{sec:ascenrg}]{gough1986a,gough1993}. In the case of red giants the conditions in the central region of the star are considerably different compared with the conditions in main-sequence stars (Section~\ref{sec:ascenrg}). Nevertheless, it seems that the asymptotic approximation and the universal pattern provide reasonable results for both models and observations.  Only for the very luminous and expanded red giants close to the tip of the RGB the universality does not hold \citep{stello2014}. For modes with low radial orders the dipole mode is located closer to the neighbouring quadrupole mode, providing a regular pattern of three modes together instead of a pattern of a cluster of two modes together (the 0-2 pair) alternating with one (dipole) mode.

\subsubsection{Large frequency separation}
\label{sect:dnu}
The \textit{large frequency separation} $\Delta\nu$ is the separation between modes of the same degree and consecutive radial orders.  $\Delta\nu$ is proportional to the inverse of the acoustic diameter, i.e. the sound travel time across a stellar diameter. Furthermore, it can be shown that $\Delta\nu$ is a direct probe of the mean density ($\overline{\rho}$) of the star \citep{ulrich1986}:
\begin{equation} 
\Delta\nu = \nu_{n\,l} - \nu_{n-1\,l} = \left(2 \int_0^R \frac{{\rm d}r}{c} \right )^{-1} \propto \sqrt{\frac{M}{R^3}} \propto \sqrt{ \overline{\rho}} ,
\label{eq:dnu}
\end{equation}
with $r$ the distance to the centre of the star, see also Eq.~(\ref{eq:larsep1}).

The near-regular pattern of the large frequency separation can be measured in a global sense from the autocorrelation of a power spectrum \citep[e.g.,][]{huber2009}, the power spectrum of a power spectrum \citep[e.g.,][]{hekker2010method}, or the mathematically equivalent autocorrelation of the timeseries \citep[EACF,][]{mosser2009} with a cosine filter or Hanning function with a full-width at half maximum (FWHM) of the order of the FWHM of the power excess. Additionally, \citet{mosser2011} have used the universal pattern (sect.~\ref{sect:pattern}) in which the known patterns in the power spectrum are used to compute templates which are convolved with observed power spectra to determine  $\Delta\nu$. Furthermore, the large frequency separation can be obtained from fits to individual frequencies \citep{kallinger2010}, from pair-wise differences, or a linear fit of the frequencies versus their radial order. It has been shown that the different determinations are consistent within their uncertainties and definitions \citep[e.g.][]{verner2011,hekker2011comp,hekker2012}. Nevertheless, some biases depending on the number of radial orders that were used in the analysis have been identified \citep{hekker2012}.

A convenient way to represent the power spectrum of solar-like oscillators is in an {\it \'echelle diagram} \citep{grec1983},
as shown for a subgiant star in Fig. \ref{fig:echelle}. This is obtained by dividing the frequency spectrum into segments of length $\Delta \nu$ and stacking the segments. According to Eq. (\ref{eq:tassoul}) this should lead to roughly vertical sequences
of points corresponding to different degrees. 
As shown in the figure this is satisfied for modes of degree $l = 0$ and $2$;
for $l = 1$ the presence of mixed modes (see Sections
\ref{sect:obsmixedmodes}, \ref{sec:themixedmodes} and \ref{sect:resmixedmodes})
causes departures from the simple behaviour.

We stress here that the actual separation between the frequencies of adjacent modes of the same degree varies as a function of frequency due to stellar internal structure properties, as also reflected by Eq.~(\ref{eq:UP}) and discussed further by \citet{mosser2013} and in Section~\ref{sec:astheory}. Structure variations that happen on scales that are comparable with or shorter than the oscillation wavelength, i.e. a glitch, cause a damped sinusoidal modulation in the frequencies (see for more details Sections~\ref{sect:glitches} and \ref{sect:glitchestheory}). Structure changes that take place over longer scales cause a gradual change, or curvature, in the large frequency separation. Therefore the value of the large frequency separation may change depending on the frequency range that is taken into account.

\subsubsection{Phase term ($\epsilon$)}
\label{sect:epsilon}
The asymptotic relation for acoustic modes (Eq.~\ref{eq:tassoul}) also contains a phase term $\epsilon$. This \textit{phase term} is a dimensionless offset of the radial modes in an \'{e}chelle diagram (see Fig.~\ref{fig:echelle}). Its value is correlated with the determination of $\Delta\nu$. The value of $\epsilon$ can be determined from the universal pattern in which it is considered to be a function of $\Delta\nu$. Additionally, a (weighted\footnote{Ideally one should apply the same weights in observations and computed oscillations. However, in observations the weights are generally derived from the uncertainties in the frequencies. When using frequencies computed from models a Gaussian weight resembling the amplitude of the oscillation modes is generally applied. Nevertheless, \citet{hekker2013} showed that there is good agreement between $\Delta\nu$ obtained from the power spectrum of the power spectrum \citep[e.g., ][]{hekker2010method} and from a weighted linear fit through a set of computed oscillations.}) least-squares fit to the radial ($l = 0$) frequencies to simultaneously determine $\Delta\nu$ and $\epsilon$ can be performed. These global methods average over the variation of  $\epsilon$ with frequency. Additionally, a `local' $\epsilon$ can be determined by only including the three central radial orders around $\nu_{\rm max}$ in the analysis \citep{kallinger2012}.

The inferred value of $\epsilon$ is generally between $0.5$ and $1.5$, with a potential `offset' of $\pm 1$.  This `offset' only reflects the observational limitations in that the radial order $n$ cannot be measured independently and is not an offset of the actual value of $\epsilon$.

The main diagnostic power of the global phase term lies in the mode identification of the different ridges in the \'{e}chelle diagram for stochastic oscillators with short mode lifetimes. In these cases the width of the oscillation signals in the Fourier space does not allow to resolve the small frequency separation (see next subsection) between the $l=0$ and $l=2$ modes and hence the $l=1$ and $l=0,2$ ridge have very similar characteristics. In these cases $\epsilon$, i.e. the location of the radial modes in an \'{e}chelle diagram, can be used to identify the ridges correctly \citep{white2012}.

The `local' phase term can be used to distinguish between different evolutionary phases \citep{kallinger2012}. This is caused by the fact that the differences in the core cause differences in the thermodynamic state of the envelope, which results in a different location of the second helium-ionisation zone for stars with an inert helium core compared with stars with helium-core burning. The location of the second helium-ionisation zone leaves a trace in small oscillatory deviations in the frequencies (see Sections~\ref{sect:glitches} and \ref{sect:glitchestheory} on glitches). This causes a difference in the `local' phase term for stars in different evolutionary phases \citep{jcd2014}.  

\subsubsection{Small frequency separation ($\delta\nu_{02}$ and $\delta\nu_{13}$)}
\label{sect:sdnu_02}
A typical separation in frequency exists between odd- and even-degree modes. This is the so-called \textit{small frequency separation} that can be approximated asymptotically, for main-sequence stars, by \citep[e.g.][see also Eq.~\ref{eq:smallsep}]{gough1986a}:
\begin{equation}
\delta\nu_{l\,l+2}(n)=\nu_{n\,l}-\nu_{n-1\,l+2} \simeq -(4l+6)\frac{\Delta\nu}{4\pi^2 \nu_{n\,l}}\int_0^R\frac{{\rm d} c}{{\rm d} r}\frac{{\rm d} r}{r}. \; 
\label{eq:smallsepd} 
\end{equation}

This parameter is generally measured from frequency differences between observed individual frequencies.
Eq.~(\ref{eq:smallsepd}) depends on the sound-speed gradient, which depends on the composition.
Hence, $\delta\nu_{l\,l+2}(n)$ provides a measure of the helium content in the core of main-sequence stars, and with that $\delta\nu_{l\,l+2}(n)$ is a diagnostic of stellar age \citep{jcd1984,ulrich1986,jcd1988}. 
It was noted by \citet{huber2010} from early {\it Kepler} observations, that for more evolved stars (subgiants and giants) the small frequency separation between modes of degree 0 and 2
essentially scales as $\Delta \nu$. This is due to the fact that these stars have a highly concentrated core such that the inner turning point of the 
{\orange pressure-mode} cavity lies outside the compact core. Therefore, this small separation does not provide a measure of the helium content in the core and hence is no longer an age diagnostic.

\subsubsection{Small frequency separation ($\delta\nu_{01}$)}
\label{sect:sdnu_01}
In a Fourier spectrum dipole modes are located approximately mid-way between radial modes as per Eq.~(\ref{eq:tassoul}). The offset from the midpoint between the radial modes and the frequency of the dipole mode can be computed from a three-point difference\footnote{\citet{roxburgh2003} noted that a smoother behaviour is obtained by defining this quantity with a five-point difference.} and indicated as $\delta\nu_{0\,1}$:
\begin{equation}
\delta\nu_{0\,1}(n)=0.5(\nu_{n\,0}-2\nu_{n\,1}+\nu_{n+1\,0}).
\label{eq:smallsepd1}
\end{equation}
In main-sequence stars $\delta\nu_{0\,1}$ is known to be sensitive to the central physical conditions. For red giants this is not the case. It has been shown that for red giants $\delta\nu_{0\,1}$ is correlated with the distance between the $l=1$ turning point and the bottom of the convective envelope. The value of $\delta\nu_{0\,1}$ takes small and negative values for stars ascending and descending the RGB where the turning points of acoustic $l=1$ modes are well within the convective envelope. Stars in the He-core burning phase have a shallower convective envelope and the turning points of $l=1$ modes are in the radiative region and $\delta\nu_{0\,1}$ generally takes positive values \citep{montalban2010}. 
As shown in Section~\ref{sec:redclump} (Figs~\ref{fig:clumpbv} and \ref{fig:clumpechl}) the evanescent region between the buoyancy and acoustic cavities is smaller in a red-clump star compared with a RGB star. This leads to a stronger coupling between modes in a red-clump star increasing the spread in the modes, which most likely leads to less regular behaviour of $\delta\nu_{0\,1}$ in clump stars compared with RGB stars. 
Hence, the value of $\delta\nu_{0\,1}$ and the regularity of the acoustic dipole spectrum can be used as a diagnostic to distinguish between different evolutionary phases \citep{montalban2010}. We note here that most of the work on $\delta\nu_{0\,1}$ for red giants was performed before the discovery of the fact that all non-radial modes are to some extent mixed modes. Hence, this influence was not  taken into account.

\begin{figure}
\centering
\begin{minipage}{0.8\linewidth}
\centering
\includegraphics[width=\linewidth]{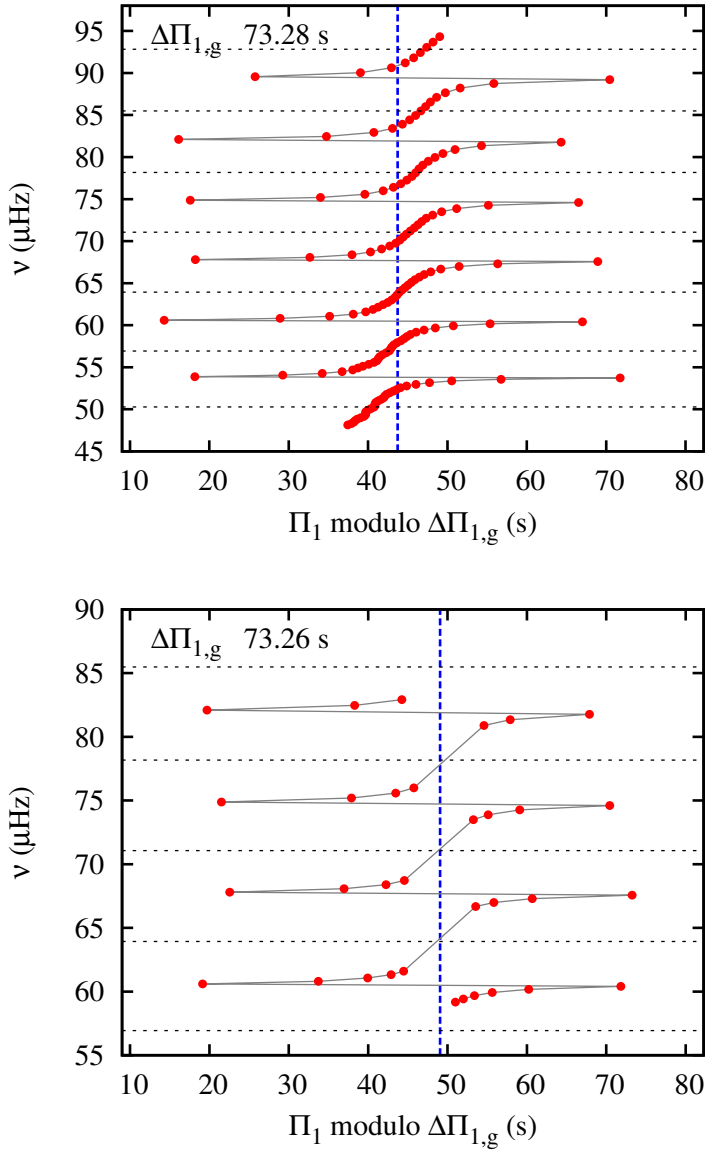}
\end{minipage}
\caption{Two period \'{e}chelle diagrams with different density of $l = 1$ modes for a $1 \Msun$ red-giant model are shown \citep[for details of the model see][]{datta2015}. The red points connected with the light grey lines are $l = 1$ modes and the horizontal dotted lines represent the frequencies of the radial modes. The blue dashed line shows the g-mode period spacing of high-order g modes. The top panel shows the period \'{e}chelle diagram for all the $l = 1$ modes in a frequency range spanning 7 radial orders, with $\Delta\Pi = 73.28$~s. The lower panel shows the period \'{e}chelle diagram for only a subset of the most p-dominated $l = 1$ modes in a smaller frequency range consisting of 5 radial orders with $\Delta\Pi = 73.26$~s. {\orange Image reproduced with permission from \citet{datta2015}, copyright by the authors.}}
\label{periodechelledatta}
\end{figure}

\subsubsection{Period spacing ($\Delta\Pi$)}
\label{sect:dP}
As mentioned above, non-radial modes in red giants all involve aspects of gravity-mode behaviour.
To analyse the relevant properties we note that
an asymptotic approximation (see Section~\ref{sec:astheory})
reveals that gravity modes of a given degree appear with near constant separation in period, the so-called \textit{period spacing} $\Delta\Pi$, as reflected in the asymptotic behaviour of their periods $\Pi_{n,l}$:
\begin{equation}
\Pi_{n,l} = {\Pi_0 \over \sqrt{l(l+1)}} (n + \epsilon_{\rm g} +1/2) \; ,
\label{eq:gasympd}
\end{equation}
with
\begin{equation}
\Pi_0 = 2 \pi^2 \left( \int_{r_1}^{r_2} N {\dd r \over r} \right)^{-1} \; ,
\label{eq:gasymp0}
\end{equation}
where $N$ is the {\orange Brunt-V\"ais\"al\"a} frequency (Eq.~\ref{eq:buoy}), $r_1$ and $r_2$ the turning points of the {\orange gravity-mode} cavity and $\epsilon_{\rm g}$ is a phase accounting for the behaviour near the turning points. This defines the period spacing:
\begin{equation}
\Delta \Pi_l = \Pi_0/\sqrt{l(l+1)}.
\end{equation}

Based on the regular nature, one can use similar techniques to determine the period spacing as used to determine $\Delta\nu$ (power spectrum of the power spectrum, auto-correlations and differences between modes with consecutive order, see Section~\ref{sect:dnu}), but now in period space and centred on regions where dipole (or quadrupole) modes are expected. It is however important to account for the fact that in solar-like oscillators no pure gravity modes can be observed, but only modes that have mixed gravity-pressure nature (see also Section \ref{sect:indf} and Section 4.2 for more details about mixed modes). Due to this mixed nature the regular spacing in period deviates from the asymptotically predicted value of `pure' gravity oscillation modes in the vicinity of `pure' pressure modes. The observed period spacing follows, in general, a predictable pattern for each acoustic radial order with an empirically determined Lorentzian shape \citep[][and Section~\ref{sec:ascenrg}]{mosser2011,mosser2012core}.

The underlying period spacing can be determined from the modulated observed period spacing using a period \'{e}chelle diagram (see Fig.~\ref{periodechelledatta}). In this diagram the period spacing can be obtained by aligning the g-dominated modes surrounded in a roughly symmetric way by the p-dominated modes \citep{bedding2011,mosser2012core,datta2015}. The patterns in the period \'echelle diagram can also be described by an analytical expression based on the coupling between the gravity and acoustic cavity as derived by \citet{mosser2012core} (see also Section~\ref{sec:ascenrg}):
\begin{equation}
\nu=\nu_{n_p,l}+\frac{\Delta\nu}{\pi}\arctan\left[q\tan\pi\left(\frac{1}{\Delta\Pi_l\nu}-\epsilon_g\right)\right] \; ,
\label{eq:mixasymp}
\end{equation}
where $q$ is the coupling strength. This is set to $q=0$ for no coupling and $q=1$ indicates maximum coupling. The period spacing and coupling strength can be determined in an iterative manner, when assuming $\epsilon_g$ is a fixed value (often zero, i.e. the pattern is assumed symmetric).  Note that \citet{buysschaert2016} have employed Eq.~(\ref{eq:mixasymp}) while leaving $\epsilon_g$ also as a free parameter. They find that although $\epsilon_g$ remains ill-defined, its determination improves the determination of the period spacing.

An alternative way to determine period spacing and to constrain the evanescent region between the p and g cavities (and thus coupling strength $q$) is through the inertia ratio of dipole and radial modes (see Section~\ref{sec:themixedmodes}). \citet{benomar2014} for the first time estimated mode inertias observationally from the measurements of mixed mode characteristics. The mode inertia ratio could develop to be a diagnostic that can potentially provide strong constraints on the stellar structure.

Recently, \citet{mosser2015,vrard2016} have proposed a way to stretch the period Fourier spectrum to remove the modulation due to the coupling between the pressure and gravity modes and obtain the regular underlying period spacing in a direct manner.

The period spacing provides a strong diagnostic on the central regions of the star. It can be used to distinguish between stars in different evolutionary phases, most notably between red giants with an inert He-core and red giants with He-core burning \citep{bedding2011, mosser2011mm,mosser2014}. Furthermore, small deviations from the regular period spacing pattern can reveal localised stellar structure changes such as the chemical discontinuity due to the first dredge-up \citep[][and Section~\ref{sect:glitchestheory}]{cunha2015}.

\subsection{Individual oscillation modes}
\label{sect:indf}
Individual oscillations modes in stochastic oscillations are characterized by their mode frequency $\nu_{\rm central}$, line width $\Gamma$ and height $H$. A resolved oscillation mode, i.e. a mode with a lifetime that is at least $\sim$10 times shorter than the timespan of the timeseries data \citep{hekker2010corot} can be modelled by a Lorentzian profile in the power spectrum $P(\nu)$:
\begin{equation}
P(\nu)=\frac{H}{\displaystyle{1+\left(\frac{2(\nu-\nu_{\rm central})}{\Gamma}\right)^2}}.
\label{eq:lorfit}
\end{equation}
The mode lifetime $t_{\rm damp}$ is directly related to the mode line width through $\Gamma=1/(\pi t_{\rm damp})$. The mode height and width are highly correlated and relate through the root-mean-square flux amplitude $A=\sqrt{\pi H\Gamma/2}$ \citep[e.g.,][]{chaplin2005} which is the area underneath the profile. Hence the amplitude is a more robust parameter. The amplitude of the modes contains information on the excitation and damping of the oscillations \citep[see e.g.][and Section~\ref{sec:excitation}]{houdek2012,samadi2012}. However, the \textit{visibility} of the modes is a combination of the intrinsic amplitude and geometrical effects, i.e., cancellation of some of the signal due to the fact that only integrated light from the visible part of stars can be observed. 
In case the oscillations are not resolved (lifetime longer than the timespan of the timeseries data) they are often approximated with a sinc function (see footnote~\ref{fn:sinc}) in which case the amplitude is $A=\sqrt{2H\delta\nu}$, where $\delta\nu$ is the frequency resolution. In intermediate cases where the modes are partly resolved a mixture of the resolved and unresolved description has to be applied (see Section~\ref{sec:modexcit}).

\subsubsection{Mixed modes}
\label{sect:obsmixedmodes}
Radial modes are always pure acoustic modes with pressure as the restoring force. Non-radial modes  in red giants, however, always have a mixed nature, i.e.~are \textit{mixed modes}, for which buoyancy acts as restoring force in the deep interior of the star and pressure acts as restoring force in the outer layers of the star. In other words a mixed mode is a single mode with different behaviour in the different regions.
Frequencies of mixed modes are shifted compared with pure acoustic or gravity modes by an amount depending on the coupling strength between the two (gravity and pressure) oscillation cavities \citep[e.g.][and Section~\ref{sec:themixedmodes}]{deheuvels2010,hekkermazumdar2014}.

Mixed modes (and hence period spacings (Section~\ref{sect:dP})) are mostly observed in dipole ($l=1$) modes as for these modes the coupling between the {\orange pressure- and gravity-mode cavities} is stronger (narrower evanescent zone between the cavities, see Fig.~\ref{fig:charfrqrg}) and also because the period spacings are larger due to the dependence on $\sqrt{l(l+1)}$ (see Eq.~\ref{eq:gasympd}), and thus better resolved.

In case there are only a few mixed modes present in for instance subgiants, \citet{deheuvels2010,deheuvels2011} showed that $l=1$ avoided crossings in subgiant stars involve more than two modes and induce a characteristic distortion in the $l=1$ ridge in the \'echelle diagram (see Fig.~\ref{fig:chenechl}). This can be used to constrain stellar models.  \citet{deheuvels2011} have done so by matching the observed large frequency separation and frequency of the avoided crossing with values obtained from models. This results in a precise age estimate given the mass and physics of the models.

Following the analysis by  \citet{deheuvels2011}, \citet{bedding2012} introduced a powerful tool for analysing these mixed modes. He showed
that by replicating the \'echelle diagram horizontally the 
full structure of the avoided crossings can be displayed.
\citet{benomar2012} subsequently used this to develop a method to fit the
avoided crossings and determine the minimum separation between the two
branches as a measure of the coupling strength (see Section~\ref{sect:thsubgiants}). This provides a useful diagnostics of the stellar mass for subgiant stars. 

\subsubsection{Mode identification} 
\label{sect:modeID}
The degree $l$ of the observed individual frequencies can be obtained from the known pattern of the stochastic oscillation (Section~\ref{sect:pattern}) through an \'echelle diagram (Fig.~\ref{fig:echelle}), where the ridges of radial and quadrupole modes are close together with the dipole ridge appearing at about a 0.5$\Delta\nu$ offset. In case the mode lifetimes are very short  the radial modes can be so broad that they overlap the quadrupole modes. In that case the phase term ($\epsilon$) can be useful to distinguish the odd and even ridge (Section~\ref{sect:epsilon}). The (acoustic) radial order $n_p$
can be estimated from Eq.~(\ref{eq:tassoul}), i.e., from the ratio of the frequency of the mode over the large frequency separation bearing in mind that $n$ is an integer and that the phase term $\epsilon$ takes values between 0.5 and 1.5. Note that for mixed modes the total radial order $n$ consists of the nodes in the acoustic cavity $n_p$ as defined here and the nodes in the buoyancy cavity. The buoyancy radial order $n_g$ is indicated by definition with a negative number and can take large values (see also Section \ref{sec:modeorder} and Fig.~\ref{fig:rgamde}). 

Mode identification can also be performed from spectroscopic data. Red- or blue-shifted parts of the surface of a star leave traces in the shape of a spectral line profile. Over the course of the pulsation the blue- and red-shifted parts change and hence the line-profile shape changes. Therefore, the amplitude and phase of the \textit{line-profile variations} at a particular frequency are fundamentally different for radial and non-radial oscillation modes. This technique has mainly been developed for stars with coherent oscillations \citep{zima2004, zima2008}, but also proved useful for providing evidence for the presence of non-radial oscillations in red giants \citep{hekker2006,hekkeraerts2010}. 

\subsubsection{Surface effect} 
\label{sect:surfaceeffect}
Due to incomplete modelling of the convective outer layers of stars and the strong non-adiabatic behaviour of the oscillations in the superficial layers there is an offset between modelled and observed frequencies, the so-called \textit{surface effect}. This offset is a function of frequency, but independent of degree at least for the Sun. Using the solar offset \citet{kjeldsen2008} proposed a widely used power-law correction that can be applied to other stars.
An alternative procedure is to directly scale the solar offset on a suitable frequency scale \citep[e.g.,][]{jcd2012}.
Additionally, \citet{aerts2010} have shown that due to their larger amplitudes in the inner regions and hence larger inertia (cf. Eq.~\ref{eq:inertia}), mixed modes in subgiants are less sensitive to the incorrect modelling of the non-adiabatic outer layers, and hence the solar calibrated offset needs to be adapted. 
\citet{ball2014}, based on \citet{gough1990}, have developed a correction based on mode inertia. So far this method has only been tested for observations of the Sun and solar-like stars as well as for models of different mass and metallicity covering a significant portion of the HR diagram from the main-sequence to red giants for models with $T_{\rm eff} < 6500$~K \citep{schmitt2015}. In their work \citet{schmitt2015} concluded that the two-term model proposed by \citet{ball2014} works much better than other models across a large portion of the HR diagram, including the red giants. 

For main-sequence stars it is possible to reduce the influence of the near-surface region by using frequency-separation ratios such as
\begin{equation}
r_{0\,2}=\frac{\nu_{n\,0}-\nu_{n-1\,2}}{\nu_{n\,1}-\nu_{n-1\,1}},
\label{fratio}
\end{equation}
which are essentially independent of the near-surface problems \citep[e.g.][]{roxburgh2003,roxburgh2005}.
As for the small frequency separations ($\delta\nu_{02}$, Section~\ref{sect:sdnu_02}) the usefulness of the frequency-separation ratios for red-giant stars is limited due to the universality of the frequency patterns \citep[e.g.][]{huber2010}. Additionally, in giants all non-radial modes are mixed modes, which have different sensitivities to the surface effect.

\subsubsection{Rotational splitting}
\label{sect:rot}
Rotation splits the non-radial modes into $2l+1$ {\orange single modes} of different azimuthal orders $m$. These incorporate modes travelling with the rotational direction (\textit{prograde modes}) and in the opposite direction (\textit{retrograde modes}) in addition to the original mode unperturbed by rotation, i.e., in cyclic frequency (see also Eq.~(\ref{eq:rotsplit}) given in angular frequency):
\begin{equation}
\nu_{n\,l\,m}=\nu_{n\,l\,0}+m\delta\nu_{n\,l\,m}
\end{equation}
In cases of slow rotation, which is in general the case for subgiants and red-giant stars, the assumption of symmetric splittings is often valid. The relative heights or \textit{visibility} of the different azimuthal orders in a multiplet are indicative of the inclination angle with respect to the rotation axis at which we view the system \citep{gizon2003}. When viewing the system pole on, only the $m=0$ mode is visible, while for an equator-on system all modes with even $l - m$ are visible.
A rotationally split mode can be fitted with a set of Lorentzian functions:
\begin{equation}
	P(\nu_{n\,l})=\sum_{m=-l}^l\frac{\Psi_{l\,m}(i)H}{\displaystyle{1+\left(\frac{2(\nu-\nu_{n\,l\,0}-m\delta\nu_{n\,l\,m})}{\Gamma}\right)^2}},
\end{equation}
where $\Psi_{l\,m}(i)$ is the visibility of the mode which depends on the inclination angle ($i$) and $\delta\nu_{n\,l\,m}$ is the rotational splitting. Here we assumed that all modes in a multiplet are excited to the same intrinsic average height, as may be reasonable for stochastically excited modes observed for a long time compared with the lifetime of the modes.
To detect the average rotational splitting one can also use the fact that the rotational splitting is approximately symmetric in slow rotators and apply for instance the EACF \citep[autocorrelation of the timeseries,][]{mosser2011} with a very narrow Hanning filter in the range of a non-radial rotationally split modes.

It has been shown that the rotational splitting of modes with mixed character (i.e. all non-radial modes in red-giant stars) is a function of the mixed character \citep[][and Section~\ref{sec:rotation}]{beck2012,mosser2012rot,goupil2013}. Hence the rotational splitting of mixed modes is a diagnostic to probe the radial internal rotation profile. 

\subsubsection{Glitches}
\label{sect:glitches}
Information on specific transition regions in a star, such as the boundaries of convective zones or ionization zones of helium or hydrogen, can be obtained from the fact that, at such boundaries, the properties of the star change on a scale substantially smaller than the local wavelength of the oscillations \citep[e.g.][and Section~\ref{sect:glitchestheory}]{vorontsov1988,gough1990}. These sharp features (also called \textit{glitches}) cause oscillatory variations in the frequencies with respect to the pattern described in Eq.~(\ref{eq:tassoul}), or the corresponding pattern satisfied by mixed modes (cf. Eq. \ref{eq:mixasymp}). The period of the variation depends on the location of the feature, while the amplitude depends on the detailed properties of the feature. Note that the diagnostic use of glitches can be done completely independent of stellar models.

For acoustic modes the oscillatory behaviour due to glitches can be measured in frequency differences ($\Delta\nu$) but more often in second differences, and can be described by a damped oscillator:
\begin{equation}
\Delta_2\nu_{n\,l} \equiv \nu_{n-1\,l}-2\nu_{n\,l}+\nu_{n+1\,l}=c_0\nu_{n\,l} e^{-c_2\nu_{n\,l}^2}\sin{(4\pi\nu_{n\,l}\tau_{\rm gl}+2\phi_{\rm gl})}
\label{eq:glitch}
\end{equation}
in the case of the glitch due to the He II ionisation zone, where $c_0$ indicates the amplitude of the oscillation and $c_2$ a characteristic width of the e-folding time of the damped oscillator.
Here $\tau_{\rm gl}$ is the \textit{acoustic depth}, i.e., the sound travel time between the surface and the glitch, and $\phi_{\rm gl}$ is a constant that accounts for the phase. Eq.~(\ref{eq:glitch}) was derived by \citet{houdek2007} and applied in various forms by e.g.~\citet{miglio2010,mazumdar2014,broomhall2014,Verma2014}. In addition to the intrinsic limits of the sharpness of the glitch compared with the local wavelength of the oscillation mode (see Section~\ref{sect:glitchestheory}), the observational data also provide natural limits. The depth in the star at which a glitch can be measured depends on the frequency range covered by the oscillations, i.e. the largest period in Eq.~(\ref{eq:glitch}) that can be measured, while the minimum period is defined by the resolution of the measured frequencies. From these limitations we find that the bottom of the convection zone is located too deep in red giants to be measured. Furthermore, the helium I and hydrogen ionisation zones are located close to the surface, which make them very challenging to measure. 

For mixed modes the effects of buoyancy glitches are described by \citet{cunha2015} and in Section~\ref{sect:glitchestheory}. However, no solid observational results on such effects have been presented so far.

\subsection{Scaling relations and grid-based modelling}
\label{sect:GBM}
The scaling relations Eqs~(\ref{eq:numax}) and (\ref{eq:dnu}) can be used to obtain the mean density and surface gravity of stars exhibiting solar-like oscillations (and from that stellar mass and radius) in a direct manner, i.e. so-called \textit{direct method}. These scaling relations are exceptionally good given that these relations do not account for metallicity differences, nor do they account for any knowledge we have about stellar evolution. To take account of this knowledge, it is also possible to compare the observables \{$\Delta\nu, \nu_{\rm max}, T_{\rm eff}, \rm{[Fe/H]}, \pi$, $\cal L$\} or a subset thereof with a grid of models; here $\rm{[Fe/H]}$ is metallicity, $\pi$ is parallax and $\cal L$ is luminosity. In this so-called \textit{grid-based modelling} one does account for knowledge of stellar structure and evolution, as well as metallicity. However, in case the scaling relations are used to determine $\Delta\nu$ and  $\nu_{\rm max}$ from the models suitable reference values (with uncertainties) to which one scales have to be adopted. Alternatively, one can compute individual frequencies for the models and derive $\Delta\nu$ from that.

\paragraph{Reference values}
Both the direct method and grid-based modelling are based on the scaling relations (Eqs~\ref{eq:numax} and \ref{eq:dnu}) which assume that the scaling is valid in a consistent way between the reference and the observed star. Often reference values based on the Sun are used; however, this may not be correct for stars with different properties, such as a different metallicity or rotation rate, or stars in different evolution phases, as their stellar internal structures are different. This has indeed been confirmed by \citet{white2011} for stars with different metallicities along the main-sequence and just beyond that. \citet{mosser2013} proposed new reference values based on relations derived using results obtained using the universal pattern \citep{mosser2011}. It is currently unclear whether these newly derived reference values are also valid when other methods are used to derive $\Delta\nu$ and $\nu_{\rm max}$. For instance, \citet{hekker2013} failed to confirm the relations quantitatively using stellar models. Furthermore, \citet{miglio2012} showed that the difference in internal temperature structure (hence sound speed) between RGB and RC stars has a significant impact on the scaling relations. Therefore, they applied a correction for red-clump stars in the open clusters NGC 6719 and NGC 6819 based on the masses of the RGB stars in the respective clusters and theoretical models. Subsequently, \citet{jcd2014} showed that the difference in the variation of the phase term $\epsilon$ (see Section~\ref{sect:epsilon}) with frequency between RC and RGB stars is related to differences in the thermodynamic state of the convection zone. This supports the findings of \citet{miglio2012} that RC and RGB stars have internal structures that are significantly different, which calls for corrections to the scaling relations. Recently, \citet{guggenberger2016} proposed a new reference function for the $\Delta\nu$ scaling relation (Eq.~\ref{eq:dnu}) that accounts for metallicity differences and is applicable for stars on the main sequence up till past the RGB bump on the red-giant branch. Additionally, \citet{sharma2016} and Serrenelli et al. (in preparation) devised methods to apply a correction between $\Delta\nu$ scaling and $\Delta\nu$ from frequencies in model calculations. 

\subsection{`Boutique' modelling}
\label{sect:boutique}
Oscillation frequencies can be compared with stellar models to infer the internal stellar structure. This is often done on a star by star basis, hence `boutique' or 'detailed' modelling. This modelling can be done in both a forward and an inverse approach.

\paragraph{Forward modelling} In \textit{forward modelling} the observed frequencies and stellar parameters such as ($\log g$, $T_{\rm eff}$, [Fe/H]) are matched with stellar models. This is mostly done using $\chi^2$ minimisation either in a direct manner or through Singular Value Decomposition (SVD) \citep[e.g.][]{brown1994}.
In this procedure the surface term is accounted for by one of the prescriptions mentioned above \citep{kjeldsen2008,ball2014}, or by using the frequency-separation ratios \citep{roxburgh2003,roxburgh2005}. Detailed descriptions of the different methods currently applied can be found in the appendix of \citet{chaplin2013obl,ballard2014,silvaaguirre2015} and references therein.

\paragraph{Inverse modelling}
\label{par:inverse}
An \textit{inverse problem} is a general framework that is used to convert observed measurements into information about a physical object or system. For stars the individual frequencies can be used to obtain information about the internal structure of stars. An inverse problem is however by nature ill-posed, and many frequencies probing the star to different depths are needed to obtain meaningful results. Most commonly inversions have been used to study the stellar rotation at different depths in the stars \citep[e.g.][]{deheuvels2012,deheuvels2014,dimauro2016}. It is however in principle also possible to perform structure inversions. These have been very powerful in determining the internal structure of the Sun
\citep[e.g.,][and references therein]{gough1996}. For other stars the lack of observed frequencies probing the star to different depth has precluded detailed structure inversions.
However, inversion techniques have been used in asteroseismic analyses
to constrain specific properties of stars \citep[e.g.][]{reese2012, buldgen2016}.
For a complete overview of stellar inversions we refer to \citet{basu2014,basu2016}

\section{Theory of stellar pulsations}
\label{sec:theory}

The general theory of stellar pulsations has been presented 
in considerable detail by, for example, \citet{unno1989} and \citet{aerts2010}.
However, oscillations of evolved stars present special properties which
are important for the understanding of the observations.
Thus, here we provide some background which is useful in the interpretation
of the observations of the oscillations of such stars, 
relating the frequencies and other aspects of the oscillations 
to the properties of the stars.

We are dealing with low-amplitude oscillations, which can be regarded
as small perturbations to the equilibrium structure.
Formally, these can be described using linearized perturbation analysis of
the general equations of hydrodynamics.
An important result concerns the geometrical properties of the modes
of spherically symmetric stars.
For the modes that are relevant the properties can be described by spherical harmonics
$Y_l^m(\theta, \phi) = P_l^m(\cos \theta) \ee^{\eye m \phi}$
as functions of co-latitude $\theta$ and longitude $\phi$;
here $P_l^m$ is a Legendre function, characterized by
the degree $l$ and the azimuthal order $m$ 
(see Sect~\ref{sect:osc} for a definition).
The time dependence of a mode is conveniently written as
$\ee^{- \eye \omega t}$ where $\omega$ is the angular frequency, which is in general complex.
The displacement vector can be written, as a function of position $\boldr$ and
time $t$,
\begin{equation}
\bolddelr(\boldr, t) = {\rm Re}\left\{ \left[\xi_r(r) Y_l^m \bolda_r
+ \xi_{\rm h}(r) \left( {\partial Y_l^m \over \partial \theta} \bolda_\theta 
+ {1 \over \sin \theta} {\partial Y_l^m \over \partial \phi} \bolda_\phi
\right) \right] \ee^{- \eye \omega t} \right\} \; ,
\label{eq:deltar}
\end{equation}
where ${\rm Re}$ denotes the real part, $\xi_r$ and $\xi_{\rm h}$ are
the radial- and horizontal-displacement
amplitude functions that depend only on the distance $r$ to the centre,
and $\bolda_r$, $\bolda_\theta$ and $\bolda_\phi$ are unit vectors
in a spherical polar coordinate system.
Other oscillating variables, such as the pressure perturbation, vary
as the real part of $Y_l^m(\theta, \phi) \ee^{-\eye \omega t}$.
We can separate the frequency into real and imaginary parts as
$\omega = \omega_{\rm r} + \eye \omega_{\rm i}$.
Then the dependence of the oscillations on longitude $\phi$ and time $t$
is essentially
\begin{equation}
\cos(m \phi - \omega_{\rm r} t) \ee^{\omega_{\rm i} t} \; .
\label{eq:timedep}
\end{equation}
Unless $m = 0$ this describes a wave running in the $\phi$ direction, 
growing or decaying with time depending on whether $\omega_{\rm i}$ is
positive or negative.
In much of the following we consider adiabatic oscillations where 
processes causing excitation or damping are neglected.
Then the frequency $\omega$ is real and we ignore the distinction 
between $\omega$ and $\omega_{\rm r}$.
We note that {\it observed} oscillations are typically discussed in terms of
the cyclic frequency $\nu = \omega_{\rm r}/2 \pi$,
as done in the previous sections.
However, for the theoretical analysis it is more convenient to use 
the angular frequency $\omega$.

After this separation of variables we are left with differential equations that
depend just on $r$.
Combined with suitable boundary conditions, this is a relatively
straightforward numerical problem, which determines 
the frequencies $\omega$ as eigenvalues.
However, the physical treatment of the
near-surface layers still suffers from substantial uncertainties,
particularly when the mode energetics is taken into account
(see Section~\ref{sect:surfaceeffect}).

To evaluate the diagnostic potential of solar-like oscillations in giant stars
and interpret the inferences that are made, an understanding of the properties
of the oscillations is required. Moreover, these properties are
fascinating in their own right.
A full utilization of the observed data requires detailed
comparison of the observed frequencies, and other properties, with computations
for stellar models.
However, a great deal of insight as well as powerful diagnostics can
be obtained from asymptotic analyses of the oscillations,
to which we turn next.

\subsection{Asymptotic theory}
\label {sec:astheory}

We first concentrate on the oscillation frequencies and
overall properties of the eigenfunctions,
assuming the oscillations to be adiabatic.
{\orange Solar-like oscillations are generally of high radial order,
such that the eigenfunctions mostly vary on a scale much shorter than 
the scale of variation of the equilibrium structure.
In this case the analysis of the 
behaviour of the modes and their relation to stellar structure
in terms of their asymptotic properties is extremely informative;
some effects of rapid variations in the model structure,
and hence departures from the asymptotic behaviour, are discussed in 
Section~\ref{sec:glitch}.
Also, owing to the high radial order it is common to ignore the perturbation
to the gravitational potential,}
in the so-called {\it Cowling approximation} \citep{cowling1941}.
We return to the limitations of this approximation towards the end of
Section~\ref{sec:astheory}.

\begin{figure}
\centering
\begin{minipage}{\linewidth}
\centering
\includegraphics[width=9cm]{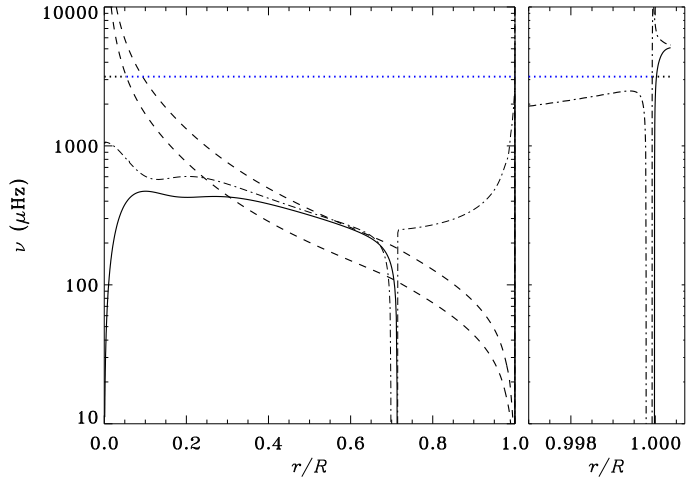}
\end{minipage}
\caption{Characteristic frequencies for a model of the Sun,
expressed in terms of cyclic frequencies, as functions of fractional radius.
The solid line shows the buoyancy frequency $N/2 \pi$ (cf. Eq.~\ref{eq:buoy}),
the dashed lines show the Lamb frequencies $S_l/2\pi$ for $l = 1$ and $2$
(cf. Eq.~\ref{eq:lamb}) with $S_1/2\pi$ having lower values than $S_2/2\pi$,
and the dot-dashed line shows the acoustic cut-off frequency 
$\omega_{\rm c}/2 \pi$ (cf. Eq.~\ref{eq:fullac}).
The horizontal dotted line indicates an estimate of the frequency of 
maximum oscillation power $\nu_{\rm max}$ (cf. Eq.~\ref{eq:numaxt}),
{\orange with the blue part marking the p-mode cavity for $l = 1$.}
The right-hand panel shows an expanded view of the near-surface region.
}
\label{fig:charfrqsun}
\end{figure}

The Cowling approximation reduces the equations of adiabatic oscillations
to a second-order system, greatly simplifying the analysis.
The equations are often expressed on the form
\begin{equation}
{\dd^2 X \over \dd r^2} = - K X \; ,
\label{eq:basasymp}
\end{equation}
for suitable choices of the dependent variable $X$ and the function $K$. The choice of $X$ and $K$ may depend on the specific properties that
are being investigated.
A convenient formulation was derived by Gough 
\citep{deubner1984, gough1993} based on an analysis by \citet{lamb1932}.
Here $X = c^2 \rho^{1/2} \div \bolddelr$,
where $c$ is the adiabatic sound speed and $\rho$ is density.
The corresponding approximation to $K$ is
\begin{equation}
K = {1 \over c^2} \left[ S_l^2 \left({N^2 \over \omega^2} - 1\right) 
+ \omega^2 - \omega_{\rm c}^2 \right] \; ,
\label{eq:kasymp}
\end{equation}
which is determined by three characteristic frequencies of the star:\newline
\begin{itemize}
\item  The {\it Lamb frequency} $S_l$, with
\begin{equation}
S_l^2 = {l(l+1) c^2 \over r^2} \; .
\label{eq:lamb}
\end{equation}
The Lamb frequency is a local characteristic frequency
of horizontally-propagating sound waves with a wavenumber
$k_{\rm h} = \sqrt{l(l+1)}/r$.\newline
\item The {\it buoyancy frequency} (or {\it Brunt-V\"ais\"al\"a frequency}) $N$,
\begin{equation}
N^2 = g \left( {1 \over \Gamma_1} {\dd \ln p \over \dd r} 
- {\dd \ln \rho \over \dd r} \right) \; ,
\label{eq:buoy}
\end{equation}
where $g$ is the local gravitational acceleration, $p$ is pressure
and $\Gamma_1 = (\partial \ln p/\partial \ln \rho)_{\rm ad}$,
the derivative being for an adiabatic process;
note that $N^2$ is negative in convectively unstable regions. The Brunt-V\"ais\"al\"a frequency is the local frequency of internal gravity waves of short horizontal wavelength.\newline
\item The {\it acoustic cut-off frequency}  $\omega_{\rm c}$, 
\begin{equation}
\omega_{\rm c}^2 = {c^2 \over 4 H^2} \left(1 - 2 {\dd H \over \dd r} \right)
 \; ,
\label{eq:fullac}
\end{equation}
where $H = - (\dd \ln \rho / \dd r)^{-1}$ is the \textit{density scale height}. %
The acoustic cut-off frequency arises from the inability of modes to propagate
when their vertical wavelength is too long compared with 
the scale of the density variation in the equilibrium structure.
This leads to reflection of the waves. 
We note that since $\omega_{\rm c}^2$
depends on the second derivative of density 
(see the definition of density scale height and Eq.~\ref{eq:fullac}), 
it varies rapidly in the
region of substantial superadiabaticity just below the surface, as shown in the right-hand panel of Figure~\ref{fig:charfrqsun}. On the other hand, in the nearly isothermal atmosphere $H \simeq H_p$ is essentially constant, and $\omega_c^2 \simeq c^2 /(4 H^2)$. Thus $\omega_{\rm c}/2 \pi$ reduces to $\nu_{\rm ac}$ (cf. Eq.~\ref{eq:nu_ac}).
\end{itemize}

According to Equations (\ref{eq:basasymp}) and (\ref{eq:kasymp}) the
behaviour of the oscillations is determined by the dependence of
the characteristic frequencies on position.
This is illustrated in Figure~\ref{fig:charfrqsun} for a model 
approximating the present Sun.
To characterize the location of typical frequencies of solar-like 
oscillations the horizontal line shows the estimated frequency $\nu_{\rm max}$
of maximum power, obtained as
\begin{equation}
\nu_{\rm max} = 0.6 \nu_{\rm ac} \; ,
\label{eq:numaxt}
\end{equation}
where $\nu_{\rm ac}$ is the isothermal acoustic cut-off frequency
(cf. Eq.~\ref{eq:nu_ac}); in the Sun $\nu_{\rm max} \simeq 3150 \muHz$.
At this frequency $\omega \gg N$ except in the atmosphere,
reducing Eq.~(\ref{eq:kasymp}) to
$K \simeq c^{-2}(\omega^2 - S_l^2 - \omega_c^2)$.
The corresponding modes are acoustic modes, or {\it p modes},
where pressure is the restoring force.
In the region where $K > 0$ the solution $X$ oscillates as a function of $r$
(see  Eq.~\ref{eq:basasymp}),
whereas $X$ behaves locally exponentially where $K < 0$. 
For acoustic oscillations the oscillatory region,
{\orange also known as the {\it p-mode cavity},
extends from the surface} to a distance $r_{\rm t}$
from the centre, approximately given by $\omega \simeq S_l$, or
\begin{equation}
{\omega \over \sqrt{l(l+1)}} = {c(r_{\rm t}) \over r_{\rm t}} \; .
\label{eq:rt}
\end{equation}
Physically this corresponds to total internal reflection of the oscillations,
described as a superposition of sound waves.
The upper turning point $R_{\rm t}$ is where 
$\omega \simeq \omega_{\rm c}(R_{\rm t})$,
which is satisfied just below the photosphere at high frequency
and somewhat deeper at lower frequency 
(see right panel of Fig.~\ref{fig:charfrqsun}).
Waves with frequencies exceeding the acoustic cut-off frequency
in the atmosphere
are free to travel outwards in the atmosphere, resulting in strong damping. 

A more quantitative analysis of Eq.~(\ref{eq:basasymp}) can be carried out
using JWKB theory \citep[see][]{gough2007}.
In general, this results in an eigenvalue condition on $\omega$ given by
\begin{equation}
\int_{r_1}^{r_2} K^{1/2} \dd r = (k - 1 / 2) \pi \; ,
\label{eq:jwkb}
\end{equation}
for integer $k$, where $r_1$ and $r_2$ are adjacent turning points 
at which $K = 0$, such that $K > 0$ between $r_1$ and $r_2$.
In the present case, for predominantly acoustic modes in main-sequence stars,
this leads to
\begin{equation}
\omega \int_{r_{\rm t}}^{R_{\rm t}} \left(1 - {\omega_{\rm c}^2 \over \omega^2}
- {S_l^2 \over \omega^2} \right)^{1/2} {\dd r \over c} 
\simeq (k - 1/2)\pi \; .
\label{eq:acujwkb}
\end{equation}
For low-degree modes, $r_{\rm t}$ is
close to the centre in main-sequence stars. Equation (\ref{eq:acujwkb}) can be reduced to
\begin{equation}
	\nu_{n l} \simeq \Delta\nu \left( n+\frac{l}{2}+\epsilon \right)
	-d_{n l} \; ,
\label{eq:tassoul1}
\end{equation}
using an expansion around $r = 0$ as well as near the surface
\citep{gough1986b, gough1993}.
For main-sequence stars $k$ can in general be related directly to the 
{\it radial order} $n$ of the mode which was therefore
used instead of $k$ in Eq.~(\ref{eq:tassoul1}).
For evolved stars the definition of mode order is more complex;
we return to this in Section~\ref{sec:modeorder}.
Eq.~(\ref{eq:tassoul1}) is the basis for the different frequency separations
\begin{equation}
\Delta\nu = \nu_{n l} - \nu_{n-1\,l} 
\simeq \left(2 \int_0^R \frac{{\rm d}r}{c} \right )^{-1} \; ,
\label{eq:larsep1}
\end{equation}
\begin{equation}
\delta\nu_{l\,l+2} (n)
= \nu_{n l}-\nu_{n-1\,l+2} \simeq -(4l+6)\frac{\Delta\nu}{4\pi^2 \nu_{n l}}\int_0^R\frac{{\rm d} c}{{\rm d} r}\frac{{\rm d} r}{r} \; ,
\label{eq:smallsep}
\end{equation}
and the phase term $\epsilon$.
As discussed in Sections~\ref{sect:dnu} -- \ref{sect:sdnu_01} 
these quantities provide
important diagnostics of stars based on their acoustic oscillations.%
\footnote{We note that the asymptotic behaviour of $\delta \nu_{l\,l+2}$
is only valid for main-sequence stars, and even here it
has limited validity; e.g., \citet{jcd1991, aerts2010}.}
From homology scaling $c^2 \propto M/R$.
It follows from Eq.~(\ref{eq:larsep1}) that $\Delta \nu$, and hence
from the leading-order first term in Eq.~(\ref{eq:tassoul1}) $\nu_{nl}$,
scale as
\begin{equation}
\nu_{nl} \propto \Delta \nu \propto \sqrt{\frac{M}{R^3}} 
\propto \sqrt{ \bar \rho}  \; .
\label{eq:acscale}
\end{equation}
From Eq.~(\ref{eq:lamb}) it follows that the same scaling applies to $S_l$.

\begin{figure}
\centering
\begin{minipage}{\linewidth}
\centering
\includegraphics[width=9cm]{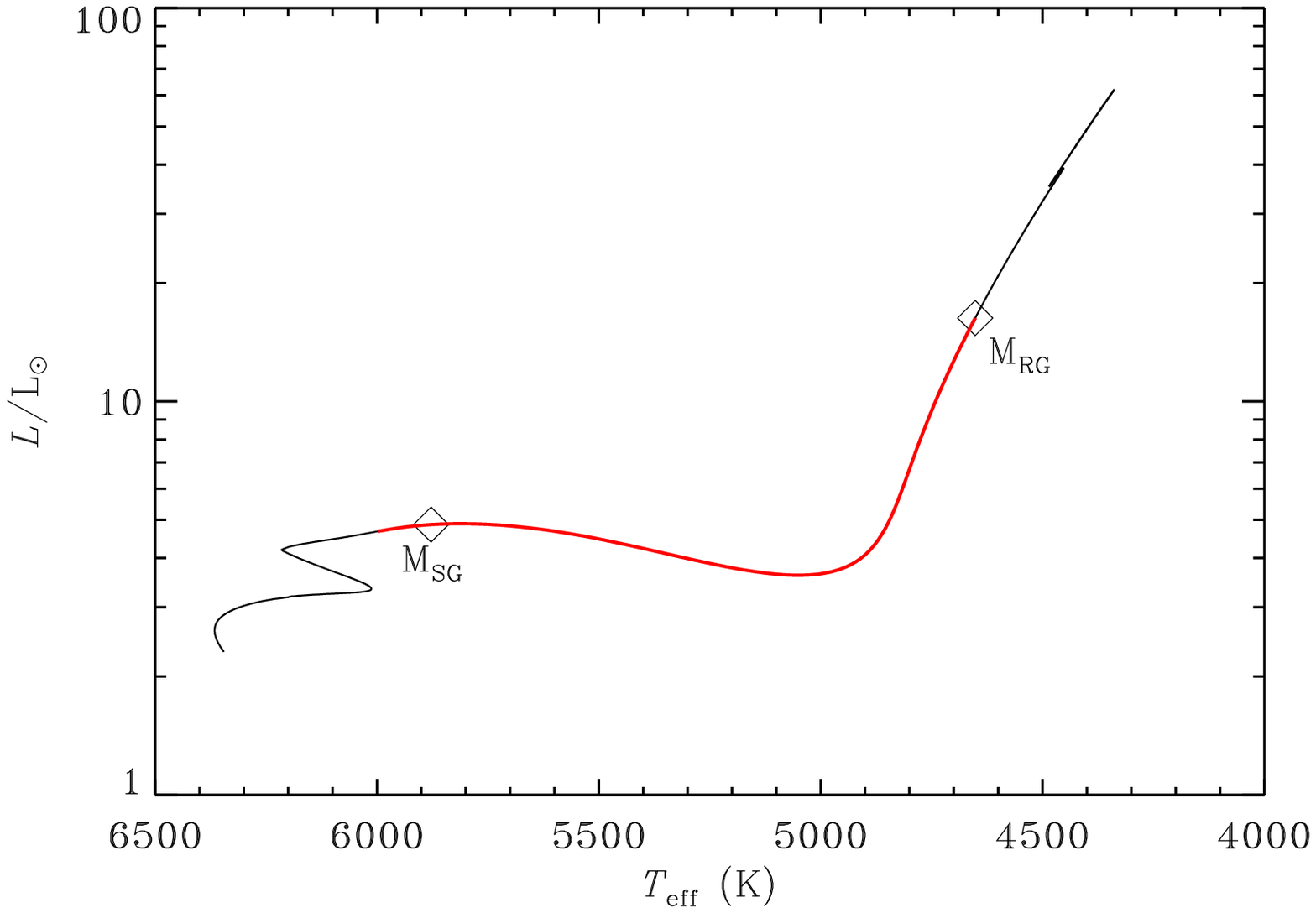}
\end{minipage}
\caption{Evolution track of a $1.3 \Msun$ star,
from \citep{jiang2014}.
The models $\modsg$ and $\modrg$ which are analysed below are
marked by diamonds, and the segment corresponding to Fig.~\ref{fig:chenfig} is indicated in red.
}
\label{fig:chenhr}
\end{figure}
\begin{figure}
\centering
\begin{minipage}{\linewidth}
\centering
\includegraphics[width=9cm]{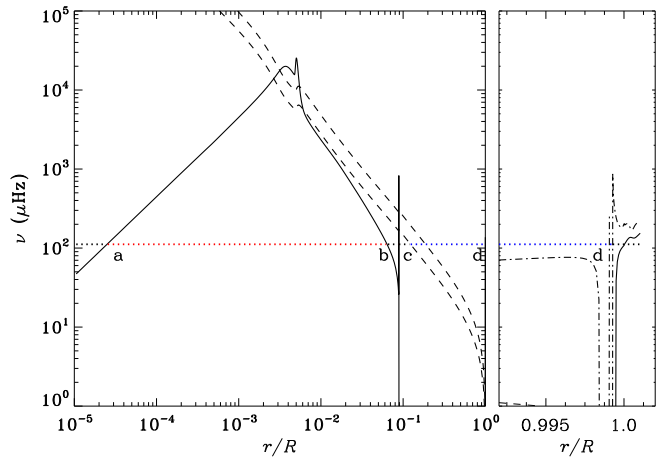}
\end{minipage}
\caption{Characteristic frequencies for a red-giant model with
mass $1.3\,\Msun$ and radius $6.2\,\Rsun$ (Model $\modrg$ in Fig.~\ref{fig:chenhr});
a -- d mark the turning points for $l = 1$
(cf. the analysis in Section~\ref{sec:ascenrg})
at the frequency indicated by the dotted line (cf. Eq.~\ref{eq:numaxt}),
{\orange with the red and blue parts marking respectively
the g-mode and p-mode cavities for $l = 1$.}
See caption to Fig.~\ref{fig:charfrqsun} for the meaning of the different linestyles.
}
\label{fig:charfrqrg}
\end{figure}

To investigate the changes in oscillation properties as a star evolves
from the main sequence through the subgiant phase to the red-giant branch
we consider a $1.3 \Msun$ evolution sequence from \citet{jiang2014}.
The evolution in the HR diagram is illustrated in Fig.~\ref{fig:chenhr}.
The behaviour of the Lamb frequency remains similar as the star evolves,
apart from the scaling with $\bar \rho^{1/2}$.
However, the behaviour of the Brunt-V\"ais\"al\"a frequency is
very different for red-giant models compared with main-sequence models.
This can be seen by approximating $N^2$ as
\begin{equation}
N^2 \simeq {g^2 \rho \over p} ( \nabla_{\rm ad} - \nabla + \nabla_\mu) \; ,
\label{eq:bvapprox}
\end{equation}
where $\nabla = \dd \ln T / \dd \ln p$, $\nabla_{\rm ad}$ is its adiabatic
value and $\nabla_\mu = \dd \ln \mu / \dd \ln p$, with $\mu$ being the mean
molecular weight.
In a red giant with a compact core $g$ reaches very high values in
the deep interior of the star and so therefore does $N$.
This is illustrated in Fig.~\ref{fig:charfrqrg} (note the logarithmic abscissa).
Given the larger radius, both $S_l$ and $\nu_{\rm max}$ are substantially
reduced. However, the most dramatic difference compared with Fig.~\ref{fig:charfrqsun}
is the very large value of $N$ in the core.
Additional features in $N$ are caused by the composition discontinuity
at $r \simeq 0.09 R$ left behind after the first dredge-up 
(see Sect.~\ref{par:1dredge})
and local maximum at $r \simeq 0.005 R$ arising from the steep abundance
gradient in the hydrogen-burning shell.
Note also the very deep convective envelope, where $N$ is imaginary and hence
not shown.

For ascending-branch red-giant models such as the one shown in 
Fig.~\ref{fig:charfrqrg} there are two regions where $K >0$, 
leading to an oscillatory behaviour of the eigenfunctions, i.e.,
with $\omega > S_l, N$ or $\omega < S_l,N$.
Of these, the outer region ($\omega > S_l, N$) corresponds essentially to the 
{\orange p-mode cavity} in main-sequence stars discussed above.
Modes trapped in this region, with the eigenfunction decreasing
exponentially below it, satisfy Eq.~(\ref{eq:acujwkb}).
We note, however, that the analysis leading to the asymptotic approximation 
for low-degree modes cannot immediately be transferred to more evolved stars,
given that the lower turning points lie outside the compact core
(see also Section~\ref{sec:ascenrg}).
Even so, both model computations and observations show that the acoustic
modes of red giants satisfy a relation very similar to Eq.~(\ref{eq:tassoul1}),
i.e., the `universal pattern'.
This was discussed in Sect.~\ref{sect:pattern},
where departures from this pattern were also mentioned.
Indeed, \citet{dziembowski2012} found from stellar models
that for the most luminous red-giant branch stars the dipolar mode frequencies
are shifted substantially relative to the location at
the mid-point between the neighbouring
radial-mode frequencies which is predicted by the leading-order term
in Eq.~(\ref{eq:acujwkb}).
This was confirmed observationally by \citet{stello2014}.

The inner oscillatory region ($\omega < S_l, N$) is in the core of the model, 
where the frequency is below the Brunt-V\"ais\"al\"a frequency.
Modes trapped in this {\orange {\it g-mode cavity}}
are standing internal gravity waves,
{\it g modes}, where buoyancy is the restoring force.
Their frequencies may be estimated
from Eq.~(\ref{eq:jwkb}), with $r_1$ and $r_2$ being approximately the points
where $\omega = N$.
Approximating $K$ by assuming that $\omega \ll S_l$ and neglecting 
$\omega_c$ we obtain
\begin{equation}
\sqrt{l(l+1)}\int_{r_1}^{r_2} \left({N^2 \over \omega^2} - 1 \right)^{1/2}
\dd r \simeq (k - 1/2) \pi \; .
\label{eq:gravjwkb}
\end{equation}
In most of the region $\omega \ll N$. With a correction for the behaviour near the turning points we can therefore approximate this further. We obtain
a relation for the period $\Pi = 2 \pi / \omega$:
\begin{equation}
\Pi_{k\,l} = {\Pi_0 \over \sqrt{l(l+1)}} (k + \epsilon_{\rm g} +1/2) \; ,
\label{eq:gasymp}
\end{equation}
where
\begin{equation}
\Pi_0 = 2 \pi^2 \left( \int_{r_1}^{r_2} N {\dd r \over r} \right)^{-1} \; ,
\label{eq:gper0}
\end{equation}
and $\epsilon_{\rm g}$ is a phase term accounting for the behaviour near the
turning points.%
\footnote{We note that we follow the notation by
\citet{mosser2012core} for $\epsilon_{\rm g}$ 
here and in the subsequent discussions.} 
Thus for modes trapped in the core of the model we obtain
oscillations with uniformly spaced periods.
These periods increase with increasing $k$, and with a spacing 
$\Pi_{k+1\,l} - \Pi_{k\,l} \simeq \Delta \Pi_l = \Pi_0 /\sqrt{l(l+1)}$
which depends on the degree of the modes. 
The diagnostic power of the period spacing is summarized
in Section~\ref{sect:dP}.
As discussed in Section~\ref{sec:modeorder} g modes are by convention
assigned negative radial orders $n_{\rm g}$, with frequency tending
to zero as $n_{\rm g}$ tends to $- \infty$.
Thus at least in simple cases of a pure g-mode spectrum 
$k$ in Eq.~(\ref{eq:gper0}) can be identified with $|n_{\rm g}|$.
We return to a more complete discussion of mode order in 
Section~\ref{sec:modeorder} and beyond.

The preceding discussion was based on assuming 
the Cowling approximation,
reducing to two the order of the equations of adiabatic pulsations.
\citet{dziembowski2012} pointed out, however, that this approximation is
questionable for dipolar modes (with $l = 1$).
Here the perturbation to the gravitational potential gives rise to a
slowly varying component of the solution which may have a significant effect
on the properties of the modes.
\citet{takata2005, takata2006} provided the basis for a more complete analysis
of these modes by introducing a change of variables that reduces the full
oscillation equations for $l = 1$ to a second-order system, 
facilitating the asymptotic analysis.
A detailed asymptotic analysis in this case was carried out by
\citet{takata2016a} and supplemented by a more physical analysis, of
broader applicability, by \citet{takata2016b}.
Qualitatively the new analyses are broadly consistent with earlier work on
which we focus here.
However, they contribute greatly to an understanding of the properties of
the oscillations and the detailed diagnostic potential of the observed
oscillation properties.
Thus Takata's results will undoubtedly play a major role in the further
development of the field.

In the analysis leading to Eqs~(\ref{eq:acujwkb}) and (\ref{eq:gravjwkb})
we assumed that the modes were completely trapped in the corresponding regions, i.e., acoustic modes with pressure as the restoring force (p modes) in the outer part of the star and gravity (g) modes with buoyancy as the restoring force in the core.
However, it is clear from Figure~\ref{fig:charfrqrg} that the evanescent 
region separating the two trapping regions is quite thin, particularly
for $l = 1$ modes.
This leads to substantial coupling between the two regions and hence
generally to a mixed character of the modes.
These mixed modes are responsible for the diagnostic richness 
of the solar-like oscillations in evolved stars.
We discuss this in the following subsection.

\subsection{Mixed modes}
\label{sec:themixedmodes}

\begin{figure}
\centering
\begin{minipage}{\linewidth}
\centering
\includegraphics[width=11cm]{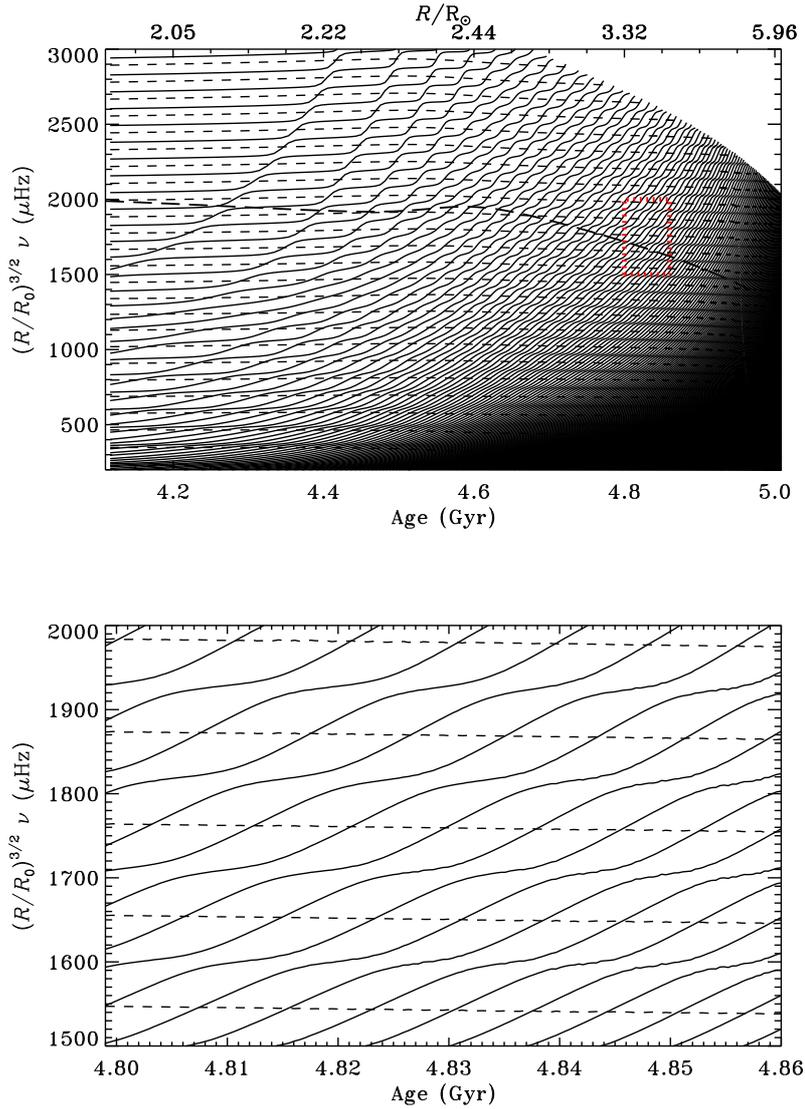}
\end{minipage}
\caption{Evolution of frequencies for a $1.3 \Msun$ stellar model, 
from the beginning of the subgiant phase through the early red-giant phase
(marked by the red part of the line in Fig.~\ref{fig:chenhr};
see Sect.~\ref{sect:Hburn}).
The frequencies have been scaled by $(R/R_0)^{3/2}$, where $R_0$ is the ZAMS
radius, to correct for the evolution with mean density of 
the acoustic-mode frequencies (cf. Eq.~\ref{eq:dnu}).
The heavy long-dashed line shows an estimate of $\nu_{\rm max}$
(cf. Eq.~\ref{eq:numaxt}), similarly scaled.
The dashed lines show radial modes while the solid lines show dipolar modes,
exhibiting avoided crossings first mentioned by \citet{aizenman1977}.
The lower panel shows a small part, marked by a red rectangle,
of the upper panel, to illustrate the behaviour in more detail.
Adapted from \citet{jiang2014}.}
\label{fig:chenfig}
\end{figure}

The first to consider non-radial mixed modes in highly-evolved stars was likely
\citet{dziembowski1971}, who analysed the oscillations of Cepheid-type stars.
Dziembowski noted that the huge values of the buoyancy frequency in the core of
such stars meant that even at high frequencies the modes behaved as standing
internal gravity waves in the inner parts of the star.
Dziembowski carried out an asymptotic analysis of these properties. This was subsequently followed by
a detailed investigation concerning the effects
of non-adiabatic properties of the modes by \citet{dziembowski1977a}.
\citet{scuflaire1974} analysed the oscillation properties of polytropes of
high polytropic index and hence centrally condensed models. Scuflaire similarly noticed
the mixed character of the modes and may have been the first to use explicitly
the term `{\it mixed modes}'.
\citet{osaki1975} followed the evolution of mixed modes in a massive
main-sequence model. He noticed the effect of the increasing frequencies 
of the gravity waves in the core that led to the characteristic behaviour
of the model frequencies with age (see Fig.~\ref{fig:chenfig}).

The overall evolution of the frequencies of mixed modes is illustrated in
Fig.~\ref{fig:chenfig},
showing results for a $1.3 \Msun$ evolution sequence
extending from just after central hydrogen exhaustion to the early part of the
red-giant ascent (cf. Fig.~\ref{fig:chenhr}).
To eliminate the dependence of the acoustic-mode frequencies
on stellar radius the frequencies have been
scaled with $(R/R_0)^{3/2}$ (cf. Eq.~\ref{eq:dnu}),
where $R_0$ is the zero-age main-sequence radius, 
such that the modes following the acoustic scaling relation,
Eq.~(\ref{eq:acscale}), appear with constant frequency.
This is approximately the case for the radial modes
(dashed lines in Fig.~\ref{fig:chenfig}), and for
$l = 1$ (solid lines in Fig.~\ref{fig:chenfig})
for those modes that are predominantly of acoustic nature.
However, there is clearly a second class of modes with scaled frequencies 
increasing with age.
These modes are predominantly of g-mode character.
Their frequencies increase
with age following the strong increase in the buoyancy frequency caused
by the contraction of the core (cf. Eq.~\ref{eq:bvapprox}).
Where the frequency of such a mode meets {\orange an acoustically} dominated mode
the frequencies do not cross but approach quite closely. This is 
followed by an exchange of mode character such that the mode 
previously of g-mode nature becomes predominantly acoustic and vice versa.
This behaviour was first noticed for evolving stars by \citet{osaki1975}.
It was also investigated by \citet{aizenman1977},
who were probably the first to use the term `{\it avoided crossing}'%
\footnote{also, more poetically, known as `mode kissing'}
to describe it in astrophysics.
Note however that similar phenomena have wide applicability to cases
of coupled oscillations, including in atomic physics.
An early illustrative analysis was provided by \citet{neuman1929}.

Even though the actual frequencies do not cross it is sometimes useful
to relate them to fictitious uncoupled gravity and acoustic modes. These are
the so-called `$\gamma$' and `$\pi$' modes, respectively, that do cross
\citep[see][]{aizenman1977}.
As discussed below the behaviour of the actual modes can then be
analysed by introducing coupling between these fictitious modes.
This also allows the definition of the numbers $N_\pi$ and $N_\gamma$
that represent the number of $\pi$ and $\gamma$ modes 
in the relevant frequency interval, e.g., corresponding to the
range of radial modes \citep[e.g.,][]{benomar2013}.
In the earlier phases of evolution $N_\gamma \ll N_\pi$,
whereas on the red-giant branch 
(towards the right edge of Fig.~\ref{fig:chenfig})
$N_\pi \ll N_\gamma$.
This variation in the overall structure of the oscillation spectrum
has a major effect on the observed Fourier spectra
(see Fig.~\ref{fig:ps_numax}).

To characterize the relative contributions of the different regions of the 
star to a mixed mode, a useful quantity is {\it normalized inertia}
$E$. This is defined by
\begin{equation}
E = {\int_V \rho |\bolddelr|^2 \dd V \over M |\bolddelr|_{\rm phot}^2 } \; ,
\label{eq:inertia}
\end{equation}
where the integral is over the volume of the star, and the normalization
uses the average squared photospheric displacement.
In addition to $E$ we also consider {\it mode mass} $M_{\rm mode} = M E$, 
where $M$ is the mass of the star.
These quantities are defined such that the average kinetic energy of 
the oscillation is
\begin{equation}
E_{\rm kin} = {1 \over 2} M_{\rm mode} V_{\rm rms}^2 \; ,
\label{eq:ekin}
\end{equation}
where $V_{\rm rms}$ is the average photospheric velocity.
For p-dominated modes $E$ is largely a function of frequency.
This frequency dependence is determined by the depth of the upper turning point $R_{\rm t}$,
and hence the decrease in 
amplitude of the eigenfunction between the oscillatory region
below $R_{\rm t}$ and the surface.
The value of $E$ can be much larger for g-dominated modes,
with a considerable amplitude in the gravity-wave propagating
region in the deep interior, than for p-dominated modes.
This is conveniently characterized by the scaled inertia
\begin{equation}
Q = {E \over \bar E_0(\omega)} \; ,
\label{eq:sclinertia}
\end{equation}
where $\bar E_0(\omega)$ is the radial (purely acoustic) mode inertia at the frequency of the mode considered.

For the analysis of red-giant oscillations it is convenient to consider the
fraction of the mode inertia that comes from the inner parts of the star 
relative to the total mode inertia.
Following \citet{goupil2013} we introduce
\begin{equation}
\zeta = {E_{\rm core} \over E} \; ,
\label{eq:zeta}
\end{equation}
where $E_{\rm core}$ is defined as in Eq.~(\ref{eq:inertia}) but 
restricting the integral to the region where $\omega < N, S_l$.
Evidently $\zeta$ is small for modes trapped in the envelope, whereas
$\zeta$ is close to one for modes trapped in the core.
We also note that, as a rough approximation,
\begin{equation}
\zeta \simeq 1 - Q^{-1} \; ,
\label{eq:qzeta}
\end{equation}
assuming that the envelope contribution to the inertia is 
similar to the inertia of a radial mode with the same frequency.%
\footnote{This neglects the contribution from the evanescent region
to the inertia in the calculation of $Q$, which may be significant.
Thus estimating $Q$ from $\zeta$ using Eq.~(\ref{eq:qzeta})
leads to an underestimate for g-dominated modes.
\label{fn:qzeta}}

For the practical evaluation and later analysis we note
that the inertia can be expressed in terms of the displacement vector
(\ref{eq:deltar}) as
\begin{equation}
E = {4 \pi \int_0^R [\xi_r(r)^2 + l(l+1) \xi_{\rm h}(r)^2] \rho r^2 \dd r
\over M [\xi_r(R_{\rm phot})^2 + l(l+1) \xi_{\rm h}(R_{\rm phot})^2]} \; ,
\label{eq:inertia_comp}
\end{equation}
where $R_{\rm phot}$ is the photospheric radius.

\subsubsection{Mode order}
\label{sec:modeorder}

The mixed nature of the modes precludes a simple identification of the 
radial order of a mode based on the number of nodes.
Using an earlier analysis by \citet{eckart1960},
\citet{scuflaire1974} and \citet{osaki1975} independently 
proposed a scheme plotting the eigenfunction in a suitable phase diagram,
e.g., in terms of $(\xi_r, \xi_{\rm h})$;
the radial order is determined by counting the zero-crossings of $\xi_r$
with a positive or negative sign depending on
whether the curve crosses the axis in the counter-clockwise or clockwise
direction in the phase diagram.
These zero-crossings are associated with the regions in the star where the
mode has a p-mode or a g-mode character, respectively.
If the number of counter-clockwise crossings is $\hat n_{\rm p}$ and the number
of clockwise crossings is $|\hat n_{\rm g}|$,%
\footnote{Here the hat is used to distinguish these numerical contributions
to the order from the asymptotic properties, discussed below.}
with $\hat n_{\rm g} < 0$, the mode order is
\begin{equation}
n = \hat n_{\rm p} + \hat n_{\rm g} \; .
\label{eq:modeorder}
\end{equation}
When the perturbation to the gravitational potential is neglected
this defines a mode order that is not changed for a given mode
as the star evolves, even though its dominant physical character may change.
For modes of degree $l \ge 2$ this property has also been found to be
satisfied for solutions of the full equations of adiabatic oscillation.
On the other hand, an application of the Eckart scheme to dipolar
modes ($l = 1$) leads to a poorly defined mode order for
evolved models \citep{lee1985, guenther1991} or centrally condensed polytropes
\citep{jcdmullan1994}.
It was shown by \citet{takata2005, takata2006} that a relation satisfied by
the eigenfunctions of dipolar modes allows the definition of a scheme
for the determination of mode order that is well-defined and invariant
under evolution.
As in Eq.~(\ref{eq:modeorder}) this is characterized by contributions
$\hat n_{\rm p}$ and $\hat n_{\rm g}$ from the p-mode and g-mode dominated
parts of the star.
We use this in the later discussion of dipolar modes.
Together with the original scheme proposed by \citet{scuflaire1974}
and \citet{osaki1975} this defines mode orders that are invariant under
evolution.
Thus the order is unchanged when following a given mode as the star
evolves in Fig.~\ref{fig:chenfig}.

Formally modes with positive $n$ may be classified as p modes and modes
with negative $n$ as g modes.
This classification largely corresponds to the physical nature of the modes
for unevolved stars.
However, as discussed below the non-radial mixed modes develop
a large number of nodes in
the g-mode propagation region as the star evolves on the red-giant branch,
and hence typically have a large negative value of $\hat n_{\rm g}$.
Thus for such stars all relevant modes have negative $n$, and other properties
of the modes must be used to characterize their physical nature.
In any case it should be kept in mind that the radial order defined here
is a purely theoretical concept, although very useful in characterizing
the oscillation modes of a given stellar model.

\subsubsection{Subgiant stars}
\label{sect:thsubgiants}

\begin{figure}
\centering
\begin{minipage}{\linewidth}
\centering
\includegraphics[width=9cm]{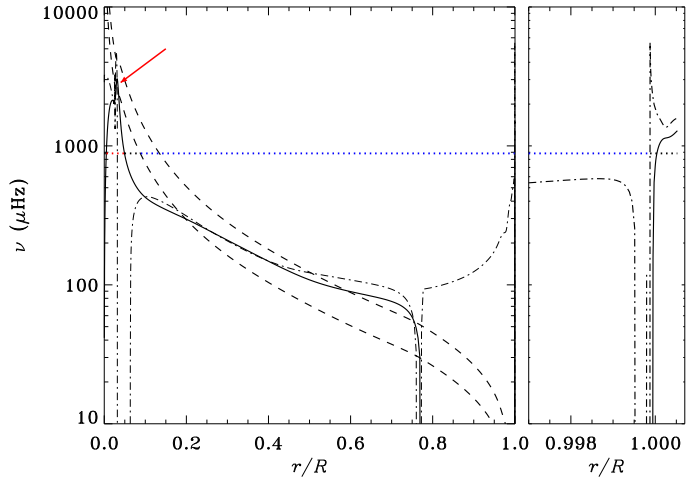}
\end{minipage}
\caption{Characteristic frequencies for the $1.3 \Msun$ subgiant model
$\modsg$ in Fig.~\ref{fig:chenhr}, of age 4.293\,Gyr
and with $T_{\rm eff} = 5887.8\,$K.
The red arrow marks the peak in the buoyancy frequency arising from
the hydrogen-burning shell.
See {\orange captions to} Figs~\ref{fig:charfrqsun}
{\orange and \ref{fig:charfrqrg}} for the meaning of the linestyles.
}
\label{fig:charfrqchen}
\end{figure}

\begin{figure}
\centering
\begin{minipage}{0.48\linewidth}
\centering
\includegraphics[width=6cm]{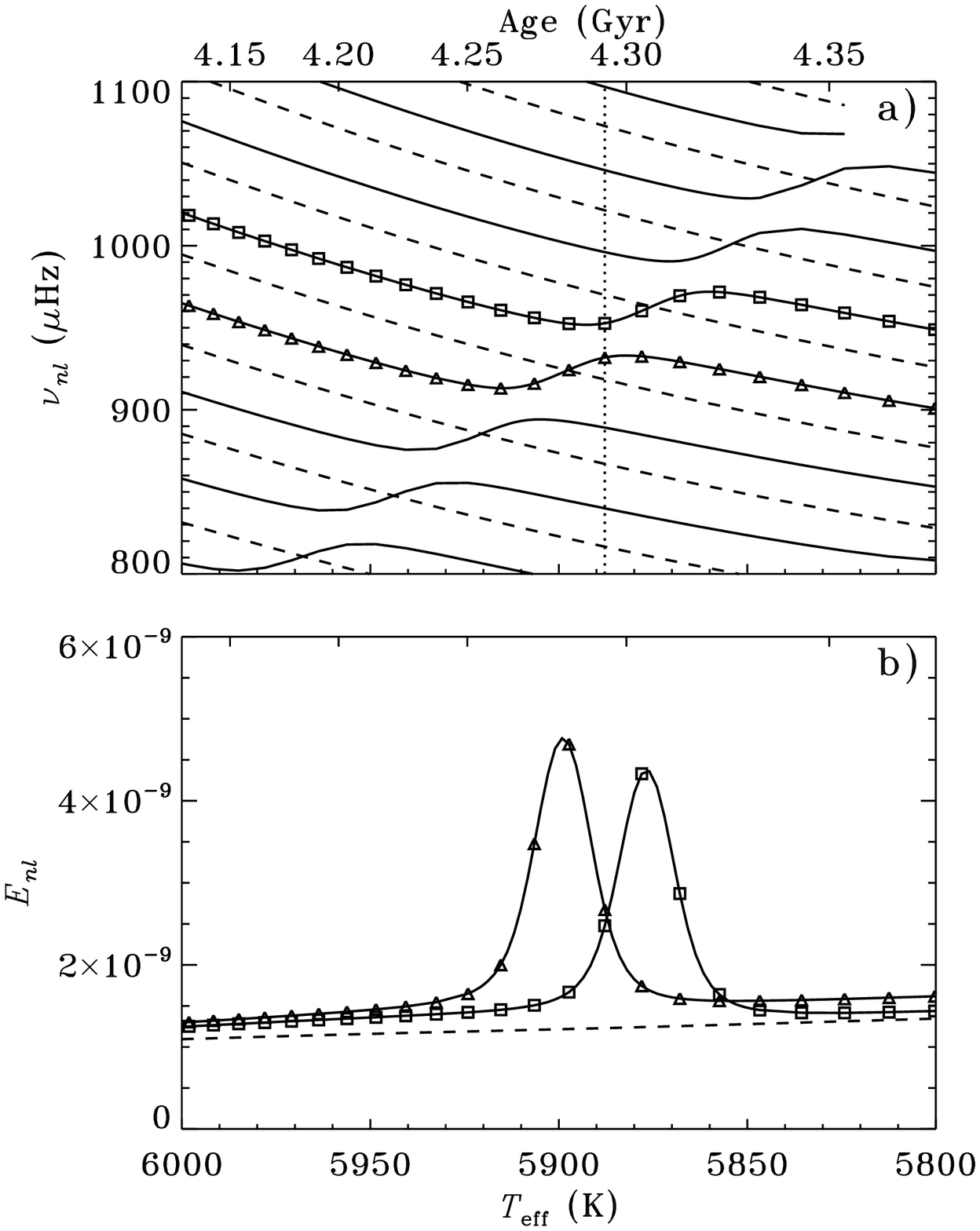}
\end{minipage}
\begin{minipage}{0.48\linewidth}
\centering
\includegraphics[width=6cm]{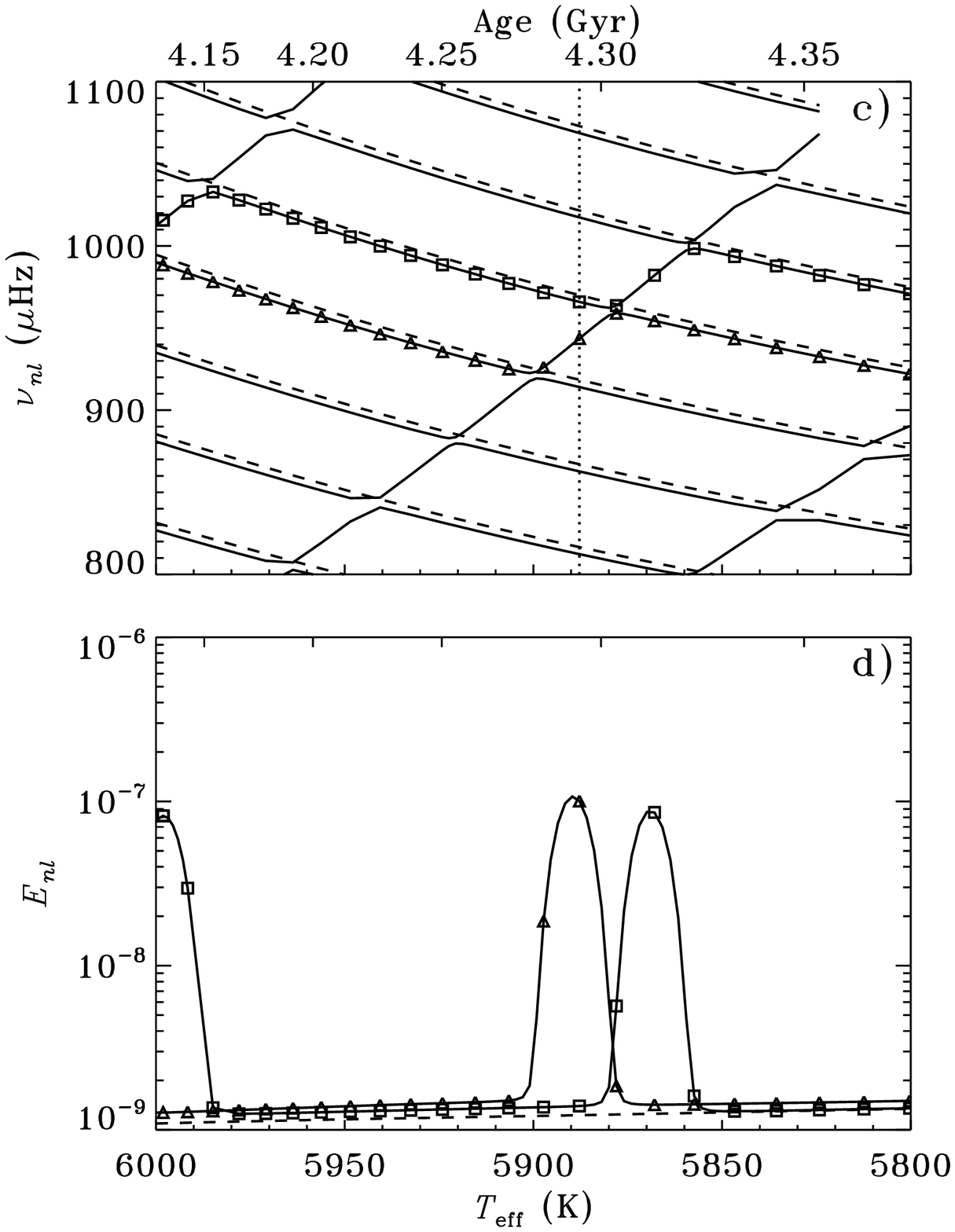}
\end{minipage}
\caption{Evolution of frequencies and mode inertias 
as a function of age (upper abscissa) and
effective temperature (lower abscissa),
for a $1.3 \Msun$ evolution sequence (cf.\ Fig.~\ref{fig:chenhr}), 
including the model $\modsg$
illustrated in Fig.~\ref{fig:charfrqchen}.
Dashed lines show modes with $l = 0$, and solid lines show modes
with $l = 1$ (left) and $l = 2$ (right).
The lower panels show the evolution of the inertia for the modes
identified by triangles and squares in the upper panels,
as well as a neighbouring radial mode (dashed line); note that the right-hand panel uses a logarithmic ordinate scale.
The vertical dotted line in the top panels 
marks the model corresponding to the \'echelle 
diagram in Fig.~\ref{fig:chenechl}.
}
\label{fig:frqchenl1}
\end{figure}

As a star evolves, the first g-dominated mixed modes start to be observable
in subgiants. 
This provides a useful illustration of the properties of mixed modes.
The characteristic frequencies for a subgiant model are shown in 
Fig.~\ref{fig:charfrqchen}.
This model has a helium core of radius $0.022 R$,
containing 5\,\% of the star's mass.
The behaviour of the acoustic frequencies differs little from those of
a main-sequence model.
However, the compact core and resulting high gravitational acceleration
give rise to a high peak in the buoyancy frequency in the core,
augmented by the sharp composition gradient in 
the hydrogen-burning shell.
Therefore, at the indicated typical oscillation frequency
(horizontal dotted line in Fig.~\ref{fig:charfrqchen})
there are two trapping regions.
The increase in the core buoyancy frequency with age
leads to avoided crossings when following the evolution of the modes with age.
This is illustrated in  the top left panel of Fig.~\ref{fig:frqchenl1},
which shows the behaviour of radial and dipolar modes.
The resulting changes in the character of the modes
in terms of the mode inertia for two of the dipolar modes,
compared with a neighbouring radial mode,
are shown in the lower left panel of Fig.~\ref{fig:frqchenl1}.
As long as the dipolar modes are predominantly acoustic their inertia is 
very similar to that of the radial mode.
As a mode undergoes an avoided crossing and takes on a substantial g-mode
character its inertia increases.
At the next avoided crossing its inertia decreases again 
and the mode returns to an acoustic character, 
exchanging character with the next mode.
At the point of closest approach the inertias of the two modes are very 
similar.

Fig.~\ref{fig:charfrqchen} shows that for dipolar modes the evanescent 
region is relatively thin.
This leads to strong coupling between the two oscillatory regions, 
a relatively large minimum separation during the avoided crossings,
and a rather modest increase in the mode inertia when the modes are 
most g-mode like.
This should be contrasted with the case of quadrupolar modes,
shown in the right panels of Fig.~\ref{fig:frqchenl1}.
Here the evanescent region is considerably thicker, the coupling consequently
weaker and the avoided crossings very sharp.
Also, the inertia increases very rapidly as a mode takes on predominantly
g-mode character.
This shows that in any given model it is unlikely to find a mode that is
not either predominantly p- or g-dominated.

The diagnostic potential of mixed modes in subgiant stars was discussed in
Section~\ref{sect:obsmixedmodes}.

\begin{figure}
\centering
\begin{minipage}{\linewidth}
\centering
\includegraphics[width=9cm]{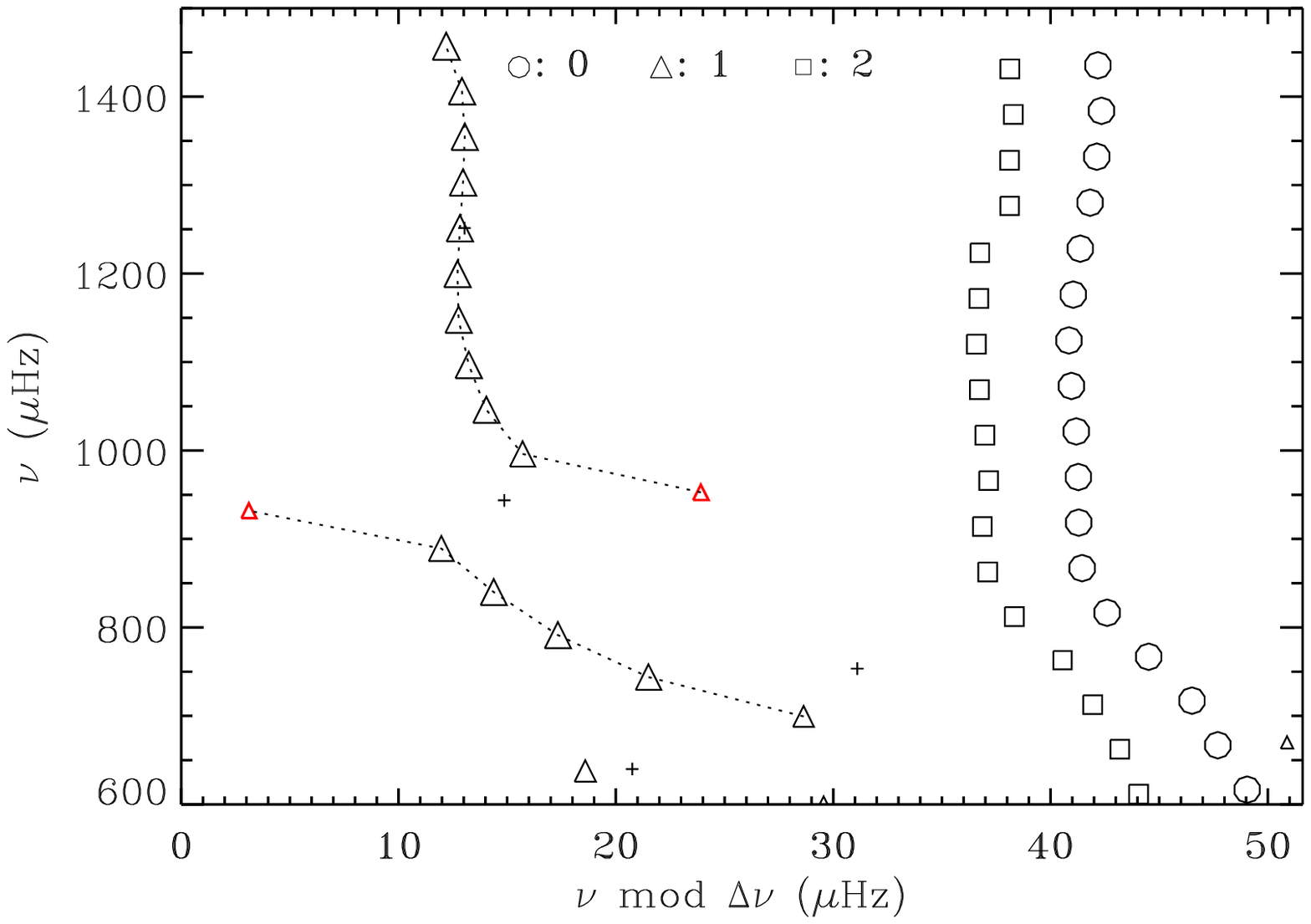}
\end{minipage}
\centering
\begin{minipage}{\linewidth}
\centering
\includegraphics[width=9cm]{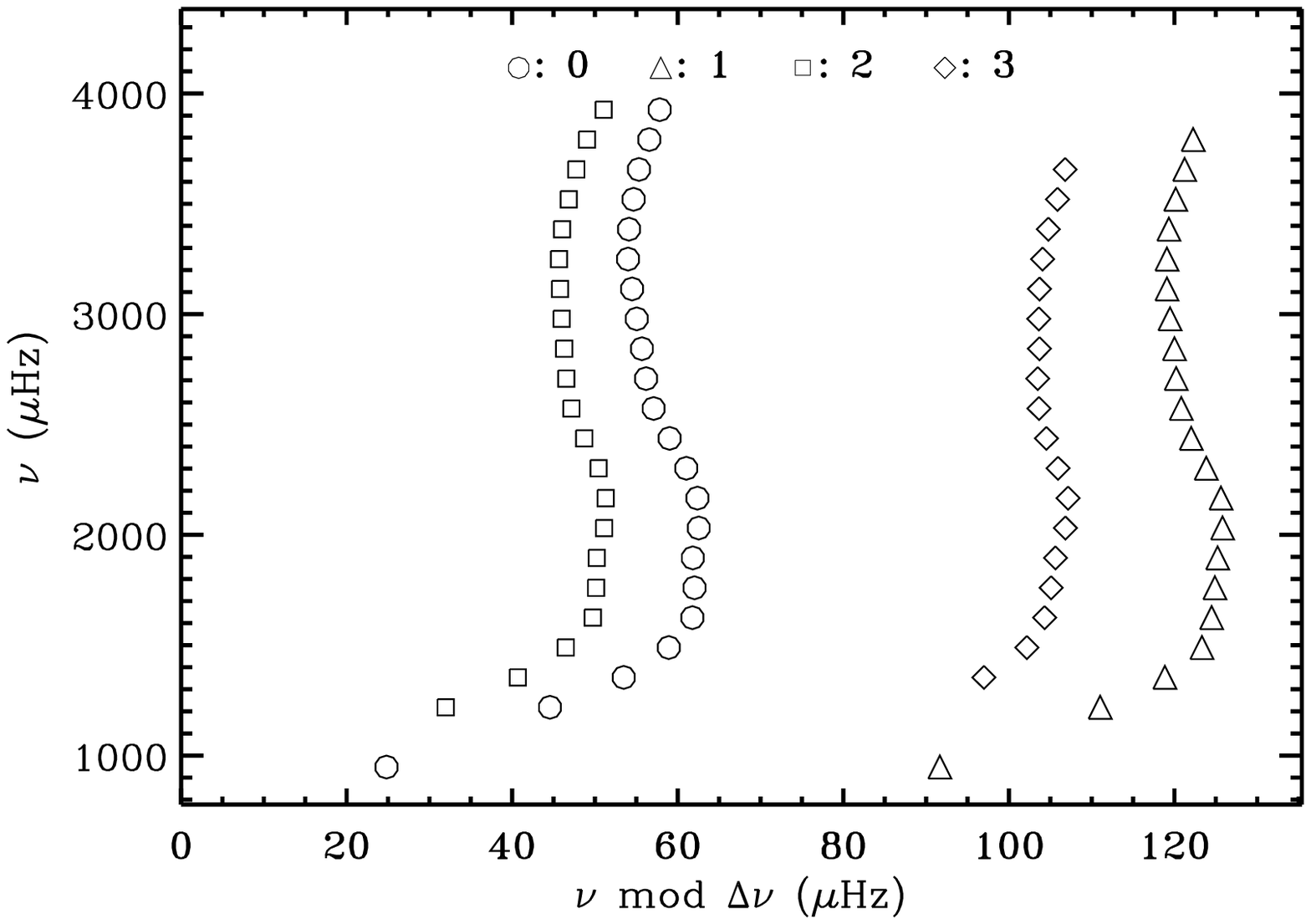}
\end{minipage}
\caption{Top: \'echelle diagram plotted with $\Delta \nu = 51.6 \muHz$,
for the $1.3 \Msun$ subgiant model $\modsg$, illustrated in Fig.~\ref{fig:charfrqchen} and
marked with a vertical dotted line in Fig.~\ref{fig:frqchenl1},
of age 4.293\,Gyr and with $T_{\rm eff} = 5887.8\,$K.
Modes with $l = 0$ are shown as circles, $l = 1$ as triangles, $l = 2$ as squares and $l = 3$ as diamonds.
The size of the symbols is proportional to $Q^{-1/2}$
(cf. Eq.~\ref{eq:sclinertia}), providing a rough estimate of mode amplitude
relative to a radial mode of the same frequency (cf. Eq.~\ref{eq:qsclampl}).
For four quadrupolar modes very small symbols have been replaced by `+'.
Dipolar modes are connected with dotted lines to highlight 
the effects of an avoided crossing;
the two dipolar modes at the centre of an avoided crossing 
in Fig.~\ref{fig:frqchenl1} are shown with red triangles.
Bottom: for comparison we also show an \'echelle diagram for the observed low-degree observations of the Sun \citep[][$\Delta \nu=135.4$~$\mu$Hz]{chaplin2002}. }
\label{fig:chenechl}
\end{figure}

The presence of mixed modes causes departures from the asymptotic behaviour
of pure acoustic modes, as illustrated in the \'echelle diagram in
the top panel of Fig.~\ref{fig:chenechl}; for comparison the bottom panel
shows the corresponding diagram for the purely acoustic modes in the Sun.
In the case of subgiant models the density of g-dominated modes is
relatively low (i.e., $N_\gamma \ll N_\pi$; see above),
and each avoided crossing effectively adds another mode to the frequency
spectrum.
The resulting changes in the oscillation spectrum were
analysed by \citet{deheuvels2010} based on a simple physical 
model of coupled oscillators. 
They considered the coupling between
a single $\gamma$ mode and several $\pi$ modes.%
\footnote{This was extended to the coupling between $N_\pi$ $\pi$
modes and $N_\gamma$ $\gamma$ modes by \citet{benomar2013}.}
The effect on the distribution of peaks in the \'echelle diagram 
is illustrated in Fig.~\ref{fig:chenechl} for the model marked by
a vertical dotted line in Fig.~\ref{fig:frqchenl1}.
At high frequency the dipolar modes have not yet been affected by
g-mode mixing, and the behaviour corresponds to the purely acoustic case.
However, as shown in Fig.~\ref{fig:frqchenl1} there is a pair of 
mixed dipolar modes with nearly the same inertia at a frequency around $930 \muHz$.
These are visible as a pair of modes on either side of the $l = 1$ ridge
in Fig.~\ref{fig:chenechl}.
At lower frequency essentially all the dipolar modes show an effect
of the avoided crossings, as also argued by \citet{deheuvels2010}.

For $l = 2$ the right-hand panels of Fig.~\ref{fig:frqchenl1} also indicate the presence of 
g-dominated modes. 
However, for these the scaled inertias $Q_{n\,l}$ are so high that they
would not be visible with the proposed scaling of symbol size 
in Fig.~\ref{fig:chenechl}.
Therefore, they are indicated by plusses.
As discussed above the high scaled inertias of mixed $l=2$ modes are a consequence of the much weaker coupling between the buoyancy and acoustic cavities for quadrupole modes.
Only mixed quadrupole modes with frequencies close to the pure acoustic $l=2$ mode can reach observable amplitudes and could thus be detected. However, such modes would be difficult to identify observationally, 
except if a clear pair of closely spaced modes is detected.

\begin{figure}
\centering
\begin{minipage}{\linewidth}
\centering
\includegraphics[width=9cm]{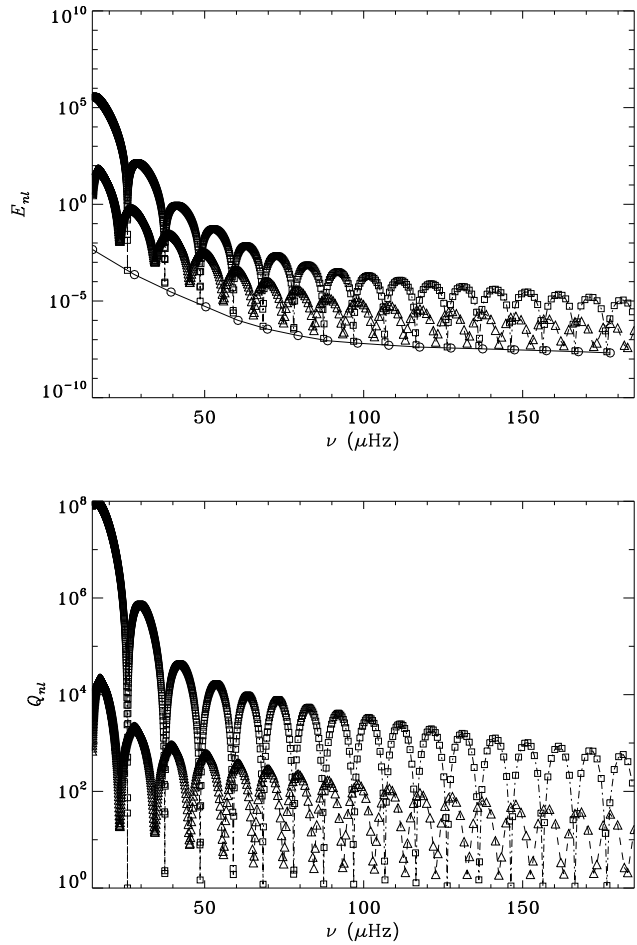}
\end{minipage}
\caption{Mode inertia for the red-giant 
model $\modrg$ in Fig.~\ref{fig:chenhr} $(1.3 \Msun \, , \; 6.2 \Rsun)$.
The upper panel shows the inertia (cf. Eq.~\ref{eq:inertia}) as a function
of frequency
for modes of degree $l = 0$ (circles, connected by a solid line), 
$l = 1$ (triangles, connected by a dashed line) and 
$l = 2$ (squares, connected by a dot-dashed line). 
The lower panel shows the normalized inertia
(cf. Eq.~\ref{eq:sclinertia}) for $l = 1$ and 2.
}
\label{fig:rginertia}
\end{figure}

\begin{figure}
\centering
\begin{minipage}{\linewidth}
\centering
\includegraphics[width=9cm]{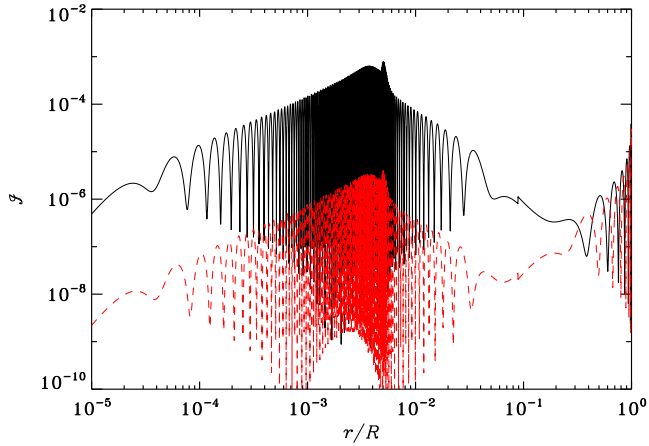}
\end{minipage}
\caption{Integrands $\CI$ for the mode inertia $E$ (cf. Eq.~\ref{eq:inertia}),
defined such that $E = \int_0^R \CI \dd \ln r$, for modes with $l = 1$
in the model $\modrg$ in Fig.~\ref{fig:chenhr} $(1.3 \Msun \, , \; 6.2 \Rsun)$.
The red dashed line shows the p-dominated mode with frequency $84.3 \muHz$ and 
the black solid line the g-dominated mode with frequency $79.1 \muHz$.
The modes are marked as red diamonds in Fig.~\ref{fig:rgperechl}.
}
\label{fig:rgamde}
\end{figure}

\begin{figure}
\centering
\begin{minipage}{\linewidth}
\centering
\includegraphics[width=9cm]{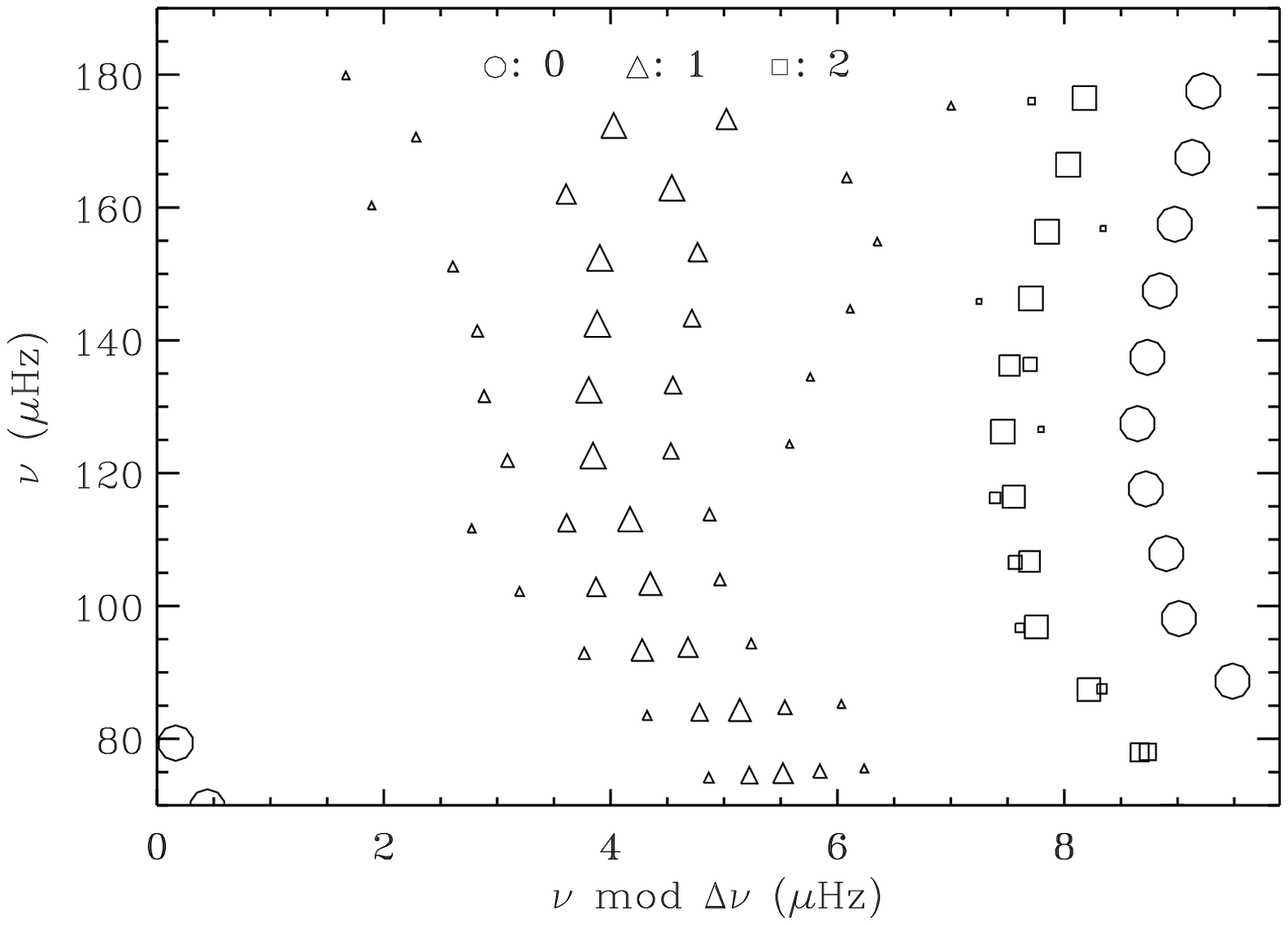}
\end{minipage}
\caption{\'Echelle diagram 
for the model $\modrg$ in Fig.~\ref{fig:chenhr} $(1.3 \Msun \, , \; 6.2 \Rsun)$.
See the caption to Fig.~\ref{fig:chenechl} for the meaning of 
symbols and symbol sizes.
}
\label{fig:rgechl}
\end{figure}

\subsubsection{Ascending-branch red giants}
\label{sec:ascenrg}

As a star evolves up the red-giant branch its internal structure
and hence its oscillation spectrum change dramatically.
As an example of an ascending-branch star we consider the most evolved 
stellar model in Fig.~\ref{fig:chenfig},
which is indicated with the upper diamond in Fig.~\ref{fig:chenhr}.
The properties of the modes of this model in terms of the mode inertia and scaled mode inertia 
(cf. Eqs~\ref{eq:inertia} and \ref{eq:sclinertia}),
are shown in Fig.~\ref{fig:rginertia}.
The $l = 1$~and~$2$ modes form a very dense spectrum with inertias 
typically exceeding the inertia of the neighbouring acoustic modes 
by several orders of magnitude.
These high-inertia modes are buoyancy-dominated modes
that are predominantly trapped in the g-mode cavity in the deep interior of the star.
Additionally, there are acoustic resonances where one or more modes have inertias 
close to the radial-mode inertia. 
This is particularly visible in the bottom panel of Fig.~\ref{fig:rginertia}, 
where low scaled inertia $Q_{n\,l}$
indicates the modes that have largest amplitude in the p-mode cavity.

To further illustrate the properties of the modes, Fig.~\ref{fig:rgamde}
shows the integrands of the inertia for the most p-dominated and 
the most g-dominated dipolar modes with a frequency near $80 \muHz$.
There is clear similarity in shape between the two curves;%
\footnote{In particular, it may be shown (see Eq.~\ref{eq:g_inertia})
that the integrand in the g-mode cavity is proportional to
the buoyancy frequency $N$; cf. Fig.~\ref{fig:charfrqrg}.}
the main difference is in the behaviour in the evanescent region near
$r/R = 0.1$, where the eigenfunction increases with depth for the g-dominated
mode (black curve) and decreases with depth for the p-dominated mode
(red curve).
As a result, the region beneath $0.1 R$ contributes 99.5\,\% to the inertia
of the g-dominated mode,
while for the most p-dominated mode 63\,\% of the inertia comes
from the g-mode cavity.
Hence, even for p-dominated modes the contribution from the g-mode cavity is significant.
Additionally, Fig.~\ref{fig:rgamde} illustrates the extremely rapid variation of the
eigenfunction that results from the high buoyancy frequency in the g-mode
cavity.
This rapid variation becomes even more extreme in more evolved red giants where 
the eigenfunctions may have thousands of nodes in the deep interior.
This places severe requirements on the numerical techniques used
to compute these modes. Techniques that explicitly take such rapid variations into
account have indeed been developed \citep{gabriel1976, townsend2013}.

\paragraph{Pressure-dominated modes}
The acoustic resonances of non-radial modes together with the radial modes 
satisfy a frequency pattern similar to the asymptotic behaviour
of acoustic modes in main-sequence stars (cf. Eq.~\ref{eq:tassoul1}).
This is evident in the \'echelle diagram shown in Fig.~\ref{fig:rgechl}
where in particular the $l = 0$~and~$2$ modes have a behaviour very similar
to what is seen in the solar case (cf. bottom panel of Fig.~\ref{fig:chenechl}).
In addition to these p-dominated modes a few additional $l = 2$ modes
are visible with the chosen scaling.
For $l = 1$ the pattern is more complicated, i.e., several mixed
modes are visible per acoustic-mode order.
Nevertheless, a dominant set of modes still follows the asymptotic expression.

This regular pattern is a universal
feature of observed (and modelled) red-giant oscillations \citep[][and Section~\ref{sect:pattern}]{mosser2011}.
However, the reason that the asymptotic relation holds is not entirely clear.
The analysis leading to Eq.~(\ref{eq:tassoul1}) is fundamentally related to 
the behaviour of the solution of the oscillation equations near the
singularity at $r = 0$ \citep[see also][]{gough1986b, gough1993}.
However, in the case of red giants the properties of the
acoustic resonances are determined by the requirement that the solution
decrease with increasing depth in the evanescent region
(see also Fig.~\ref{fig:rgamde}). 
Here conditions are very different from the conditions in the central regions
of a main-sequence star.
Despite these different conditions the universal pattern seems to hold. A better physical understanding of this behaviour of the frequency
pattern of the acoustically dominated modes may lead to additional
diagnostic potential of the observations.

\begin{figure}
\centering
\begin{minipage}{\linewidth}
\centering
\includegraphics[width=9cm]{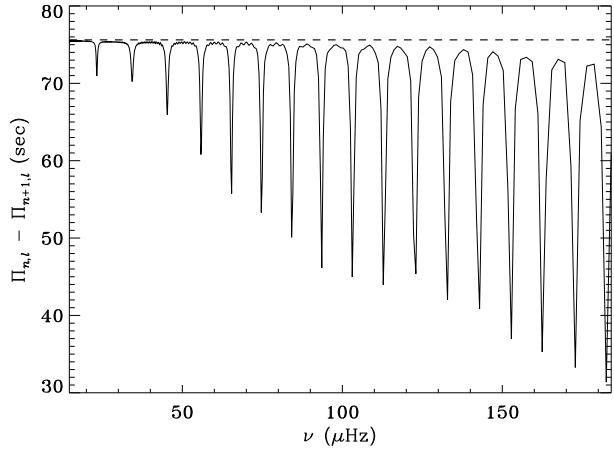}
\end{minipage}
\caption{Period spacing for $l = 1$ modes, as a function of frequency,
in the model $\modrg$ in Fig.~\ref{fig:chenhr} $(1.3 \Msun \, , \; 6.2 \Rsun)$.
The heavy dashed line shows the asymptotic period spacing
$\Pi_0/\sqrt{2}$ (cf. Eqs~\ref{eq:gasymp} and \ref{eq:gper0}).
For clarity we do not show the individual modes.
}
\label{fig:rgperspac}
\end{figure}

\begin{figure}
\centering
\begin{minipage}{\linewidth}
\centering
\includegraphics[width=11cm]{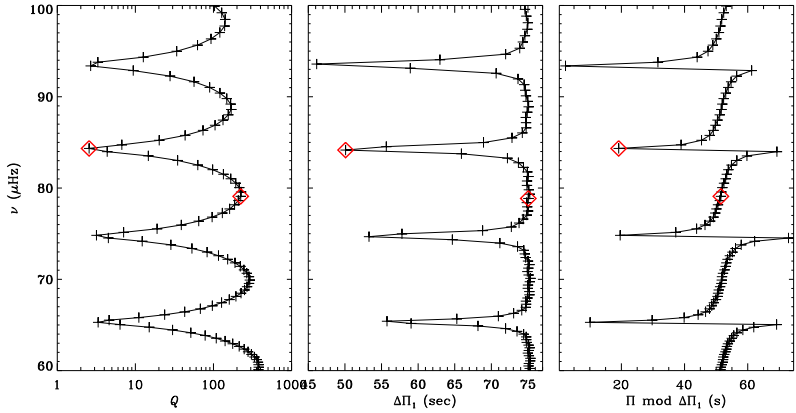}
\end{minipage}
\caption{Frequency as a function of scaled mode inertia $Q$ (left), period spacing (centre) and period modulo period spacing 
($\Delta \Pi_1 = 75.4 \muHz$), i.e., period \'echelle diagram (right)
for $l = 1$ modes in the model $\modrg$ in Fig.~\ref{fig:chenhr} $(1.3 \Msun \, , \; 6.2 \Rsun)$.
The red diamonds mark the modes illustrated in Fig.~\ref{fig:rgamde} or,
in the central panel, the period spacings between those marked modes and 
the adjacent mode with higher period.
}
\label{fig:rgperechl}
\end{figure}

\paragraph{Gravity-dominated modes}
The properties of gravity-dominated modes are to a large extent 
determined by the structure of the core.
Hence, these oscillations have a large potential for the study 
of the internal structure of red giants.
It was found by \citet{beck2011} that the observed spectrum of a red giant
showed additional peaks which were identified as coming
from mixed modes with a stronger g-dominated component
(see Fig.~\ref{fig:ts+ps}).
The behaviour of these mixed modes is dominated by the g-mode asymptotic
relation, Eq.~(\ref{eq:gasymp}). According to this relation the gravity modes are 
uniformly spaced in period.
This is illustrated by Fig.~\ref{fig:rgperspac} showing the period spacing
$\Pi_{n\,l} - \Pi_{n+1\,l}$ for $l = 1$,
as a function of frequency.
At relatively low frequency there are modes with nearly
constant spacing, while the spacing shows a characteristic decrease
around the more p-dominated modes.
The actual period spacing is in excellent agreement, particularly at low
frequency, with the asymptotic period spacing marked by the horizontal 
dashed line.
Thus observations of low-frequency g-dominated modes provide a direct measure of
the integral of the buoyancy frequency in the core of the star.

The fact that several dipolar mixed modes are visible 
in the vicinity of the acoustically dominated modes, as observed
by \citet{beck2011}, might indeed have been expected from 
the rough amplitude scaling in an \'echelle diagram such as shown
in Fig.~\ref{fig:rgechl}.
In fact, as discussed in Section~\ref{sec:energetics}, 
even fully g-dominated
modes may have sufficient visibility to be observed in very long time series,
such as those that were obtained by {\it Kepler}'s nominal 4-year mission.
An example is shown in fig. 1b of \citet{stello2013} for a red-clump star.

Following \citet{bedding2011} we introduced period \'echelle diagrams in 
Section~\ref{sect:dP}, based on the uniform period spacing of g-dominated modes.
Figure~\ref{fig:rgperechl} shows such a diagram for
a part of the modes shown in Fig.~\ref{fig:rgperspac}.
For comparison we also show the scaled mode inertia and the period spacing.
Gravity-dominated modes with high inertia fall approximately on a vertical line in the period \'echelle diagram.
This is as expected from the asymptotic behaviour.
Around this vertical line departures are visible near the acoustic resonances.
It may be noticed that the variations near the resonances are
qualitatively similar to
the variations induced by g-dominated modes in the frequency \'echelle diagram
for a subgiant star (cf. Fig.~\ref{fig:chenechl}).

\paragraph{Coupling of buoyancy and acoustic cavities}
Asymptotic analysis of the mixed modes in red giants must take into account
the coupling between the buoyancy and acoustic cavities.
A full analysis of this problem
has been carried out by \citet{shibahashi1979}
\citep[see also][]{unno1989}.%
\footnote{Shibahashi used a slightly different form of the asymptotic equation,
replacing $K$ in Eq.~(\ref{eq:kasymp}) by
\[
K = {1 \over c^2} \left[ S_l^2 \left({N^2 \over \omega^2} - 1\right) 
+ \omega^2 - N^2 \right] 
= {\omega^2 \over c^2} \left({S_l^2 \over \omega^2} -1 \right) 
\left({N^2 \over \omega^2} - 1\right) \; .
\]
This essentially only differs from Eq.~(\ref{eq:kasymp}) in the near-surface
layers and hence affects, e.g., the phase $\epsilon$ in Eq.~(\ref{eq:tassoul1}).
The following analysis follows \citet{shibahashi1979}.
\label{fn:shibk}
}
The analysis is based on simplified equations of the form given in
Eq.~(\ref{eq:basasymp}),
and has been implemented for the red-giant case by
Goupil (private communication) and \citet{mosser2012core}.
Shibahashi showed that continuous matching of the solutions between 
the two cavities leads to the resonance condition 
\begin{equation}
\cot \left( \int_{r_{\rm a}}^{r_{\rm b}} K^{1/2} \dd r - \phi_{\rm g} \right)
\tan \left( \int_{r_{\rm c}}^{r_{\rm d}} K^{1/2} \dd r - \phi_{\rm p} \right)
= q \; ,
\label{eq:mixasymp0}
\end{equation}
implicitly determining the eigenfrequencies.
In Eq.~(\ref{eq:mixasymp0}) $r_{\rm a}$, $r_{\rm b}$, $r_{\rm c}$ and $r_{\rm d} \simeq R$
are the four turning points where $K$ changes sign 
(cf. Fig.~\ref{fig:charfrqrg}), and $\phi_{\rm g}$ and $\phi_{\rm p}$ are phases
depending on the properties of the turning points.
Additionally, 
\begin{equation}
q = {1 \over 4} \exp\left( -2 \int_{r_{\rm b}}^{r_{\rm c}}
|K|^{1/2} \dd r \right)
\label{eq:coupling}
\end{equation}
is a measure of the strength of the coupling between the two cavities.

\citet{takata2016b} carried out a very illuminating physical analysis of
the properties of mixed modes in terms of the reflection and transmission
of waves at the evanescent region,
with a more rigorous analysis for dipolar modes provided by \citet{takata2016a}.
He pointed out that Eq.~(\ref{eq:coupling}) is only valid in the case of
weak coupling and provided expressions of more general validity.
In particular, with strong coupling the value of $q$ may substantially
exceed the upper limit of $1/4$ indicated by Eq.~(\ref{eq:coupling}).
However, the resonance condition, Eq.~(\ref{eq:mixasymp0}), on
which the following discussion is largely based, remains valid.

To analyse Eq.~(\ref{eq:mixasymp0}) it is convenient to introduce
\begin{equation}
\theta_{\rm g} = \int_{r_{\rm a}}^{r_{\rm b}} K^{1/2} \dd r - \phi_{\rm g} \; ,
\quad
\theta_{\rm p} = \int_{r_{\rm c}}^{r_{\rm d}} K^{1/2} \dd r - \phi_{\rm p} 
\end{equation}
\citep[e.g.][]{unno1989, mosser2012core},
so that the equation becomes
\begin{equation}
\tan \theta_{\rm p} \cot \theta_{\rm g} = q \; .
\label{eq:mixasymp1}
\end{equation}
We approximate $K$ by 
$K \simeq l(l+1) r^{-2} \, N^2/\omega^2$ in the buoyancy cavity,
$[r_{\rm a}, r_{\rm b}]$
and by
$K \simeq \omega^2 / c^2$ in the acoustic cavity, $[r_{\rm c}, r_{\rm d}]$.
Then we obtain
\begin{equation}
\theta_{\rm p} \simeq \omega / \omega_{\rm p} - \phi_{\rm p} \; , \quad
\theta_{\rm g} \simeq \omega_{\rm g} / \omega - \phi_{\rm g} \; ,
\label{eq:phases}
\end{equation}
where
\begin{equation}
\omega_{\rm p} = \left( \int_{r_{\rm c}}^{r_{\rm d}} {\dd r \over c} 
\right)^{-1} \simeq 2 \Delta \nu 
\label{eq:omega_p}
\end{equation}
(cf. Eq.~\ref{eq:larsep1}) and
\begin{equation}
\omega_{\rm g} = 
\sqrt{l(l+1)} \int_{r_{\rm a}}^{r_{\rm b}} {N \over r} \dd r 
\simeq {2 \pi^2 \over \Delta \Pi_l} \;  
\label{eq:omega_g}
\end{equation}
(cf. Eq.~\ref{eq:gper0}).
In Eq.~(\ref{eq:omega_p}) we neglected the small contribution to 
$\Delta \nu$ from the core region.

To interpret Eq.~(\ref{eq:mixasymp1}) we first consider the uncoupled
case, i.e. $q = 0$.
This should in principle yield Eq.~(\ref{eq:tassoul1}) and Eq.~(\ref{eq:gasymp})
for the uncoupled $\pi$ and $\gamma$ modes, respectively.
In the uncoupled case Eq.~(\ref{eq:mixasymp1}) has one set of solutions
with $\tan \theta_{\rm p} = 0$,
or, equivalently,
\begin{equation}
\nu = {\omega \over 2 \pi} \simeq \Delta \nu (n_{\rm p} + \epsilon_{\rm p}) 
\label{eq:mixacous}
\end{equation}
for integer $n_{\rm p}$, and $\epsilon_{\rm p} = \phi_{\rm p}/\pi$.
This indeed superficially recovers the leading-order term in the
acoustic-mode asymptotic relation (\ref{eq:tassoul1})
for the uncoupled $\pi$ modes,
although with $\phi_{\rm p}$ and $\epsilon_{\rm p}$ being independent of $l$.
In contrast, the observed and computed p-dominated frequencies 
satisfy Eq.~(\ref{eq:tassoul1}) which includes an $l$ dependence.
However, the analysis leading to Eq.~(\ref{eq:mixacous})
neglected the dependence on $l$ of $K$ 
in the acoustic cavity and the detailed behaviour near the lower turning point
$r_{\rm c}$.
To compensate for this we
simply replace $\epsilon_{\rm p}$ by $\epsilon_{{\rm p}\,l}$
(and similarly for $\phi_{\rm p}$) with
$\epsilon_{{\rm p}\,l} = \epsilon_{\rm p\,0} + l/2$, such that we recover 
Eq.~(\ref{eq:tassoul1}).
A full understanding of the universal pattern (Eq.~\ref{eq:UP})
requires a more complete analysis extending
the one leading to Eq.~(\ref{eq:mixasymp0}), which has so far not 
been carried out.
However, it was pointed out by \citet{mosser2012core},
\citet{deheuvels2015} and \citet{mosser2015} that expressing
$\theta_{\rm p}$ as
\begin{equation}
\theta_{\rm p} = {\pi \over \Delta \nu} 
\left(\nu - \nu_{n_{\rm p}\,l}^{\rm (p)}\right) \; ,
\label{eq:theta_p}
\end{equation}
in terms of purely acoustic frequencies $\nu_{n_{\rm p}\,l}^{\rm (p)}$,
such as the universal acoustic-mode pattern, Eq.~(\ref{eq:UP}),
allows the use of a more complete description of the location of
the acoustic resonances than provided by Eq.~(\ref{eq:mixacous}).

The second set of solutions for an uncoupled ($q=0$) case satisfy
$\cot \theta_{\rm g} = 0$ or, equivalently,
\begin{equation}
\Pi = {2 \pi \over \omega} = 
\Delta \Pi_l ( |n_{\rm g}| + \epsilon_{\rm g} + 1/2) \; ,
\label{eq:mixbuoy}
\end{equation}
where $n_{\rm g}$ is an integer and
$\epsilon_{\rm g} = \phi_{\rm g}/\pi$.
In accordance with the discussion in Section~\ref{sec:modeorder}
we have chosen $n_{\rm g} < 0$, 
using $n_{\rm g}$ rather than $\hat n_{\rm g}$ in this asymptotic expression.
We discuss the relation between $n_{\rm g}$ and $\hat n_{\rm g}$ below.
In this case we indeed recover Eq.~(\ref{eq:gasymp}),
defining the uncoupled $\gamma$ modes.
Corresponding to Eq.~(\ref{eq:theta_p}), we express $\theta_{\rm g}$ as
\begin{equation}
\theta_{\rm g} \equiv {\omega_{\rm g} \over \omega} - \phi_{\rm g}
\simeq \pi \left( {1 \over \Delta \Pi_l \nu} - \epsilon_{\rm g} \right)
= {\pi \over \Delta \Pi_l} 
\left({1 \over \nu} - {1 \over \nu_{n_{\rm g}\,l}^{\rm (g)}} \right) 
+ {\pi \over 2} \; ,
\label{eq:theta_g}
\end{equation}
where we introduced the uncoupled g-mode frequency
as per Eq.~(\ref{eq:mixbuoy}):
\begin{equation}
\nu_{n_{\rm g}\,l}^{\rm (g)} =
\left[\Delta \Pi_l ( |n_{\rm g}| + \epsilon_{\rm g} + 1/2)\right]^{-1} \; .
\label{eq:nu_g}
\end{equation}
We note that, in the last equalities in Eqs~(\ref{eq:theta_p})
and (\ref{eq:theta_g}),
we ignored the contributions $n_{\rm p} \pi$ and $|n_{\rm g}| \pi$
which have no effect in Eq.~(\ref{eq:mixasymp1}).

\begin{figure}
\centering
\begin{minipage}{\linewidth}
\centering
\includegraphics[width=9cm]{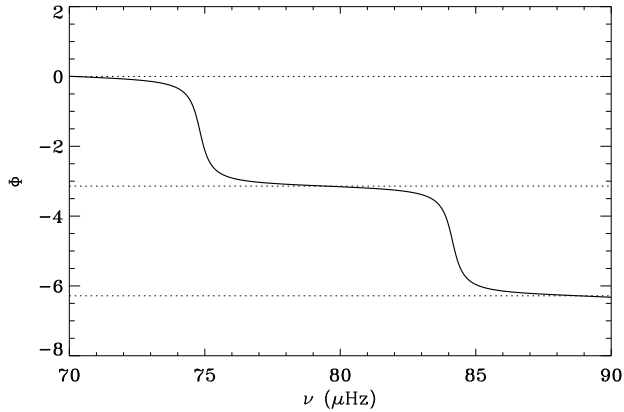}
\end{minipage}
\caption{The effective phase $\Phi$ (cf. Eq.~\ref{eq:mixphi})
as a function of frequency for $l = 1$, in the model $\modrg$ in Fig.~\ref{fig:chenhr} $(1.3 \Msun \, , \; 6.2 \Rsun)$.
The horizontal dotted lines indicate multiples of $\pi$.
Here we used
$\theta_{\rm p} = \pi (\nu / \Delta \nu - \epsilon_{{\rm p}\,l})$, 
with $\Delta \nu = 9.32 \muHz$ and $\epsilon_{{\rm p}\,l} = 1.528$.
}
\label{fig:asympphi}
\end{figure}

To analyse the coupled case, i.e. Eq.~(\ref{eq:mixasymp1}),
\citet{mosser2012core} rewrote it
by moving the g-mode-related part
to the right-hand side as $\tan \theta_{\rm g}$, using 
Eqs~(\ref{eq:theta_p}) and (\ref{eq:theta_g}) and
taking the inverse tangent, to obtain 
\begin{equation}
\nu = \nu_{n_{\rm p}\,l}^{\rm (p)} + {\Delta \nu \over \pi}
\arctan \left[ q \tan\pi\left({1 \over \Delta \Pi_l \nu} - \epsilon_{\rm g}
\right) \right] \; .
\label{eq:mixmosser}
\end{equation}
This provides an implicit equation for $\nu$, which can be determined
iteratively, given the other parameters of the equation. 
Also, it provides the basis for determining these parameters
through fits to observed frequencies.
As discussed in Sections~\ref{sect:dP} and \ref{sect:resmixedmodes} this results in powerful
diagnostics for stellar structure and evolution.

The detailed relation between the numerical order $n$ and its components
$\hat n_{\rm p}$ and $\hat n_{\rm g}$ (cf. Section~\ref{sec:modeorder})
and the asymptotic indices $n_{\rm p}$ and $n_{\rm g}$ has apparently
not been fully explored.
Passing through an acoustic resonance $\nu = \nu_{n_{\rm p}\,l}^{\rm (p)}$ is
associated with the increase by one in the number of nodes in the 
p-mode region of the star, and hence with an increase by one 
in $\hat n_{\rm p}$.
In fact it seems that we can identify $\hat n_{\rm p}$ with $n_{\rm p}$
for the acoustically-dominated modes.
Similarly, there is a close relation between $\hat n_{\rm g}$ and
$n_{\rm g}$ characterizing the asymptotic properties of the g-dominated modes,
at least for stars on the ascending red-giant branch.
The situation is probably more complicated in more complex stars, possibly
with several separate regions of g-mode propagation and/or buoyancy glitches
(see also Sections~\ref{sec:glitch} and \ref{sec:redclump}).

As noted by \citet{mosser2012core} a useful quantity is the estimate
of the number of g-dominated modes associated with a given acoustic resonance,
which can be obtained as
\begin{equation}
\CN = {\Delta \nu \over \Delta \Pi_l {\nu_{n_{\rm p}\,l}^{\rm (p)}}^2} \; .
\label{eq:gnumber}
\end{equation}
Adding the mode at the acoustic resonance the total number of modes in
a $\Delta \nu$ interval around the resonance is $\CN + 1$.

An alternative way to analyse Eq.~(\ref{eq:mixasymp1}) was proposed by
\citet{jcd2012hakone}%
\footnote{We note that the analysis by \citet{jcd2012hakone} (CD12) suffers
from two sign mistakes which fortuitously cancel.
One arose from the neglect of a singularity in the evanescent region
in the asymptotic expression (CD12, eq.~1). 
The second is a simple sign error in the analysis leading to CD12, eq.~(22),
such that that equation, and the remaining analysis, is correct.}
and developed further by \citet{jiang2014}.
They rewrote Eq.~(\ref{eq:mixasymp1}) as
\begin{equation}
\sin \theta_{\rm p} \cos \theta_{\rm g} 
- q \cos \theta_{\rm p} \sin \theta_{\rm g} =0 \; .
\end{equation}
This can be rewritten as 
\begin{equation}
\cos [\omega_{\rm g} / \omega + \Phi(\omega) - \phi_{\rm g}] = 0 \; ,
\end{equation}
where $\Phi$ satisfies
\begin{equation}
\tan \Phi = q \cot \theta_{\rm p}
= q \cot\left[ {\pi \over \Delta \nu} \left(
\nu - \nu_{n_{\rm p}\,l}^{\rm (p)} \right) \right] \; .
\label{eq:mixphi}
\end{equation}
By choosing the appropriate integer multiples of $\pi$ we can define 
$\Phi$ as a continuous function of $\omega$.
Thus the eigenfrequencies satisfy
\begin{equation}
\omega_{\rm g}/\omega + \Phi(\omega) - \phi_{\rm g} = (k + 1/2) \pi \; ,
\end{equation}
for integer $k$, or equivalently
\begin{equation}
\Psi(\nu) \equiv {1 \over \Delta \Pi_l \nu} + \pi^{-1} \Phi(2 \pi \nu)
- \epsilon_{\rm g} -1/2 = k \; .
\label{eq:mixfreq}
\end{equation}

\begin{figure}
\centering
\begin{minipage}{\linewidth}
\centering
\includegraphics[width=9cm]{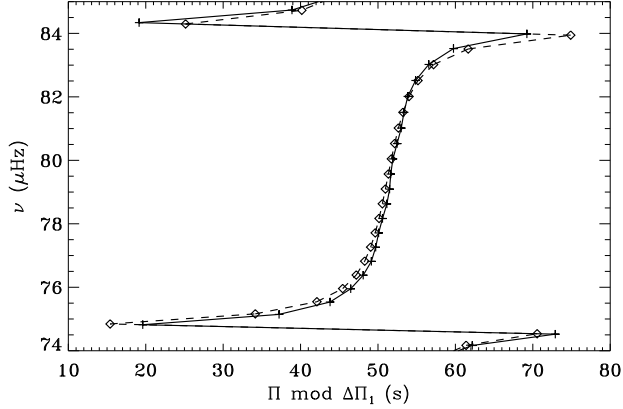}
\end{minipage}
\caption{Period \'echelle diagram for $l = 1$, in the model $\modrg$ in Fig.~\ref{fig:chenhr} $(1.3 \Msun \, , \; 6.2 \Rsun)$.
Here a frequency range encompassing two pressure-dominated modes is shown.
The numerically computed periods are shown by a solid line with plus
symbols.
The solution from Eq.~(\ref{eq:mixfreq}), based on $\Phi$ as shown in
Fig.~\ref{fig:asympphi}, is depicted by a dashed line and diamonds.
}
\label{fig:compperechl}
\end{figure}

The function $\Phi$ is illustrated in Fig.~\ref{fig:asympphi}. 
To understand its properties it is convenient to rewrite Eq.~(\ref{eq:mixphi})
as $\Phi = \arctan[ q \cot\pi (\nu - \nu_{n_{\rm p}\,l}^{\rm (p)})/\Delta \nu]$.%
\footnote{Combining this equation with Eq.~(\ref{eq:mixfreq}) leads to
an equation analogous to Eq.~(\ref{eq:mixmosser}), 
although perhaps
of doubtful usefulness in the analysis of the properties of the modes}.
The value of $q$ is typically relatively small. Therefore, for most frequencies
$\Phi$ is close to a multiple of $\pi$. This is illustrated by the dotted lines
in Fig.~\ref{fig:asympphi}.
However, near the acoustic resonances,
where $\nu \simeq \nu_{n_{\rm p}\,l}^{\rm (p)}$,
the argument to $\arctan$ goes through a singularity. This leads to the rapid
variation of $\Phi$.

We note that Eq.~(\ref{eq:mixfreq}) provides an alternative way 
to determine the frequencies of mixed modes,
given $\Delta \nu, \nu_{n_{\rm p}\,l}^{\rm (p)}, q, \Delta \Pi_l$
and $\epsilon_{\rm g}$: 
$\Phi$ can be directly computed from Eq.~(\ref{eq:mixphi}), defining $\Psi$ as
a function of frequency from Eq.~(\ref{eq:mixfreq});
the frequencies of the mixed modes are then simply obtained as those
values where $\Psi$ takes on consecutive integer values.
This is numerically straightforward, given that $\Psi$ is a 
monotonic function of $\nu$, and relates the frequencies directly
to the mode order in the integer $k$.

In Eq.~(\ref{eq:mixfreq}) increasing $k$ leads to decreasing frequency,
corresponding to decreasing $n$, where $n$ is
the numerical mode order introduced in Section~\ref{sec:modeorder};
also, $\Phi$ is only defined from Eq.~(\ref{eq:mixphi})
to within an integer multiple of $\pi$.
It appears that this can be chosen such that $k = -n$.
This reflects the fact that Eq.~(\ref{eq:mixfreq}) defines the complete
spectrum of mixed modes for the given degree.

The behaviour of $\Phi$ near the resonances is directly related to the
variation in the period spacing (cf. Fig.~\ref{fig:rgperspac}).
Writing Eq.~(\ref{eq:mixfreq}) as
\begin{equation}
\Pi_{n\,l} = \Delta \Pi_l [ k + \epsilon_{\rm g} + 1/2
- \pi^{-1} \Phi(2 \pi \nu_{n\,l}) ] \;, 
\label{eq:mixper}
\end{equation}
with $\nu_{n\,l} = 1/\Pi_{n\,l}$,
and making a Taylor expansion of
$\Phi(2 \pi \nu_{k+1}) - \Phi(2 \pi \nu_{k})$ we obtain
\begin{equation}
\Pi_{n\,l} - \Pi_{n+1\,l} \simeq \Delta \Pi_l
\left(1 - {2 \Delta \Pi_l \over \Pi_{n\,l}^2} {\dd \Phi \over \dd \omega }
\right)^{-1} \; .
\label{eq:mixperspac}
\end{equation}
Thus for the g-dominated modes, where $\Phi$ is nearly constant,
we recover $\Delta \Pi_l$. At the same time the period spacing decreases near the acoustic resonances (note that $\dd \Phi/\dd \omega < 0$).
\citet{jiang2014} presented an approximation to Eq.~(\ref{eq:mixperspac})
and discussed the use of this to estimate $q$ from the variation in
the period spacings.

The quality of the asymptotic fit derived above (Eq.~\ref{eq:mixfreq}) 
is illustrated in Fig.~\ref{fig:compperechl} for model $\modrg$
in Fig.~\ref{fig:chenhr}.
For the g-dominated modes the relative differences between the computed
and asymptotic frequencies are less than $10^{-4}$.
The relative difference increases to
$5 \times 10^{-4}$  for frequencies undergoing acoustic resonances.
However,
these differences have a far stronger effect on the period spacings.

The coupling constant $q$ depends on the properties of the evanescent region
and hence in principle provides further diagnostics of the stellar interior.
This is reflected by Eq.~(\ref{eq:coupling})
that was already obtained by \citet{shibahashi1979}.
The values of $q$ determined from Eq.~(\ref{eq:coupling})
and from fits to numerical frequencies were compared by \citet{jiang2014},
who found substantial differences between these values.
This appears in part to be due to the neglect of the perturbation to the
gravitational potential in the asymptotic analysis,
and could, for dipolar modes, be avoided by using the formulation
of \citet{takata2006} which avoids this approximation.
Furthermore, the approximation to Eq.~(\ref{eq:mixperspac}) by \citet{jiang2014} neglects potential singularities, particularly in the evanescent region.
These effects require further analysis.
Finally, as mentioned above, \citet{takata2016a, takata2016b}
showed that
Eq.~(\ref{eq:coupling}) is only valid for weak coupling, which may not 
be appropriate in the cases considered by \citet{jiang2014}.

From an observational point of view \citet{mosser2012core} determined
coupling strengths for a substantial number of red-giant and clump stars
observed with {\it Kepler}; they noted that clump stars showed 
larger values of $q$ than did the red giants, in many cases exceeding 
the upper limit of $0.25$ predicted by Eq.~(\ref{eq:coupling}). A very large number of {\it Kepler} targets
was analysed by \citet{mosser2017b},
providing a detailed overview of the dependence of $q$ on stellar parameters.

The phase $\epsilon_{\rm g}$ in principle provides further information
about stellar structure, particularly the regions at the turning points of
the g-mode cavity.
\citet{buysschaert2016} carried out fits to data for three red giants obtained
by {\it Kepler}, including a variable $\epsilon_{\rm g}$ as one of the 
fitted parameters (see Section~\ref{sect:dP}).
They found that $\epsilon_{\rm g}$ was only weakly constrained by the
observations; 
however, allowing $\epsilon_{\rm g}$ to vary resulted in a clearer 
definition of the values of the period spacing and its uncertainty 
obtained in the fit. 
From a theoretical point of view, \citet{takata2016a} provided
some insight into the dependence of $\epsilon_{\rm g}$ on the
properties of the model.
These issues also deserve further studies.

\paragraph{Asymptotic properties of mode inertia}
It is instructive to consider the asymptotic properties of the mode
inertia, in particular the relative contribution $\zeta$ from the core
(cf. Eq.~\ref{eq:zeta}). The value of $\zeta$ is closely related
to the effects of rotation on the oscillation frequencies 
and the damping of the modes.
The following is based on the detailed analysis
by \citet[S79]{shibahashi1979},
which was further developed by \citet{goupil2013} and  \citet{deheuvels2015}.

S79 expressed the asymptotic analysis in terms of
two functions $v(r)$ and $w(r)$ related to $\xi_r$ and $\xi_{\rm h}$ by
\begin{eqnarray}
\label{eq:asympvar}
v &=& \rho^{1/2} c r 
\left(\left|1 - {S_l^2 \over \omega^2} \right| \right)^{-1/2}
\xi_r \\
w &=& \rho^{1/2} \omega r^2 
\left( \left| {N^2 \over \omega^2} -1\right| \right)^{-1/2} \xi_{\rm h} \; .
\nonumber
\end{eqnarray}
In the outer parts of the star the modes are predominantly acoustic, with
$|\xi_r| \gg |\xi_{\rm h}|$.
In this case
the integral for the inertia (Eq.~\ref{eq:inertia_comp}) is dominated
by the term in $\xi_r$.
From the asymptotic expression for $v$ (eq. 28 of S79, applied
in the acoustic outer region) we here obtain
\begin{equation}
\xi_r \simeq C \rho^{-1/2} c^{-1/2} \omega^{-1/2} r^{-1}
\cos \left( \int_r^R K^{1/2} \dd r - \phi_{\rm p}^\prime \right) \; ,
\label{eq:asympsol_p}
\end{equation}
where $C$ is a normalization constant and $\phi_{\rm p}^\prime$ is a phase.
As noted above, we use the definition of $K$ given in footnote~\ref{fn:shibk}.
As in the derivation of Eq.~(\ref{eq:omega_p}),
we neglected $S_l^2/\omega^2$ in the amplitude function 
in Eq.~(\ref{eq:asympvar}) compared with 1,
which is a good approximation except close to the turning point $r_{\rm c}$.
Using the same approximation, we obtain from Eq.~(\ref{eq:asympsol_p}) 
the contribution of the outer parts of the star to the numerator in 
Eq.~(\ref{eq:inertia_comp})
\begin{equation}
I_{\rm p} = \int_{r_{\rm c}}^R \xi_r^2 \rho r^2 \dd r
\simeq C^2 \omega^{-1} \int_0^{\tau_{\rm c}} 
\cos^2(\omega \tau - \phi_{\rm p}^\prime) \dd \tau \;,
\end{equation}
where
\begin{equation}
\tau = \int_r^R {\dd r \over c}
\label{eq:acdepth}
\end{equation}
is the acoustic depth, and $\tau_{\rm c}$ is the acoustic 
depth of the turning point $r_{\rm c}$.
For a mode of high acoustic order the integral over $\cos^2$ can be
replaced by $1/2 \, \tau_{\rm c}$.
Using Eq.~(\ref{eq:omega_p}) we obtain
\begin{equation}
I_{\rm p} \simeq {1 \over 4} C^2 {1 \over \omega \Delta \nu} \; .
\label{eq:inertia_p}
\end{equation}

In the inner parts of the star the mode is buoyancy dominated, with
$|\xi_{\rm h}| \gg |\xi_r|$.
Here, the inertia integral is dominated by the term in $\xi_{\rm h}$.
In this case the relevant asymptotic solution, from the
relevant expression for $w$ (eq. 29 of S79, in the g-mode cavity) is
\begin{equation}
\xi_{\rm h} \simeq A \rho^{-1/2} \omega^{-3/2} r^{-3/2} L^{-1/2} N^{1/2}
\sin \left( \int_{r_{\rm a}}^r K^{1/2} \dd r - \phi_{\rm g}^\prime \right) \; ,
\label{eq:asympsol_g}
\end{equation}
where $L^2 = l(l+1)$, $A$ is a normalization constant and
$\phi_{\rm g}^\prime$ is a phase.
As in the derivation of Eq.~(\ref{eq:omega_g}) we neglected $1$ compared with
$N^2 / \omega^2$.
From Eq.~(\ref{eq:asympsol_g}) we then obtain for the contribution of
the g-mode region to the numerator in Eq.~(\ref{eq:inertia_comp})
\begin{equation}
I_{\rm g} = L^2 \int_{r_{\rm a}}^{r_{\rm b}} \xi_{\rm h}^2 \rho r^2 \dd r
\simeq A^2 \omega^{-3} L \int_0^{r_{\rm b}} N
\sin^2 \left( {L \over \omega} \int_0^r N {\dd r' \over r'} 
- \phi_{\rm g}^\prime \right) {\dd r \over r} \; ,
\label{eq:g_inertia}
\end{equation}
where we extended the integrals to $r = 0$.
It should be noticed that the amplitude of the integrand scales like $N$,
when integrating with respect to $\ln r$.
This is indeed confirmed by comparing the integrand shown in
Fig.~\ref{fig:rgamde} with the buoyancy frequency in Fig.~\ref{fig:charfrqrg}.
Introducing the buoyancy radius
\begin{equation}
\upsilon = \int_0^r N {\dd r \over r}
\label{eq:buoyrad}
\end{equation}
we obtain
\begin{equation}
I_{\rm g} \simeq A^2 \omega^{-3} L \int_0^{\upsilon_{\rm b}}
\sin^2 \left( {L \over \omega} \upsilon - \phi_{\rm g}^\prime \right) \dd \upsilon
\simeq {1 \over 2} A^2 \omega^{-3} L \upsilon_{\rm b} \; ,
\end{equation}
where $\upsilon_{\rm b}$ is the buoyancy radius of the turning point
$r_{\rm b}$,
and replacing as before the integral over $\sin^2$ by
$1/2 \, \upsilon_{\rm b}$.
Using Eq.~(\ref{eq:omega_g}) yields
\begin{equation}
I_{\rm g} \simeq A^2 {\pi^2 \over \Delta \Pi_l \omega^3} \; .
\label{eq:inertia_g}
\end{equation}
Finally, by combining Eqs~(\ref{eq:inertia_p}) and (\ref{eq:inertia_g}),
we obtain
\begin{equation}
{I_{\rm p} \over I_{\rm g}} \simeq 
{C^2 \over A^2} {\Delta \Pi_l \nu^2 \over \Delta \nu} \; .
\label{eq:inertia_rat}
\end{equation}

To complete the analysis we need the ratio $C/A$ between the amplitudes of 
the eigenfunctions in Eqs~(\ref{eq:asympsol_p}) 
and (\ref{eq:asympsol_g}).
This ratio is determined by the coupling across the evanescent region,
$[r_{\rm b}, r_{\rm c}]$.
From eq. (30) of S79, and using the dispersion
relation (Eq.~\ref{eq:mixasymp1}), it can be shown that
\begin{equation}
{C \over A} = q^{-1/2} {\cos \theta_{\rm g} \over \cos \theta_{\rm p}} \; .
\end{equation}
Thus
\begin{equation}
{I_{\rm p} \over I_{\rm g}} \simeq 
q^{-1} {\cos^2 \theta_{\rm g} \over \cos^2 \theta_{\rm p}}
{\Delta \Pi_l \nu^2 \over \Delta \nu} \; ,
\end{equation}
or, using Eqs~(\ref{eq:theta_p}) and (\ref{eq:theta_g}), 
\begin{equation}
{I_{\rm p} \over I_{\rm g}} \simeq 
q^{-1} {\Delta \Pi_l \nu^2 \over \Delta \nu}
{\displaystyle
\cos^2 \left[ \pi \left( {1 \over \Delta \Pi_l \nu} - \epsilon_{\rm g}
\right) \right] \over \displaystyle
\cos^2 \left[ { \pi( \nu - \nu_{n_{\rm p}\,l}^{\rm (p)} ) \over \Delta \nu}
\right] }
\; .
\end{equation}
It follows that the fractional contribution of
the g-mode region to the inertia is obtained as
\begin{eqnarray}
\zeta &\simeq& {I_{\rm g} \over I_{\rm p} + I_{\rm g}}
= \left(1 + {I_{\rm p} \over I_{\rm g}} \right)^{-1} \nonumber \\
&\simeq& \left\{ 1 + 
q^{-1} {\Delta \Pi_l \nu^2 \over \Delta \nu} 
{\displaystyle
\cos^2 \left[ \pi \left( {1 \over \Delta \Pi_l \nu} - \epsilon_{\rm g}
\right) \right] \over \displaystyle
\cos^2 \left[ { \pi( \nu - \nu_{n_{\rm p}\,l}^{\rm (p)} ) \over \Delta \nu}
\right] } \right\}^{-1} \equiv \zeta_{\rm as} \; 
\label{eq:as_zeta_I}
\end{eqnarray}
\citep{deheuvels2015}.
They also demonstrated
that $\zeta_{\rm as}$ provides a remarkably good fit to the numerically computed $\zeta$.
By using Eq.~(\ref{eq:mixasymp1}) this can be further simplified as
\begin{equation}
\zeta_{\rm as} = 
\left\{ 1 + q {\Delta \Pi_l \nu^2 \over \Delta \nu} 
{1 \over \displaystyle q^2 + (1 - q^2)
\sin^2 \left[ { \pi( \nu - \nu_{n_{\rm p}\,l}^{\rm (p)} ) \over \Delta \nu}
\right]} \right\}^{-1} \; .
\label{eq:zeta_asymp}
\end{equation}

\begin{figure}
\centering
\begin{minipage}{\linewidth}
\centering
\includegraphics[width=9cm]{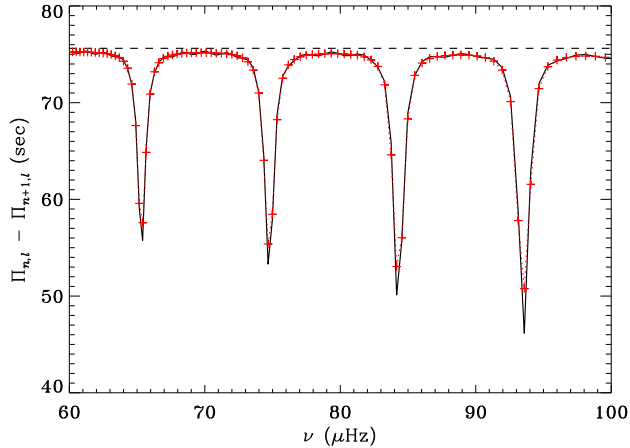}
\end{minipage}
\caption{Enlargement of the period spacing for $l = 1$ modes, 
as a function of frequency, shown in Fig.~\ref{fig:rgperspac},
in the model $\modrg$ in Fig.~\ref{fig:chenhr} $(1.3 \Msun \, , \; 6.2 \Rsun)$.
The heavy dashed line shows the asymptotic period spacing
$\Delta \Pi_1 = \Pi_0/\sqrt{2}$,
and the red dotted line and symbols show $\zeta \Delta \Pi_1$
(cf. Eq.~\ref{eq:perspac_zeta}),
where $\zeta$ was obtained from the numerically computed eigenfunctions.
}
\label{fig:rgperspac_zeta}
\end{figure}

\citet{mosser2015} pointed out that Eq.~(\ref{eq:zeta_asymp}) shows a 
very interesting relation between $\zeta$ and the actual period spacing
between mixed modes.
To see this, we return to Eq.~(\ref{eq:mixperspac}).
Differentiating Eq.~(\ref{eq:mixphi}) we obtain
\begin{eqnarray}
{\dd \Phi \over \dd \omega} &=& - q \omega_{\rm p}^{-1} 
{1 \over (q^2 \cot^2 \theta_{\rm p} + 1) \sin^2\theta_{\rm p}} \nonumber \\
&=& - q \omega_{\rm p}^{-1} 
{1 \over q^2 + (1 - q^2) \sin^2\theta_{\rm p}} \; .
\label{eq:dphidomega}
\end{eqnarray}
Substituting this into Eq.~(\ref{eq:mixperspac}), using 
Eqs~(\ref{eq:omega_p}) and (\ref{eq:theta_p}), 
yields
\begin{equation}
\Pi_{n\,l} - \Pi_{n+1\,l} \simeq \zeta_{\rm as} \Delta \Pi_l \;,
\label{eq:perspac_zeta}
\end{equation}
with $\zeta_{\rm as}$ given by Eq.~(\ref{eq:zeta_asymp}).
This is the result obtained by \citet{mosser2015}.
Equation~(\ref{eq:perspac_zeta}) is 
satisfied to high accuracy by computed values of $\zeta$ and 
period spacing for red-giant models.
As an example, Fig.~\ref{fig:rgperspac_zeta} shows part of the
period spacing from Fig.~\ref{fig:rgperspac}, but including also
$\zeta \Delta \Pi_1$, with $\zeta$ computed from the numerical eigenfunctions.

We note that the derivation of $\zeta_{\rm as}$, and hence the relation 
in Eq.~(\ref{eq:perspac_zeta}), assumes the validity of the comparatively simple
asymptotics employed by \citet{shibahashi1979}, which requires a slow
variation of the equilibrium quantities. 
This approximation is not valid in the case of sharp features in the buoyancy
frequency, such as is the case for Model 1a of \citet{cunha2015}
(see also Fig.~\ref{fig:perspac_glitch}).
In these cases the more complex analysis of Cunha et al 
would be required.

\begin{figure}
\centering
\begin{minipage}{\linewidth}
\centering
\includegraphics[width=9cm]{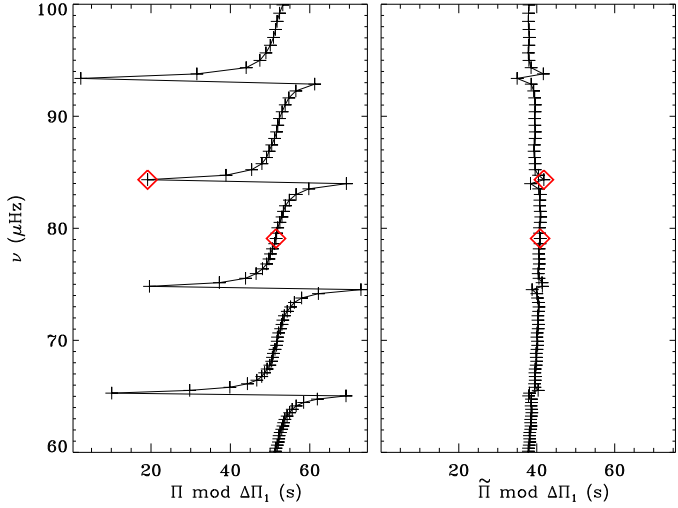}
\end{minipage}
\caption{Left: period \'echelle diagram with
$\Delta \Pi_1 = 75.4 \muHz$
for $l = 1$ modes, in the model $\modrg$ in Fig.~\ref{fig:chenhr} $(1.3 \Msun \, , \; 6.2 \Rsun)$,
as also shown in the right panel of Fig.~\ref{fig:rgperechl}.
Right: period \'echelle diagram using stretched periods $\tilde\Pi$,
with $\Delta \Pi_1 = 75.6 \muHz$.
The red diamonds mark the modes illustrated in Fig.~\ref{fig:rgamde}.
}
\label{fig:compperechl_stretch}
\end{figure}

From Eq.~(\ref{eq:perspac_zeta}) 
\citet{mosser2015} and \citet{vrard2016} noted that the properties of
$\zeta_{\rm as}$ (or $\zeta$)%
\footnote{In the following we only distinguish between $\zeta$ 
and $\zeta_{\rm as}$ when directly relevant.}
are largely controlled by the properties of the acoustic resonances,
with little dependence on the period spacing.
As a result they were able to develop a technique for determining $\zeta$
from the observations in the vicinity of the acoustic resonances.
Given $\zeta$, they introduced a stretching function $\CP(\nu)$%
\footnote{\citet{mosser2015} used the opposite sign and furthermore
denoted the stretched variable `$\tau$'; we change the notation to
avoid confusion with the acoustic depth, and change the sign to
obtain a quantity more closely related to the period.}
by
\begin{equation}
\dd \CP(\nu) = - {1 \over \zeta} {\dd \nu \over \nu^2} \; ,
\label{eq:stretch}
\end{equation}
and the stretched periods $\tilde \Pi_{nl} = \CP(\nu_{nl})$.
Replacing $\zeta$ by $\zeta_{\rm as}$ and relating 
$\zeta_{\rm as}^{-1}$ to $\dd \Phi / \dd \omega$
using Eqs~(\ref{eq:zeta_asymp}) and (\ref{eq:dphidomega})
yield
\begin{equation}
\dd \CP(\nu) = \left( - {1 \over \nu^2} + {\Delta \Pi_l \over \pi}
{\dd \Phi \over \dd \nu} \right) \dd \nu \; ,
\end{equation}
or, with suitable choice of integration constant,
\begin{equation}
\CP = {1 \over \nu} + {\Delta \Pi_l \over \pi} \Phi
= \Delta \Pi_l \left [ \Psi(\nu) + \epsilon_{\rm g} + 1/2 \right] \; ,
\end{equation}
using the definition of $\Psi$ in Eq.~(\ref{eq:mixfreq}).
From Eq.~(\ref{eq:mixfreq}) it follows that the stretched periods of the 
modes satisfy
\begin{equation}
\tilde \Pi_{n\,l} = \Delta \Pi_l (k + \epsilon_{\rm g} + 1/2) \; .
\label{eq:mixbuoy_str}
\end{equation}
Thus we formally recover the relation (\ref{eq:mixbuoy}) for the
uncoupled g modes. However, this relations is now valid for all modes.
As shown by \citet{mosser2015} this greatly simplifies the 
period \'echelle diagram when expressed in terms of the stretched period.
This is illustrated in Fig.~\ref{fig:compperechl_stretch};
as suggested by Eq.~(\ref{eq:mixbuoy_str}) all modes fall essentially
on a straight line.

\citet{vrard2016} developed the stretching technique
into an automated method for
analysing the red-giant mixed-mode spectra of large samples of stars;
this is particularly powerful in dealing with the complications of 
rotational splitting.
We return to this in Section~\ref{sec:rotation}.

\subsubsection{Effects of glitches}
\label{sec:glitch}
\label{sect:glitchestheory}

Much of the preceding discussion of the properties of red-giant oscillations
was based on the asymptotic properties of the modes.
This conveniently allows characterizing the oscillation spectrum 
by a relatively small
number of parameters, such as $\Delta \nu$, $\Delta \Pi_1$, etc.
However, the asymptotic approximation also indicates
a limitation of the diagnostic potential of the observations,
as far as the detailed internal properties of the stars are concerned.
It should be kept in mind that the asymptotic analysis is based on
the assumption that the underlying stellar structure varies on a scale
that is long compared with 
the local wavelength of the oscillations.
Structure features on a smaller scale introduce perturbations to the
frequencies which may provide diagnostics about these structures.
This was pointed out in the case of p modes in the Sun by \citet{gough1990},
who noted that at the base of the convection zone in normal solar models
the second derivative of the sound speed is discontinuous.
In addition,
the relatively rapid variation of $\Gamma_1$, and hence the sound speed,
in the second helium ionization zone may also cause departures from
the asymptotic behaviour.
\citet{vorontsov1991} used the signature of helium ionization in observed
solar frequencies to infer the solar envelope helium abundance.
Such features in the sound speed were denoted {\it acoustic glitches}
by \citet{gough2002}.
Similarly, sharp features in the buoyancy frequency 
\citep[{\it buoyancy glitches}, ][]{cunha2015} may affect g-dominated modes.

\begin{figure}
\centering
\begin{minipage}{\linewidth}
\centering
\includegraphics[width=9cm]{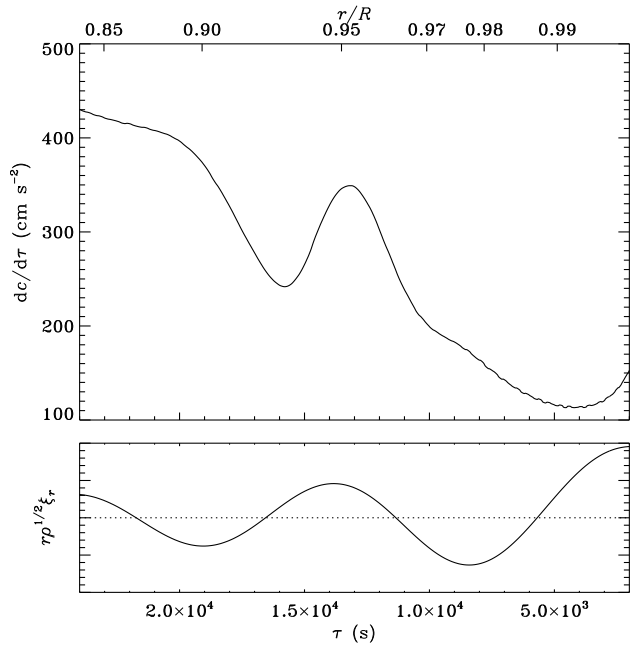}
\end{minipage}
\caption{Top panel: gradient of sound speed with respect to acoustic 
depth $\tau$ (cf. Eq.~\ref{eq:acdepth}) in the red-giant model $\modrg$ in Fig.~\ref{fig:chenhr}
$(1.3 \Msun \, , \; 6.2 \Rsun)$.
Lower panel: scaled displacement eigenfunction, on arbitrary scale,
for a radial mode with frequency $\nu = 98.1 \muHz$ in the same model.
The lower abscissa shows acoustic depth and the upper abscissa the 
corresponding relative distance to the centre.
}
\label{fig:heglitch}
\end{figure}

The base of the convective envelope in red giants is typically too deep to
induce a significant effect on the p-dominated modes \citep{miglio2010}.
An example of the acoustic glitch associated with the helium ionization zone
in a red-giant star
is illustrated in Fig.~\ref{fig:heglitch}, showing the derivative
of sound speed in terms of acoustic depth $\tau$ (cf. Eq.~\ref{eq:acdepth}).
Here the dominant feature is the dip in $\dd c /\dd \tau$ near
$\tau = 1.5 \times 10^4 \, {\rm s}$ which is associated with the second
helium ionization zone.
For comparison the lower panel shows the scaled displacement eigenfunction
of a radial mode of frequency $98.1 \muHz$, with a variation at the helium
feature which is substantially slower than the structural variation.
Thus the helium dip does act as an acoustic glitch.

\begin{figure}
\centering
\begin{minipage}{\linewidth}
\centering
\includegraphics[width=9cm]{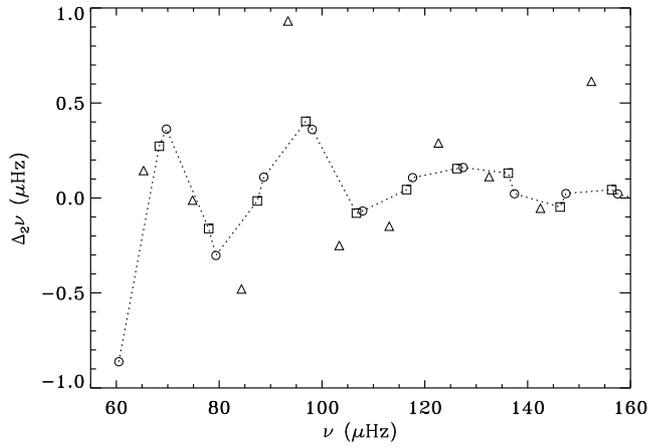}
\end{minipage}
\caption{Second frequency differences (cf. Eq.~\ref{eq:secdif})
in the model illustrated in Fig.~\ref{fig:heglitch}
for $l = 0$ (circles) and the most acoustically dominated modes with
$l = 1$ (triangles) and $l = 2$ (squares).
For clarity modes of degree 0 and 2 are connected by a dotted line.}
\label{fig:secdif}
\end{figure}

\citet{gough1990} showed that the effect of an acoustic glitch is to 
introduce a variation $\delta \nu^{\rm (gl)}$ of the form
\begin{equation}
\delta \nu^{\rm (gl)} \propto 
\sin(4 \pi \nu \tau_{\rm gl} + 2 \phi_{\rm gl}) \; 
\end{equation}
(see also Eq.~\ref{eq:glitch}),
where $\tau_{\rm gl}$ is the acoustic depth of the glitch and $\phi_{\rm gl}$
is a phase.
Thus at fixed $\tau_{\rm gl}$ we expect an oscillatory behaviour of
$\delta \nu^{\rm (gl)}$ as a function of cyclic frequency, 
with a period of around $(2 \tau_{\rm gl})^{-1}$.
\citet{gough1990} furthermore proposed isolating the effect of the glitch
from other possible
slower variations of the frequencies with radial order by considering
second differences at fixed $l$,
\begin{equation}
\Delta_2 \nu_{n\,l} = \nu_{n-1\,l} - 2 \nu_{n\,l} + \nu_{n+1\,l} \; .
\label{eq:secdif}
\end{equation}
This is illustrated in Fig.~\ref{fig:secdif} for the same model as shown
in Fig.~\ref{fig:heglitch}.
The effect of the helium glitch is only relevant for the acoustic properties of
the modes, and hence we have selected just those modes with $l = 1$ and $2$
with the lowest inertia in each radial-mode frequency interval; modes with degree 0 and 2 are connected by a dotted line.
The figure shows the expected oscillatory behaviour, 
although modes with $l = 1$ that are more strongly affected by the 
mixed nature show some scatter around the general trend and hence have not been connected.
The period of the oscillation is roughly consistent with the 
acoustic depth of the glitch. 
The amplitude of the oscillation depends on the strength of the glitch,
which is determined by the magnitude of the variation in $\Gamma_1$
and hence provides a diagnostic of the helium abundance
in the convective envelope of the star.
With increasing frequency the local wavelength of the eigenfunction gets
shorter, making the glitch appear less sharp and hence reducing the  
amplitude of the variation in the second difference
(see also Section~\ref{sect:glitches}).

\citet{houdek2007} carried out a detailed analysis of the effects
of the acoustic glitches in the first and second 
helium ionization zones and the
base of the convective envelope in the Sun.
This led to an expression already summarized in Eq.~(\ref{eq:glitch}). 
They noted the importance of a proper definition of the {\it acoustic
surface}, i.e., the zero point of the acoustic depth, in the 
interpretation of the analysis of acoustic glitches.
The potential uncertainty involved in the definition of the
acoustic surface, and the effects of the near-surface errors in
stellar modelling (cf. Section~\ref{sect:surfaceeffect}),
can in principle be avoided by converting the
acoustic depth to an acoustic distance from the centre 
\citep[e.g.,][]{ballot2004}; 
this uses the fact that the asymptotic large frequency separation
$\Delta \nu$ is related to the total acoustic radius of the star
(cf. Eq.~\ref{eq:larsep1}).
In practice the acoustic depth of a feature and the large frequency
separation may respond differently to the near-surface uncertainties,
in particular since $\Delta \nu$ is typically determined from a suitable
fit or average over the observed modes.

\begin{figure}
\centering
\begin{minipage}{\linewidth}
\centering
\includegraphics[width=9cm]{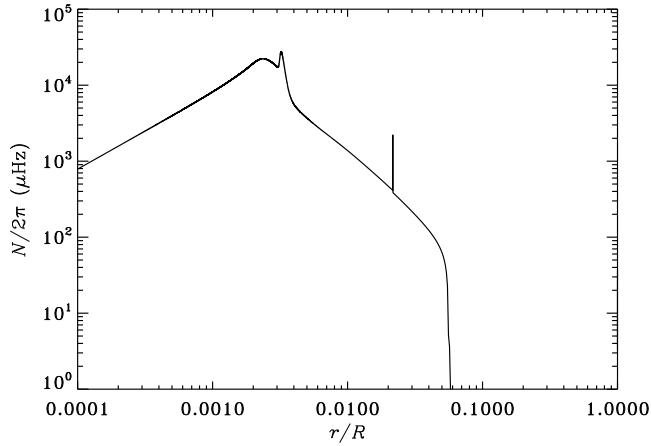}
\end{minipage}
\caption{Buoyancy frequency in $1 \Msun$ model with a radius of $9.7 \Rsun$
and an effective temperature of $4439 \, {\rm K}$
\citep[adapted from][]{cunha2015}.
}
\label{fig:bvrg_cunha}
\end{figure}

\begin{figure}
\centering
\begin{minipage}{\linewidth}
\centering
\includegraphics[width=9cm]{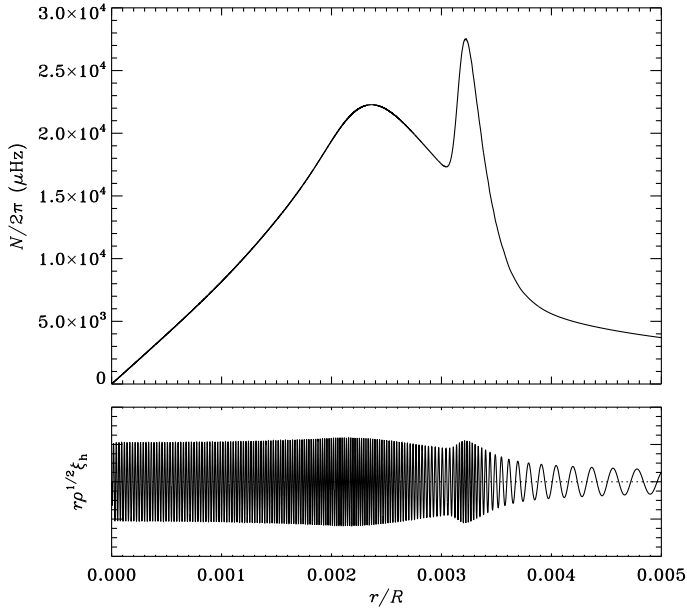}
\end{minipage}
\caption{Top panel:
Buoyancy frequency in the core of the model illustrated in
Fig.~\ref{fig:bvrg_cunha} \citep{cunha2015}.
Bottom panel: scaled horizontal displacement (on arbitrary scale) for 
a mode in this model with $l = 1$, $\nu = 43.41 \muHz$.
}
\label{fig:bvglitch}
\end{figure}

Buoyancy glitches, i.e., sharp features in the buoyancy frequency,
are of great potential interest owing to their sensitivity to the
composition structure.
Analysis of buoyancy glitches 
caused by the composition discontinuities established through 
gravitational settling
has played a major role in the study of white-dwarf oscillations
as important diagnostics of
the internal structure of the stars \citep[e.g.,][]{winget1994}.
Additionally, in a detailed analysis \citet{miglio2008} demonstrated the 
diagnostic potential of the buoyancy glitch at the edge of convective
cores in more massive main-sequence pulsators, 
leading to oscillations with mode order in the period spacings of
high-order g modes.
Evidence for this behaviour was found by \citet{degroote2010} in
a slowly pulsating B star observed by CoRoT,
indicating effects of processes smoothing the composition gradient
at the edge of the core.
Thus it is of great interest to consider the potential for studies of
buoyancy glitches in evolved stars.

Figure~\ref{fig:charfrqrg} shows the presence of potential glitches in
the buoyancy frequency in red-giant models.
These were analysed in considerable detail by \citet{cunha2015}, 
including an asymptotic analysis of the effect of a discontinuity in
composition and a resulting singularity in the buoyancy frequency
(cf. Eq.~\ref{eq:bvapprox}).
Here we discuss aspects of these issues using as an example 
a model corresponding
to Model 1a of \citet{cunha2015}, but with a density discontinuity at the
base of the dredge-up region.
The buoyancy frequency for this model is shown in Fig.~\ref{fig:bvrg_cunha}.
The model is in a somewhat later evolutionary stage,
just before the red-giant `bump', than the model 
shown in Fig.~\ref{fig:charfrqrg}.
Here the convective envelope has retracted in mass since the maximal
extent of the first dredge-up (see Section~\ref{sect:Hburn}),
leaving behind a composition and density discontinuity within the
g-mode propagation region, here shown as a sharp spike in $N$ 
at $r/R \simeq 0.02$.
This is obviously a glitch.%
\footnote{As discussed by \citet{cunha2015} this feature 
remains sharp compared with the local wavelength
even if some numerical (or physical) diffusion is allowed,
smoothing the composition gradient.
This deserves further study.}
The second relatively sharp feature is associated 
with the hydrogen-burning shell.
As illustrated in Fig.~\ref{fig:bvglitch} this is not a glitch, however,
since the eigenfunction varies extremely rapidly in this region.
This is an immediate consequence of the asymptotic behaviour
in Eq.~(\ref{eq:asympsol_g}),
according to which the local radial wave number scales as $N/r$ and hence
becomes very large at the peak in the buoyancy frequency.

\begin{figure}
\centering
\begin{minipage}{\linewidth}
\centering
\includegraphics[width=9cm]{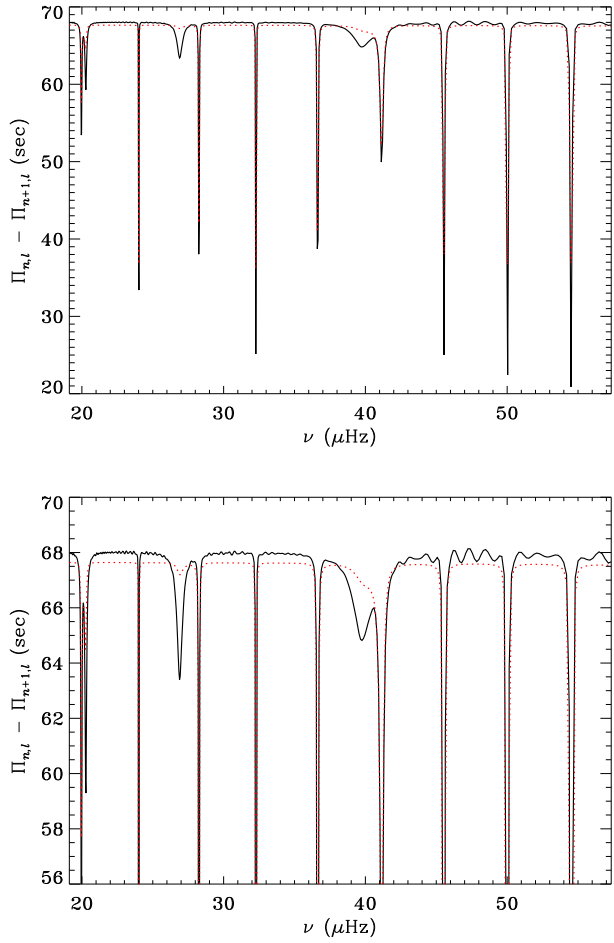}
\end{minipage}
\caption{Period spacings
for modes with $l = 1$, in a $1 \Msun$ model at the bump, with buoyancy glitch
\citep[adapted from][]{cunha2015}.
The red dotted line shows $\zeta \Delta \Pi_{1,{\rm as}}$.
The lower panel provides a blow-up of the region near the asymptotic period
spacing.
}
\label{fig:perspac_glitch}
\end{figure}

The analysis by \citet{cunha2015} shows that, for a pure g mode,
the presence of a buoyancy glitch gives rise to an oscillatory behaviour
such as was found by, e.g., \citet{miglio2008}, with a period depending
on the location of the glitch within the g-mode cavity.
Taking into account also the mixed nature of the modes, with the coupling
to the p-mode behaviour,
leads to a complex variation in the frequencies and the period spacings,
as illustrated in Fig.~\ref{fig:perspac_glitch}.%
\footnote{We note that this is rather more complex than the simple model
of the effects of a buoyancy glitch considered by \citet{mosser2015}.}
The presence of a glitch clearly invalidates the derivation of 
the asymptotic relation in Eq.~(\ref{eq:perspac_zeta}) between
$\zeta$ and the period spacing, as also illustrated 
in Fig.~\ref{fig:perspac_glitch}.
This shows a very interesting diagnostic potential for the
characterization of the chemical profile left by the dredge-up 
in red giants just below the bump;
however, since the effects are subtle a more detailed analysis is required
to ascertain whether the effect can reliably be isolated in the 
existing data.
\citet{cunha2015} proposed a diagnostic of the irregularity of the 
period spacings, based on a Fourier transform of period spacing as a
function of period,
which in the model results provides a strong indication 
of glitch effects.
This also demonstrated that the effects in red-giant models are closely
linked to the red-giant bump.

As discussed in the next section buoyancy glitches play a dominant
role in the oscillation spectrum of core helium-burning stars.

\subsubsection{Stars on the red clump}
\label{sec:redclump}

A major early realization in the analysis of space-based asteroseismic
observations of evolved stars was that there is a sharp distinction
between stars on the ascending red-giant branch, with hydrogen fusion 
around the helium core, and stars in the red clump where in addition
there is helium fusion near the centre
\citep{bedding2011, mosser2011mm} (see also Section~\ref{sect:dP}).
The g-mode period spacing is substantially smaller in the red giants than
in clump stars.
This is a straightforward consequence of the difference in the internal
structure between these different stars (at fixed surface radius, say),
given the asymptotic expression for the period spacing, Eq.~(\ref{eq:gper0}),
and the expression for the
buoyancy frequency, Eq.~(\ref{eq:bvapprox}).
The core helium-burning stars have a convective core
(see Figs~\ref{fig:kip1} and \ref{fig:kip3}),
which restricts the g-mode cavity and contributes to reducing
the integral over the buoyancy frequency in Eq.~(\ref{eq:gper0}),
hence increasing the asymptotic period spacing
\citep[e.g.,][]{jcd2014iac}.
In addition, with ignition of helium burning the core expands,
reducing the local gravitational acceleration and hence the buoyancy
frequency (cf. Eq.~\ref{eq:bvapprox}),
further increasing the period spacing.

\begin{figure}
\centering
\begin{minipage}{\linewidth}
\centering
\includegraphics[width=9cm]{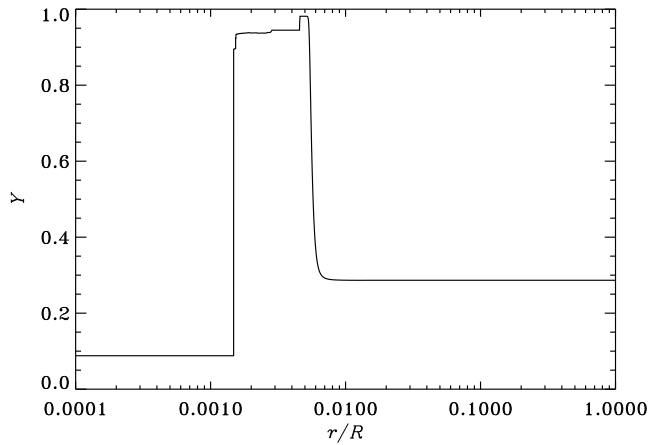}
\end{minipage}
\caption{Helium abundance as a function of fractional radius in
a model in the core helium-burning phase,
with initial mass $1.5 \Msun$, radius $12.2 \Rsun$ and
effective temperature $4648\,{\rm K}$. 
}
\label{fig:clumphe}
\end{figure}

To illustrate these effects we consider models discussed by \citet{jcd2014}.
The evolution sequence, extending from the pre-main sequence through the central
helium-burning phase, was computed with the GARSTEC code \citep{weiss2008}.
This includes full treatment of the helium flash and the subsequent
sub-flashes (see Section~\ref{sect:He-flash}).
The resulting helium-abundance profile $Y$ in a model with initial
mass $1.5 \Msun$ near the end of central helium burning
is illustrated in Fig.~\ref{fig:clumphe}.
The region of constant $Y$ inside $0.0015\, R$ is convective.
As discussed in Section~\ref{sect:Hecoreburn} 
a discontinuity in $Y$, and hence in density, may be established
at the edge of the convective core.
In the rest of the helium core the smaller steps in composition 
reflect the convective mixing associated with the initial helium flash
and the sub-flashes.

\begin{figure}
\centering
\begin{minipage}{\linewidth}
\centering
\includegraphics[width=9cm]{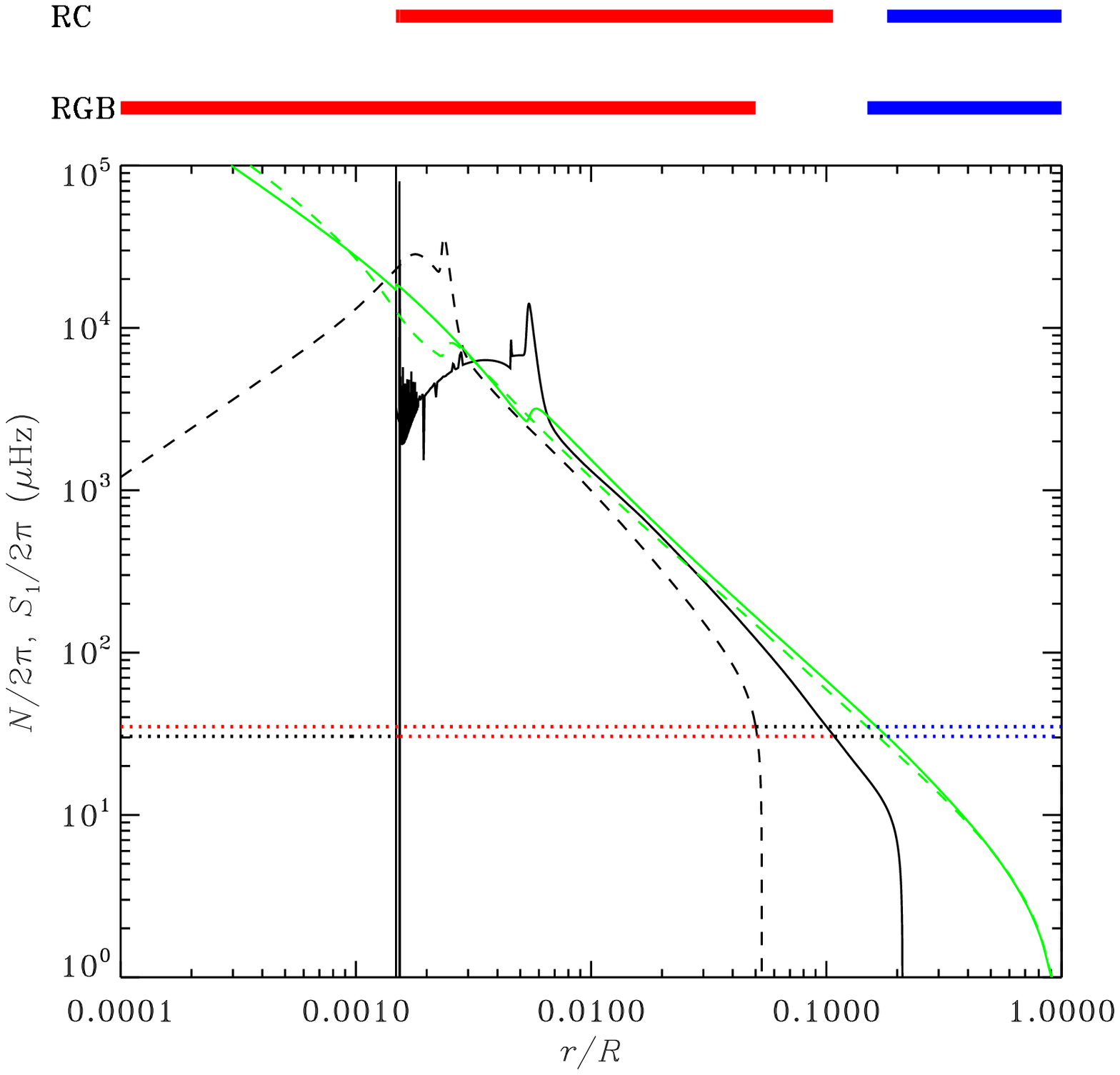}
\end{minipage}
\caption{The solid curve shows the buoyancy frequency in the 
core helium-burning model illustrated in Fig.~\ref{fig:clumphe}.
For comparison, the dashed curve shows the buoyancy frequency
in a red-giant model on the same $1.5 \Msun$ initial mass 
evolution sequence, with a radius of $12.0 \Rsun$ and effective 
temperature $4532 \,{\rm K}$.
The solid and dashed {\orange green} lines show the Lamb frequency $S_{\!1}$ for the
clump and red-giant models, respectively. 
The dotted lines indicate
the frequencies $\nu_{\rm max}$ of maximum oscillation power
(cf. Eq.~\ref{eq:numaxt}),
{\orange with the red and blue parts marking respectively
the g-mode and p-mode cavities for $l = 1$;
for clarity the g- and p-mode cavities are also shown, respectively, 
by the red and blue bars above the plot, for the red-giant (RGB) and clump (RC)
models.}
}
\label{fig:clumpbv}
\end{figure}

The buoyancy frequency of this model is illustrated in Fig.~\ref{fig:clumpbv}
and compared with that of a red-giant model of the same initial mass and
approximately the same radius and slightly lower effective temperature.
This immediately shows the two effects mentioned above: 
the presence of a convective core and
the lower general level of the buoyancy frequency
in the core helium-burning model.
As a result the asymptotic dipolar period spacing in the core helium-burning
model is $\Delta \Pi_1 = 244.2 \, {\rm s}$ while in the red-giant model 
$\Delta \Pi_1 = 56.9 \, {\rm s}$.
This is the effect found by \citet{bedding2011} and \citet{mosser2011mm}
and extensively used since then to characterize evolved stars
\citep[e.g.,][see Fig.~\ref{fig:dPvsdnu}]{stello2013apj, mosser2014}.

\begin{figure}
\centering
\begin{minipage}{\linewidth}
\centering
\includegraphics[width=9cm]{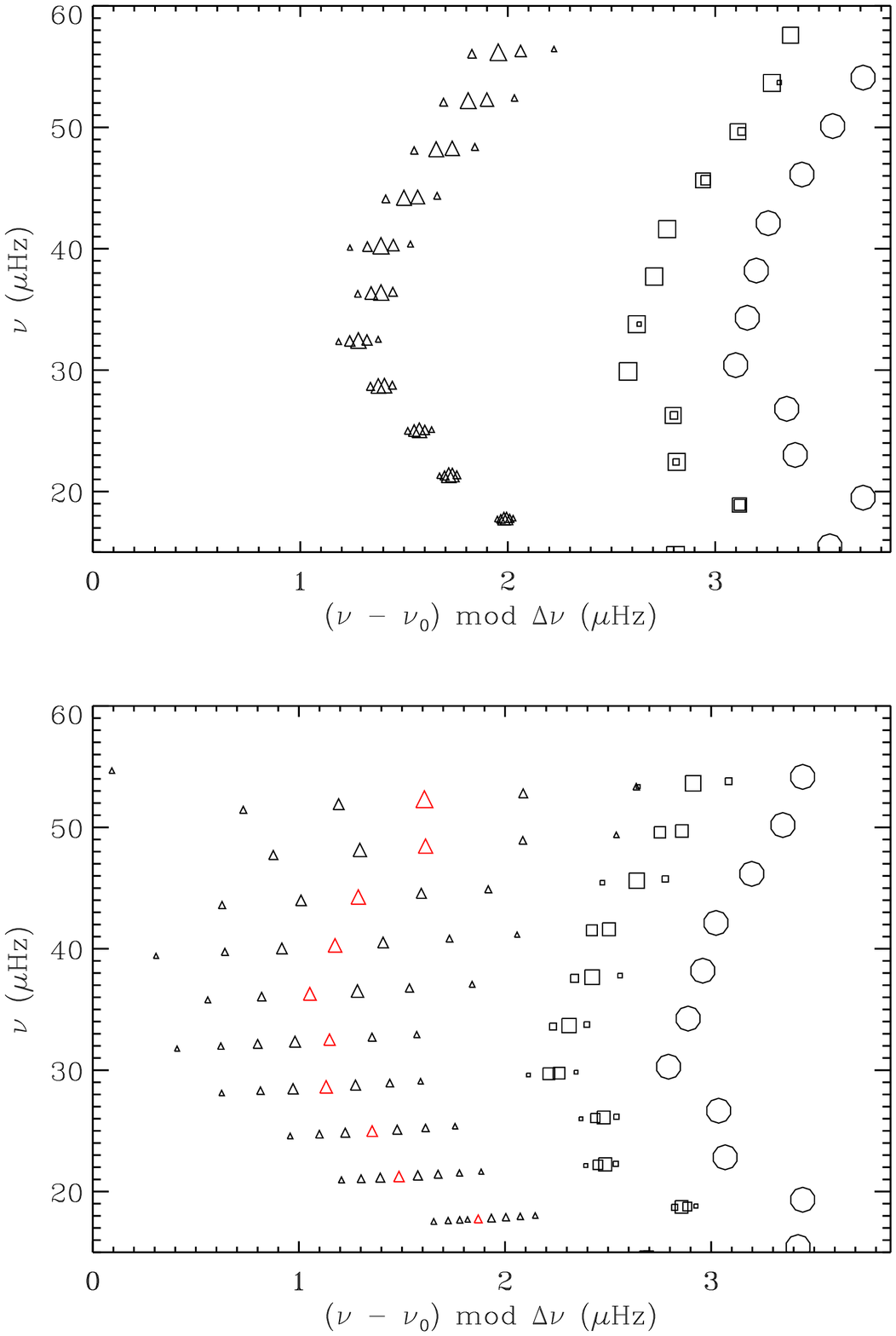}
\end{minipage}
\caption{Frequency \'echelle diagrams for the red-giant (top) and clump
(bottom) models illustrated in Fig.~\ref{fig:clumpbv}. 
In the red-giant model $\Delta \nu = 3.845 \muHz$, and in the clump model
$\Delta \nu = 3.870 \muHz$.
For clarity the frequencies have been shifted by $\nu_0 =0.4 \muHz$
in both panels.
Circles, triangles and squares show modes of degree 0, 1 and 2, respectively.
As in Fig.~\ref{fig:chenechl} the symbol size for the non-radial modes
have been scaled with $Q^{-1/2}$, where $Q$ is the inertia ratio
(Eq.~\ref{eq:sclinertia}), providing a rough indication of the expected
amplitude of the modes.
The dipolar modes with lowest inertia in each radial-mode frequency interval
are shown in red in the bottom panel.
}
\label{fig:clumpechl}
\end{figure}

A second clear difference between the red-giant and clump models in
Fig.~\ref{fig:clumpbv} is the extent of the evanescent region between 
the buoyancy and acoustic cavities, which is much smaller in the clump than 
in the red-giant model, leading to a stronger coupling.
The effect of this is illustrated in frequency \'echelle diagrams
in Fig.~\ref{fig:clumpechl} where,
as in Fig.~\ref{fig:chenechl}, the symbol size provides an indication of
the expected amplitude of the modes.
It is evident that the stronger coupling in the clump model leads to a
much broader spread of modes, particularly for $l = 1$.
This is probably the reason for the less regular behaviour of the 
dipolar mode separations found by \citet{montalban2010} who
identified the most acoustic modes as those modes having the smallest 
inertia in a given radial-mode interval (see also Section~\ref{sect:sdnu_01}).
To illustrate this, these modes are shown in red in Fig.~\ref{fig:clumpechl}.

\begin{figure}
\centering
\begin{minipage}{\linewidth}
\centering
\includegraphics[width=9cm]{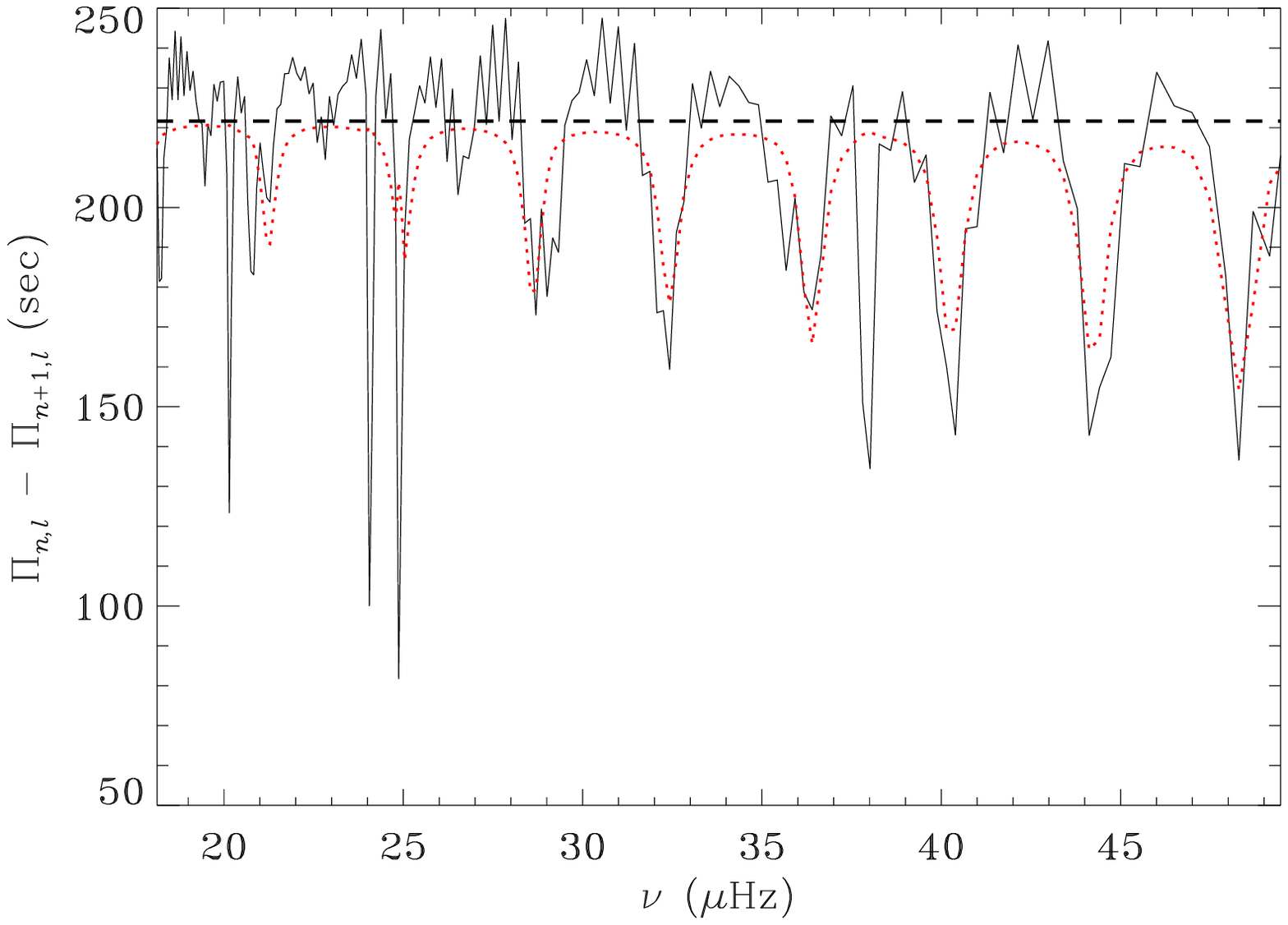}
\end{minipage}
\caption{Dipolar period spacings {\orange derived from the differences in the periods of the computed modes (solid line)} in the
core helium-burning model illustrated in Fig.~\ref{fig:clumphe}.
The horizontal dashed line shows
the asymptotic period spacing $\Delta \Pi_1$;
{\orange the red dotted line shows the period spacing $\zeta \Delta \Pi_1$
predicted from the comparatively simple asymptotic analysis 
in Section~\ref{sec:ascenrg} (cf. Eq.~\ref{eq:perspac_zeta}),
although} using $\zeta$ computed from the numerical eigenfunctions.
}
\label{fig:clumpperspac}
\end{figure}

The diagnostics of clump stars has so far predominantly been based on 
the observed and asymptotically calculated period spacings.
However, the actual behaviour of the model frequencies is rather more
complicated, owing to the composition discontinuities and resulting
spikes in the buoyancy frequency, also visible in Fig.~\ref{fig:clumpbv}.
These act as glitches and lead to partial reflection of the g modes.
The specific behaviour depends strongly on fine details of the
model calculation, such as whether diffusion (numerical or physical) is
included and whether the composition profile and buoyancy frequency are
properly resolved in the numerical computations.
Figure~\ref{fig:clumpperspac} shows the computed dipolar period spacings
in the core helium-burning model in Fig.~\ref{fig:clumpbv},
compared with the asymptotic period spacing and the period spacing 
expected from the 
asymptotics in Eq.~(\ref{eq:perspac_zeta}), 
but using the numerically computed $\zeta$.
The computed period spacing is clearly related to the expected behaviour
in this case, reflecting the location of the acoustic resonances shown by
$\zeta$, but with substantial variations caused by the glitches
in the buoyancy frequency.%
\footnote{We note that the behaviour illustrated here is insensitive to
the details of the frequency calculation, provided this is done with
sufficient precision.
Thus it is a property of the structure of the stellar model,
although not necessarily of a star!}

\begin{figure}
\centering
\begin{minipage}{\linewidth}
\centering
\includegraphics[width=9cm]{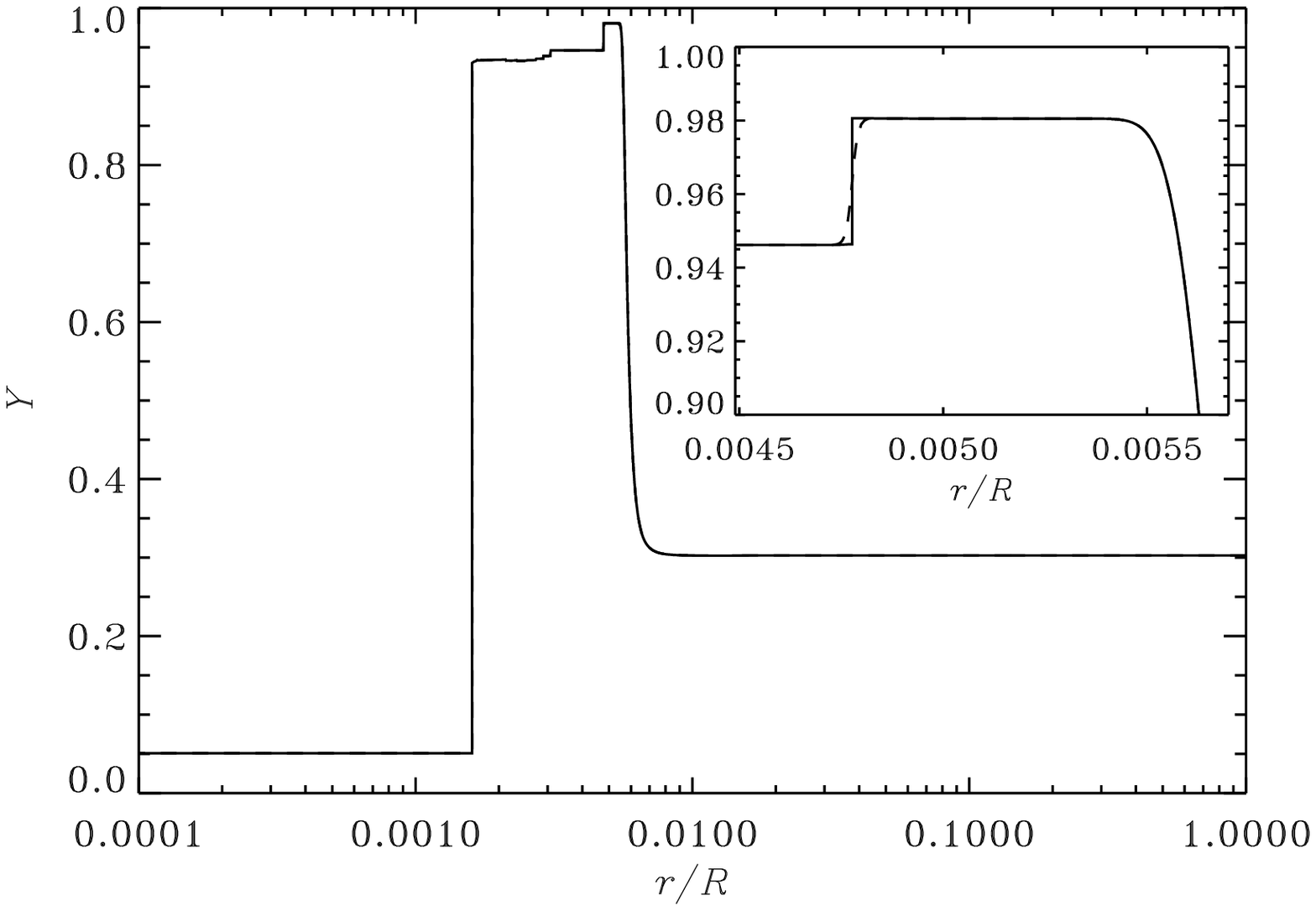}
\end{minipage}
\caption{Helium abundance as a function of fractional radius in
two models in the core helium-burning phase,
with initial mass $1.0 \Msun$, radius $12 \Rsun$ and
effective temperature $4699\,{\rm K}$.
The inset shows the region near the sharp decrease in $Y$ 
at $r/R \simeq 0.0047$ caused by
convection associated with the helium flash at the start of helium burning.
One model (solid) has a composition discontinuity, whereas in the second
model (dashed) this has been slightly smoothed by diffusion.
(Models courtesy of Marcelo Miguel Miller Bertolami.)
}
\label{fig:clumplpthe}
\end{figure}

\begin{figure}
\centering
\begin{minipage}{\linewidth}
\centering
\includegraphics[width=9cm]{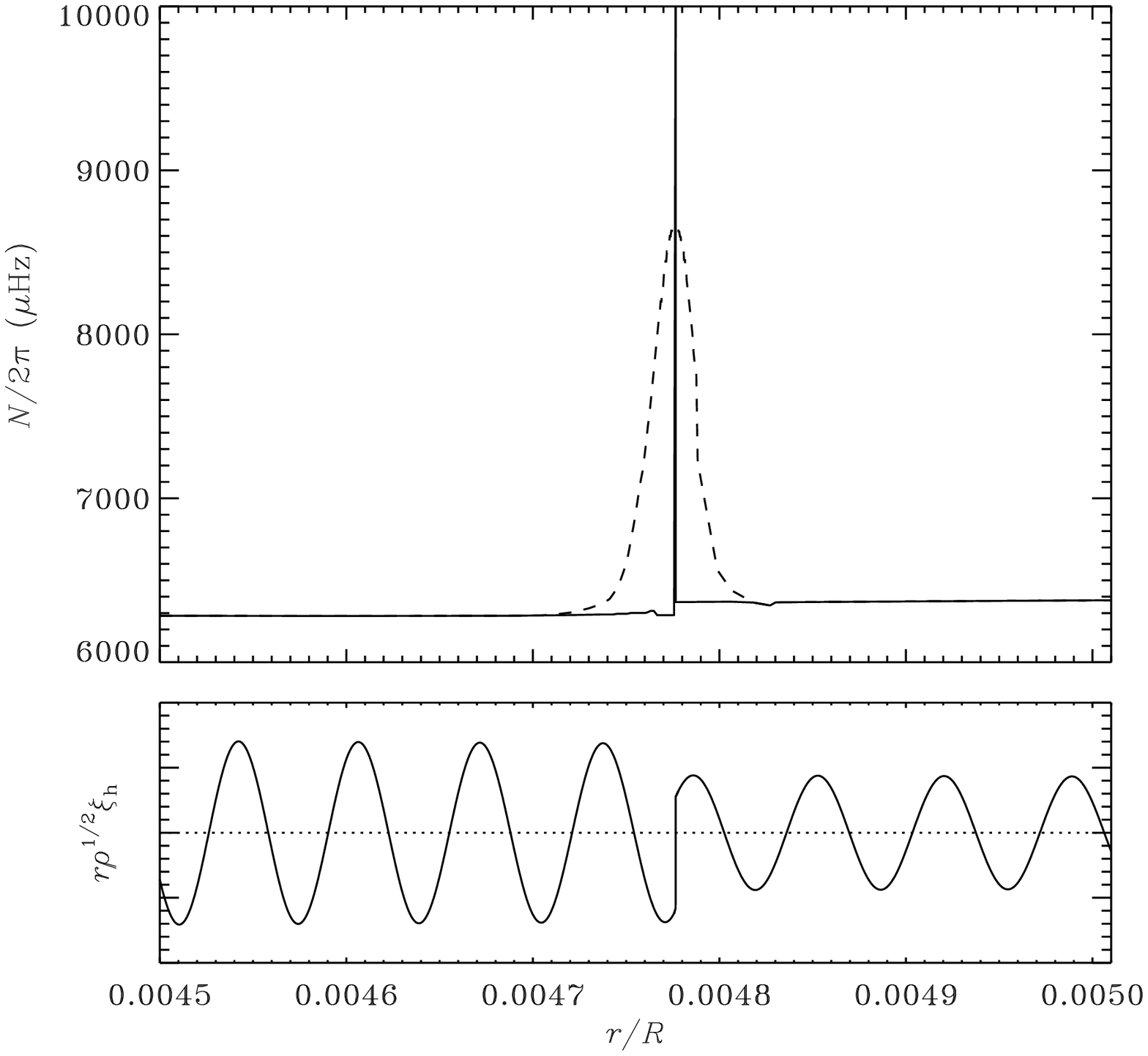}
\end{minipage}
\caption{Top panel:
Buoyancy frequency in the vicinity of the sharp decrease in the helium
abundance in the models illustrated in Fig.~\ref{fig:clumplpthe}.
For the nearly discontinuous helium profile the buoyancy frequency is
essentially a delta function (solid line).
The buoyancy frequency for the slightly smoothed model is shown by a dashed
line.
Bottom panel: scaled horizontal displacement (on arbitrary scale) for 
a mode in the model with the nearly discontinuous helium profile,
with $l = 1$, $\nu = 19.9 \muHz$.
(Models courtesy of Marcelo Miguel Miller Bertolami.)
}
\label{fig:clumplptbv}
\end{figure}

To illustrate further the sensitivity of the clump-model oscillations to
details of the model structure we consider two models 
with initial mass $1 \Msun$ computed with the LPCODE \citep{althaus2005}.
Their helium-abundance profiles $Y$ are illustrated 
in Fig.~\ref{fig:clumplpthe}.
The models only differ in the sharp decrease in $Y$ caused by the off-centre 
onset of the helium flash and the associated convective mixing.
In one model this is essentially discontinuous, whereas in the second model
the composition has been slightly smoothed by diffusion.
The resulting buoyancy frequencies in the vicinity of this glitch
are shown in Fig.~\ref{fig:clumplptbv},
which also shows the scaled horizontal displacement for a dipolar mode in the
`sharp' model.
For the model with the sharp helium profile this variation in
the buoyancy frequency is clearly a glitch.
For the slightly smoothed model the scales of the buoyancy frequency and
the eigenfunction are comparable, and hence one may expect a smaller effect
of the local model structure.%
\footnote{It might also be noticed that $\xi_{\rm h}$ is discontinuous
at the glitch: this is a consequence of the singularity in the
buoyancy frequency.}

\begin{figure}
\centering
\begin{minipage}{\linewidth}
\centering
\includegraphics[width=9cm]{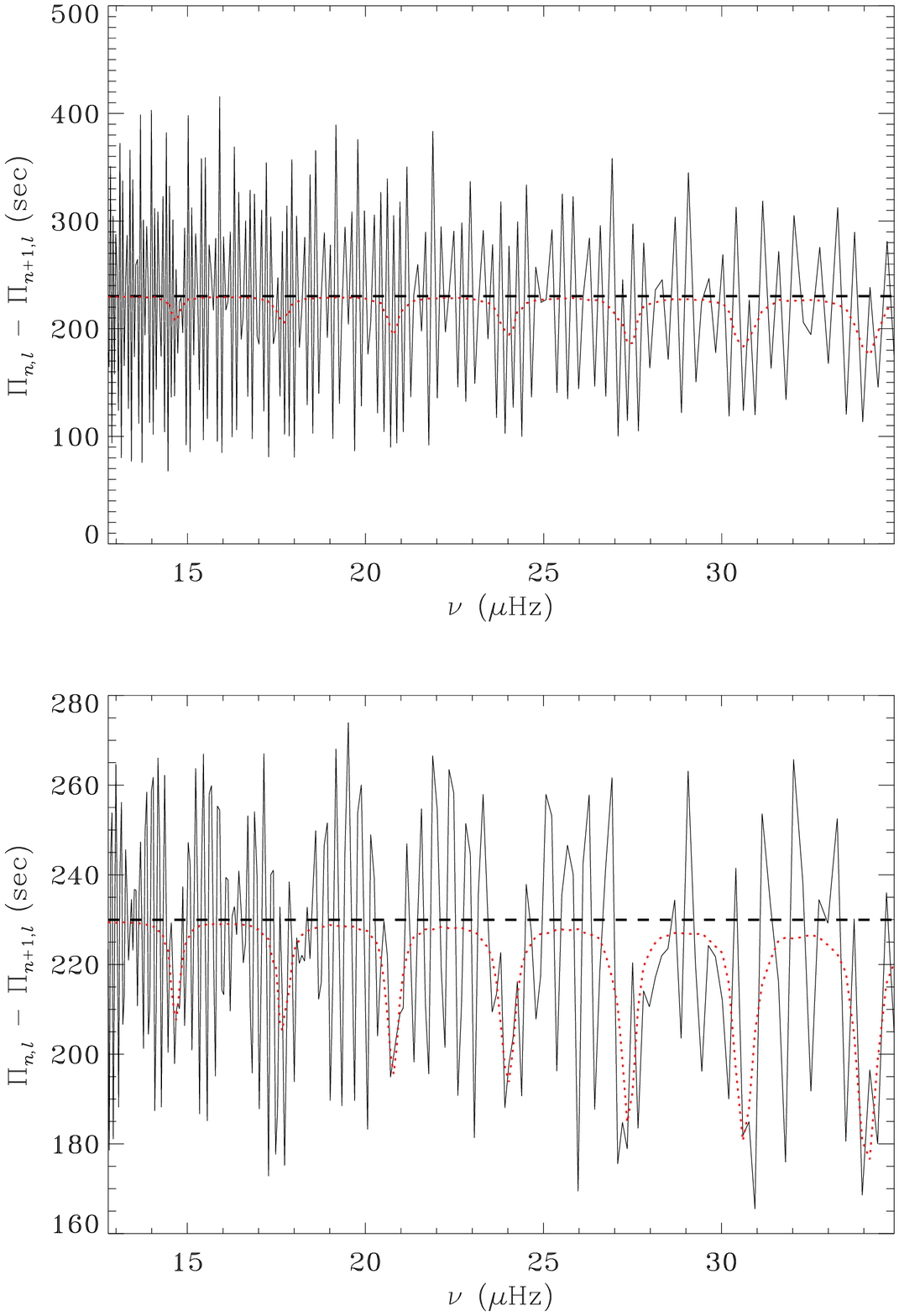}
\end{minipage}
\caption{Dipolar period spacings {\orange derived from the differences in the periods of the computed modes (solid line)} in the
core helium-burning models illustrated in Fig.~\ref{fig:clumplpthe}.
The top panel shows results for the model with a sharp helium profile
from the onset of the helium flash,
while the lower panel is for the model with a slightly smoothed profile.
The horizontal dashed lines show
the asymptotic period spacing $\Delta \Pi_1$,
while the red dotted lines show
$\zeta \Delta \Pi_1$ 
{\orange (see caption to Fig.~\ref{fig:clumpperspac})}.
(Models courtesy of Marcelo Miguel Miller Bertolami.)
}
\label{fig:clumplptperspac}
\end{figure}

This is indeed confirmed by the dipolar period spacings shown in 
Fig.~\ref{fig:clumplptperspac}.
For the `sharp' model the period spacings vary wildly, with little 
indication of the effects of the acoustic resonances, as reflected by
the period spacing expected from asymptotics (cf. Eq.~\ref{eq:perspac_zeta})
shown by the dotted red curve.
In the slightly smoothed model (bottom panel) 
the variations in the period spacing are
less dramatic and there are some indications of the decreases in the
period spacing expected from the asymptotic behaviour,
shown by the red dotted curve.
However, even in this case a proper analysis of the variation in the period
spacing, and a determination of its asymptotic value, would be difficult.

The behaviour of the internal structure during the phases of the helium flash
and sub-flashes was analysed by \citet{bildsten2012}.
They found strong changes in the asymptotic period spacing and the coupling
between the buoyancy and acoustic propagation regions, noting that this
may provide opportunities for the detailed diagnostics of this evolution
phase.
Even though it is brief, the large number of clump stars for which 
oscillation data are available from {\it Kepler} may make it realistic
that at least a few stars are in this phase; interestingly, \citet{mosser2014} do in fact identify 
several such stars in their $(\Delta \nu, \Delta \Pi_1)$ diagram
(see Fig.~\ref{fig:dPvsdnu}).
\citet{cunha2015} pointed out that the expansion of the helium core at
the onset of helium burning compresses the hydrogen-burning shell, so that
it acts as a buoyancy glitch during this phase (unlike on the red-giant
branch; cf. Fig.~\ref{fig:bvglitch}).
This would add to the diagnostic potential of the observed oscillations.
We note, however, that the early helium-burning model considered 
by \citet{cunha2015} did not have discontinuities in the composition
and hence did not suffer from the strong glitch effects illustrated 
in Fig.~\ref{fig:clumplptperspac}.

\citet{constantino2015} carried out an extensive analysis of the oscillation
properties of stars in the clump phase, varying also convective-core 
overshoot and other mixing processes at the edge of the convective core.
This clearly showed the very rich pulsational behaviour that may be 
found in these stars.
An interesting issue raised was the relation to the g-mode oscillations of the
core helium-burning subdwarf B stars which essentially correspond to
the naked core of the clump stars \citep[e.g.,][]{Ostensen2014}.
Constantino et al also addressed the fact that computed
asymptotic period spacings for clump models tend to be lower
than the observed spacings (see also Fig.~\ref{fig:dPvsdnu}).
A possible solution, although perhaps with limited physical justification,
would be the effect of what Constantino et al call the `maximum-overshoot'
scheme.
Similarly, \citet{bossini2015} considered various mixing schemes at the
edge of the helium-burning convective core and their effects on the 
period spacing.
They also pointed out that combining observed period spacings with the
location of the so-called {\it AGB bump} in stellar clusters,
resulting from non-monotonic
luminosity evolution at the onset of helium shell burning 
(see Section~\ref{sect:AGB}), 
would provide further observational constraints on these mixing processes.
Further studies are certainly needed on the apparent discrepancies between
the theoretically predicted detailed properties and the observations, and
of the consequences for the interpretation of the observations in terms
of simple diagnostics such as the asymptotic period spacing.

\subsection{Energetics of stellar oscillations}
\label{sec:energetics}
\label{sec:excitation}

It is generally accepted that solar-like oscillations, including those in
the Sun, are intrinsically damped \citep[but see][]{xiong2007}
and excited stochastically by 
the near-surface convection whose near-sonic speed makes the gas motions
efficient in generating acoustic noise \citep{stein1968}.
This is confirmed by analysis of the statistical properties of the 
variations in solar-oscillation amplitudes \citep{chaplin1997},%
\footnote{A similar analysis of the extended {\it Kepler} observations
would be very interesting.}
which follow the pattern expected from stochastic excitation
\citep{kumar1988, chang1998}.
Also, there is a striking similarity of the variation of amplitudes
with frequency
from the main sequence to red-giant branch, apart from the scaling
of the frequency of maximum power with surface gravity 
\citep [e.g.][]{deridder2009, stello2010}.
Furthermore, there is strong evidence that the variability seen in very 
evolved stars represents an extension of the red-giant solar-like oscillations
\citep{dziembowski2010, mosser2013agb}.
This possibly includes semiregular variables,
where analysis of up to century-long series of amateur observations
has revealed statistical properties
matching those of solar-like oscillations \citep{jcd2001}.
A concise review of energetics of red-giant oscillations was provided
by \citet{dupret2012}.

The general theory of stochastic forcing of a damped oscillator
was described by \citet{batchelor1953}.
It was applied to the analysis of observed solar modes by \citet{jcd1989}.
The resulting average power spectrum of a single mode,
with angular frequency $\omega_0$ and damping rate $\eta = - \omega_{\rm i}$
(cf. Eq.~\ref{eq:timedep}), has the form 
\begin{equation}
P(\omega) \simeq {1 \over 4 \omega_0^2}
{P_{\rm f} (\omega) \over (\omega - \omega_0)^2 + \eta^2} \; ,
\label{eq:stochpow}
\end{equation}
where $P_{\rm f}$ is the average power spectrum of the forcing function,
which varies relatively slowly with frequency.
Consequently the spectrum is a Lorentzian with a full width at half maximum
in angular frequency of $2 \eta$;
in terms of cyclic frequency the full width at half maximum is
\begin{equation}
\Gamma = {\eta \over \pi} \; .
\end{equation}
Measurement of the width of the observed peaks therefore provides a measure 
of the damping rates of the modes
(see also Section~\ref{sect:indf}, in particular Eq.~\ref{eq:lorfit}).
In addition to the damping rate it is convenient to characterize the damping
by the mode {\it damping time} $t_{\rm damp}$,
defined by the time required to reduce the
mode amplitude by a factor e and given by
\begin{equation}
t_{\rm damp} = \eta^{-1} \; .
\end{equation}

\subsubsection{Properties of the damping rate}

To study the energetics of the modes the full set of non-adiabatic
equations must be solved.
These include also perturbations to the
energy transport and the energy equation and result in
a determination of the complex frequency
$\omega = \omega_{\rm r} + \eye \omega_{\rm i}$ as an eigenvalue,
and hence the damping rate $\eta = - \omega_{\rm i}$.
However, to analyse $\eta$ 
it is convenient to express it in terms of the {\it work integral}
\cite[e.g.,][see also \citealt{aerts2010}]{baker1962, cox1967}.
Considering just perturbations to thermodynamic quantities the result is
\begin{equation}
\eta = \eta_{\rm gas} = - {1 \over 2 \omega_{\rm r}^2}
{\displaystyle {\rm Re} \left[\int_V {\delta \rho^* \over \rho} (\Gamma_3 - 1)
\delta (\rho \varepsilon - \div \boldF) \dd V \right] \over
\int_V \rho |\bolddelr |^2 \dd V} \; ,
\label{eq:workintg}
\end{equation}
where $\varepsilon$ is the rate of energy generation per unit mass,
$\boldF$ is the flux of energy,
\hbox{$\Gamma_3 - 1 = (\partial \ln T / \partial \ln \rho)_{\rm ad}$} and
the star indicates the complex conjugate;
also, $\delta$ denotes the Lagrangian perturbation, i.e., the perturbation
following the motion (see also footnote~\ref{fn:lagrangian}).
The integral in the numerator of Eq.~(\ref{eq:workintg}) reflects the
operation of a heat engine, with $\delta \rho$ defining compression,
and $\delta (\rho \varepsilon - \div \boldF)$ defining heating.
If the integral is positive, $\omega_{\rm i}$ is positive and
the mode is excited.
However, in the case discussed here of solar-like oscillations
the integrated effect is negative and the mode is damped.

Equation (\ref{eq:workintg}) is generally valid,
including for fully non-adiabatic solutions.
A major source of uncertainty in the calculation of the work
integral is the treatment of
the perturbations to the convective contribution to the flux,
particularly near the surface.
An additional complication in the near-surface region is the effect
of turbulent pressure $p_{\rm t}$ which makes a significant contribution to
the total pressure in the outermost parts of the convection zone 
and hence also affects the pulsations.
As a consequence, the full expression of the damping rate becomes
\begin{equation}
\eta = \eta_{\rm gas} + \eta_{\rm t} \;, 
\label{eq:workint}
\end{equation}
where $\eta_{\rm gas}$ is given by Eq.~(\ref{eq:workintg}) and
\begin{equation}
\eta_{\rm t} = 
{1 \over 2 \omega_{\rm r}} 
{\displaystyle {\rm Im} \left(\int_V {\delta \rho^* \over \rho} 
\delta p_{\rm t} \dd V \right) \over
\int_V \rho |\bolddelr |^2 \dd V} 
\label{eq:workturbp}
\end{equation}
\citep[e.g.,][]{balmforth1992a, houdek2015}, Im denoting the imaginary part.
We return to these convective effects below.

In much of the star where the oscillations are essentially adiabatic,
the contribution to the work integral
can be estimated from the {\it quasi-adiabatic approximation}, 
calculating all terms from the adiabatic eigenfunctions.
In red giants a potentially substantial contribution to the damping
comes from the buoyancy-dominated region in the core,
where the very high radial order of the g-mode behaviour causes a strong
diffusive damping.
In the outer parts of the star, on the other hand, the contributions
to the damping are dominated by the near-surface layers, where the properties
of the oscillations are independent of degree.
To analyse this it is therefore convenient to separate the work integral
into contributions from the core and the envelope.
We introduce
\begin{equation}
\CD = - {1 \over 2 \omega_{\rm r}^2}
{\rm Re} \left[{\delta \rho^* \over \rho} (\Gamma_3 - 1)
\delta (\rho \varepsilon - \div \boldF) \right] 
+ {1 \over 2 \omega_{\rm r}} 
{\rm Im} \left( {\delta \rho^* \over \rho} \delta p_{\rm t} \right) \; ,
\label{eq:workintegrand}
\end{equation}
such that the full expression for the work integral can be written
\begin{equation}
\eta = { \int_V \CD \dd V \over \int_V \rho |\bolddelr |^2 \dd V}
= { \int_{\rm env} \CD \dd V \over \int_V \rho |\bolddelr |^2 \dd V}
+ { \int_{\rm core} \CD \dd V \over \int_V \rho |\bolddelr |^2 \dd V} \; .
\end{equation}
Here, 
as in the determination of $\zeta$ (Eq.~\ref{eq:zeta}), we have separated
the star into the envelope and the core, the latter being defined as the
region where $\omega < N, S_l$. 
In the envelope contribution we can replace $\CD$ by the 
function $\bar \CD_0$ for radial modes, interpolated to the frequency
of the mode considered.
Also, the denominator is closely related to the mode inertia
(Eq.~\ref{eq:inertia});
using this we replace the denominator in the first term by
$Q$ times the corresponding integral for radial modes, similarly interpolated,
where $Q$ is the scaled inertia (Eq.~\ref{eq:sclinertia}).
Thus the first term becomes $Q^{-1} \bar \eta_0$, where $\bar \eta_0$ is
the interpolated damping rate for radial modes.
In the second term we replace the denominator by an integral just over the 
core, using Eq.~(\ref{eq:zeta}).
Thus we finally obtain the full damping rate as
\begin{equation}
\eta = Q^{-1} \bar \eta_0
+ \zeta { \int_{\rm core} \CD \dd V \over
\int_{\rm core} \rho |\bolddelr |^2 \dd V} 
\equiv Q^{-1} \bar \eta_0 + \zeta \eta_{\rm core} 
\label{eq:combeta}
\end{equation}
\citep[see also][]{grosjean2014}.
As discussed below,
$\bar \eta_0$ can be obtained from solving the full non-adiabatic 
equations for radial oscillations.
However, we first consider the estimate of the second term from
the asymptotic properties of the eigenfunction.

In the core there is no contribution to $\CD$ from the turbulent pressure.
For simplicity we neglect the term in the energy generation in 
Eq.~(\ref{eq:workintegrand}), and we assume that the flux is purely radiative.
Also, in the core we can use the quasi-adiabatic approximation, so that 
the eigenfunctions are real, and no complex conjugate is needed,
and we replace $(\Gamma_3 - 1)\delta \rho/\rho$ by $\delta T/T$.
Thus we obtain, for the contribution from the core in Eq.~(\ref{eq:combeta}),
\begin{equation}
\eta_{\rm core} = {1 \over 2 \omega_{\rm r}^2 }
{ \displaystyle \int_{\rm core} {\delta T \over T} \delta (\div \boldF) \dd V
\over \int_{\rm core} \rho |\bolddelr |^2 \dd V} \; .
\label{eq:etacore}
\end{equation}
The flux is obtained as
\begin{equation}
\boldF = - {4 a \tilde c T^4 \over 3 \kappa \rho} \nabla \ln T
\equiv -\CK \nabla \ln T \; ,
\label{eq:flux}
\end{equation}
defining the conductivity $\CK$;
here $a$ is the radiation density constant, $\tilde c$ is the speed of light
and $\kappa$ is the opacity.
In $\delta (\div \boldF)$ the radial component of $\boldF$ dominates, 
since it involves the second derivative of the rapidly varying eigenfunction.
Thus, we neglect the tangential component of the flux and obtain,
using also the neglect of the nuclear term,
\begin{equation}
\delta( \div \boldF) = {\CL \over 4 \pi r^2} {\dd \over \dd r} 
\left( {\delta \CL \over \CL} \right) \; ,
\label{eq:divf}
\end{equation}
where $\CL = 4 \pi r^2 F_r$ is the luminosity, $F_r$ being the radial component
of the flux.
Here, using Eq.~(\ref{eq:flux}),
\begin{equation}
{\delta \CL \over \CL} = 2 {\xi_r \over r} + {\delta F_r \over F_r}
= 2 {\xi_r \over r} + {\delta \CK \over \CK} + {1 \over \dd \ln T/\dd r}
\delta \left({\dd \ln T \over \dd r}\right) \; , 
\end{equation}
or, expanding the last term%
\footnote{Here we use relations such as
$\delta T = T' + \xi_r \dd T/\dd r$
between the Lagrangian perturbation $\delta T$ and the Eulerian (local)
perturbation $T'$;
note that, unlike the Lagrangian perturbation, the Eulerian perturbation
commutes with the radial derivative.
\label{fn:lagrangian}}
\begin{equation}
{\delta \CL \over \CL} = 
2 {\xi_r \over r} + {\delta \CK \over \CK} - {\dd \xi_r \over \dd r}
+ {1 \over \dd \ln T/\dd r} {\dd \over \dd r} \left( {\delta T \over T}\right)
\; .
\label{eq:lumper}
\end{equation}

In the asymptotic analysis of this equation for the extreme g-mode
behaviour in the core we follow \citet{godart2009}
\citep[see also][]{dziembowski1977a}.
From the oscillation equations it may be shown that
\begin{equation}
\left| {p' \over \xi_r \dd p/\dd r} \right| = \CO(\omega/S_l) \ll 1 \; ,
\end{equation}
and hence, according to footnote~\ref{fn:lagrangian},
\begin{equation}
{\delta p \over p} \simeq \xi_r {\dd \ln p \over \dd r} \; .
\label{eq:delp}
\end{equation}
It follows that $\delta p /p$ is small compared with terms involving
derivatives of $\xi_r$ and, given the quasi-adiabatic approximation,
the same is true of $\delta \rho/\rho$ and $\delta T/T$.
From the equation of continuity,
\begin{equation}
{\delta \rho \over \rho} = - \div (\bolddelr)
= - {1 \over r^2} {\dd \over \dd r} (r^2 \xi_r) 
+ {l(l+1) \over r^2} \xi_{\rm h} \; ,
\end{equation}
it then follows that
\begin{equation}
{\dd \xi_r \over \dd r} \simeq l(l+1) {\xi_{\rm h} \over r} \; ,
\label{eq:dxirapp}
\end{equation}
where in addition we neglected $2 \xi_r/r$ compared with $\dd \xi_r / \dd r$.
In Eq.~(\ref{eq:lumper}) $2 \xi_r/r$ and $\delta \CK/\CK = \CO(\delta p/p)$
can be neglected compared with $\dd \xi_r/\dd r$.
Using also
\begin{equation}
{\delta T \over T} = \nabla_{\rm ad} {\delta p \over p} 
\label{eq:delt}
\end{equation}
and Eqs~(\ref{eq:delp}) and (\ref{eq:dxirapp}) we obtain
\begin{equation}
{\delta \CL \over \CL} \simeq {1 \over \dd \ln T/\dd r}
\nabla_{\rm ad} {\dd \xi_r \over \dd r} {\dd \ln p \over \dd r}
- {\dd \xi_r \over \dd r} \simeq 
\left( {\nabla_{\rm ad} \over \nabla} - 1 \right) l(l+1) {\xi_h \over r} \; .
\end{equation}
Using Eqs~(\ref{eq:divf}), (\ref{eq:delp}), (\ref{eq:delt}) and the
equation of hydrostatic support we can finally write the numerator
in Eq.~(\ref{eq:etacore}) as
\begin{equation}
 \int_{\rm core} {\delta T \over T} \delta (\div \boldF) \dd V
 = - \int_{\rm core} \nabla_{\rm ad} \xi_r {g \rho \over p}
\left( {\nabla_{\rm ad} \over \nabla} - 1 \right) l(l+1) 
{1 \over r} {\dd \xi_h \over \dd r} \CL \dd r \; .
\label{eq:numcore}
\end{equation}

To approximate the integrand in Eq.~(\ref{eq:numcore}) we use the
g-mode asymptotic eigenfunction in Eq.~(\ref{eq:asympsol_g}).
From Eq.~(\ref{eq:dxirapp}) we furthermore have, to leading order,
\begin{equation}
{1 \over r} {\dd \xi_{\rm h} \over \dd r} 
\simeq L^{-2} {\dd^2 \xi_r \over \dd r^2} \simeq -L^{-2} K \xi_r  \; ,
\label{eq:derxih}
\end{equation}
where again we introduced $L^2 = l(l+1)$.
Here we used that, to leading order, $\xi_r$ satisfies an equation
of the form in Eq.~(\ref{eq:basasymp});
in the g-mode cavity $K$ can be approximated by
\begin{equation}
K \simeq {L^2 \over r^2} \left( {N^2 \over \omega^2} - 1 \right) \; .
\end{equation}
Differentiating Eq.~(\ref{eq:asympsol_g}), Eq.~(\ref{eq:derxih}) also yields
\begin{equation}
\xi_r \simeq -A \rho^{-1/2} \omega_{\rm r}^{-1/2} r^{-3/2} L^{1/2} N^{-1/2}
\cos \left( \int_{r_{\rm a}}^r K^{1/2} \dd r - \phi_{\rm g}^\prime \right) \; .
\end{equation}
Thus the integrand in Eq.~(\ref{eq:numcore}) becomes
\begin{equation}
- A^2 \omega_{\rm r}^{-3} L^3 
\nabla_{\rm ad} \left( {\nabla_{\rm ad} \over \nabla} - 1 \right) 
{g \over p} N r^{-5} \CL
\cos^2 \left( \int_{r_{\rm a}}^r K^{1/2} \dd r 
- \phi_{\rm g}^\prime \right) \; .
\end{equation}
The integral in the denominator of Eq.~(\ref{eq:etacore}) is evaluated
essentially as in Eq.~(\ref{eq:g_inertia}),
yielding
\begin{equation}
\int_{\rm core} \rho |\bolddelr |^2 \dd V
\simeq 4 \pi A^2 \omega^{-3} L \int_0^{r_{\rm b}} N
\sin^2 \left( {L \over \omega} \int_0^r N {\dd r' \over r'} 
- \phi_{\rm g}^\prime \right) {\dd r \over r} \; .
\label{eq:denomcore}
\end{equation}
Thus we finally obtain, replacing $\cos^2$ and $\sin^2$ by their
average values $1/2$,
\begin{equation}
\eta_{\rm core} \simeq {L^2 \over 8 \pi \omega_{\rm r}^2}
{\displaystyle \int_{\rm core}
\nabla_{\rm ad} \left( {\nabla_{\rm ad} \over \nabla} - 1 \right) 
{N g \over p r^4} \CL \dd r/r \over
\int_{\rm core} N \dd r /r } \; .
\label{eq:coredamp}
\end{equation}

\begin{figure}
\centering
\begin{minipage}{\linewidth}
\centering
\includegraphics[width=9cm]{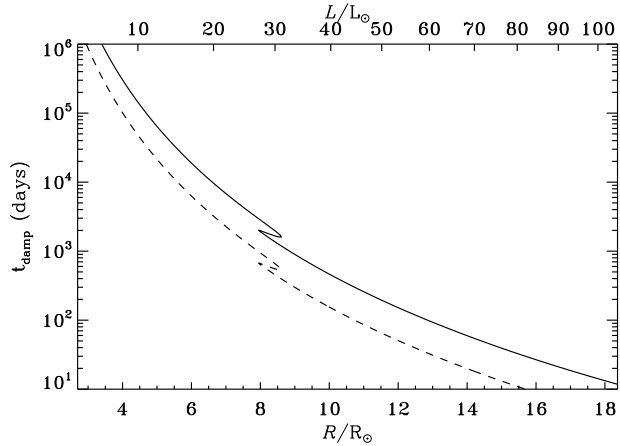}
\end{minipage}
\caption{Mode damping time $t_{\rm damp} = 1/\eta_{\rm core}$ corresponding 
to the asymptotic core damping rate (Eq.~\ref{eq:coredamp}),
evaluated at the frequency $\nu_{\rm max}$ of 
maximum oscillation power estimated as in Eq.~(\ref{eq:numaxt}).
The results were obtained for
a $1 \Msun$ evolution sequence
and are shown 
against surface radius (lower abscissa) and luminosity (upper abscissa)
in solar units, for $l = 1$ (solid) and $l = 2$ (dashed).
}
\label{fig:coredamp}
\end{figure}

As a star evolves up the red-giant branch, increasing the luminosity and
the mass of the core, the core gravitational acceleration and buoyancy
frequency increase.
According to Eq.~(\ref{eq:coredamp}) these effects all contribute to 
increasing the core damping rate.
This is illustrated in Fig.~\ref{fig:coredamp}, showing the corresponding
damping time in a $1 \Msun$ evolution sequence.
On the low red-giant branch the core damping is small and unlikely to affect
the observed properties of the modes, while around the bump, clearly 
reflected in the figure, the core damping rate has increased
to a level where the corresponding damping time is comparable with or
smaller than the duration of the nominal {\it Kepler} mission,
such that significant effects can be expected
(see also Fig.~\ref{fig:grosjean} below).

Returning to the near-surface contributions to the damping of the modes,
a full non-adiabatic treatment of the oscillations is required,
taking into account the pulsation-induced perturbations to the
convective properties.
These issues were discussed in detail by \citet{houdek2015}.
Time-dependent generalizations of mixing-length theory were developed
by \citet{unno1967} and \citet[based on earlier work in 1965]{gough1977},
while \citet{xiong1977} used a Reynolds-stress model to treat
convection in pulsating stars.

Unno's theory was further developed by \citet{gabriel1996} and
\citet{grigahcene2005}.
With a suitable adjustment of parameters it provides a relatively reasonable
fit to the detailed observations of solar oscillation linewidths
\citep{dupret2006}.
This was subsequently used in the analysis of red-giant mode
energetics by \citet{dupret2009} and \citet{grosjean2014}.

Gough's theory was generalized to included nonlocal effects by
\citet{balmforth1992a}, based on an analysis by \citet{gough1977b}.
Specifically, local mixing-length theory implicitly makes the assumption
that the relevant convective scales are much smaller than the scale of
variation of stellar structure, whereas in fact the typical 
mixing-length scale is of order a pressure scale height.
The nonlocal analysis involves an average over the extent of the motion of
convective eddies and over the ensemble of eddies at any given location;
this also gives rise to limited convective overshoot and circumvents 
mathematical problems that occur in a fully local convection formulation 
when consistently including turbulent pressure in the equation of
hydrostatic support.
The theory has so far only been developed for radial oscillations;
however, given that the relevant effects are concentrated 
in the superficial layers of the stars, the results are expected
to be representative also for non-radial oscillations of low degree,
at least in the frequency range of solar-like oscillations.
With suitable, and plausible, choice of parameters characterizing the
nonlocality this formulation results in oscillation line widths in
reasonable agreement with solar observations
\citep[e.g.,][]{balmforth1992a, houdek2001, chaplin2005}.

\begin{figure}
\centering
\begin{minipage}{\linewidth}
\centering
\includegraphics[width=9cm]{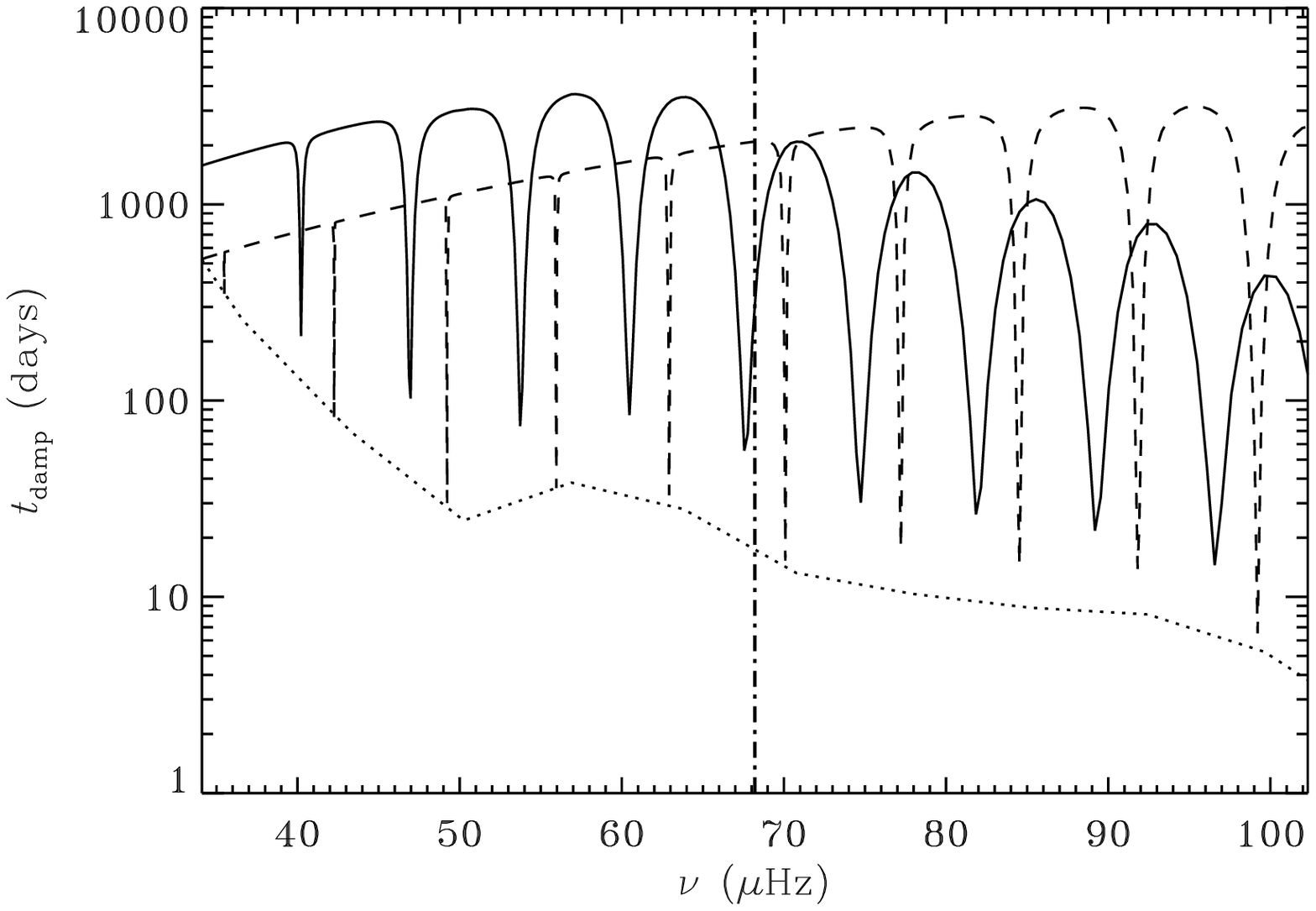}
\end{minipage}
\caption{Mode damping times $t_{\rm damp} = 1/\eta$ in the
$7 \Rsun$ red-giant model from the $1 \Msun$ 
sequence illustrated in Fig.~\ref{fig:coredamp}.
The dotted curve shows radial-mode damping times computed
using a nonlocal time-dependent convection formulation
\citep{houdek1999},
and the solid ($l = 1$) and dashed ($l = 2$)
curves show the results of combining
the radial-mode damping rates with the asymptotic estimate
(Eq.~\ref{eq:coredamp}), as in Eq.~(\ref{eq:combeta}).
The vertical dot-dashed line marks an estimate of the frequency
$\nu_{\rm max}$ of maximum oscillation power (cf. Eq.~\ref{eq:numaxt}).
(Radial-mode damping rate courtesy G\"unter Houdek.)
}
\label{fig:aslifetimes}
\end{figure}

Given the fully non-adiabatic results for radial modes,
including the effects of turbulent pressure,
we can estimate the damping times of the mixed modes
by combining the radial-mode damping rate%
\footnote{We note that the properties of radial modes,
including the damping rate, can be computed as a continuous function
of $\omega_{\rm r}$ by imposing a no-work boundary condition 
at the bottom of the computational domain \citep[e.g.,][]{jcd_sf1983rad},
thus avoiding the need for interpolation.}
with the asymptotic treatment of the damping in the core
(cf. Eq.~\ref{eq:combeta}).
An example is shown in Fig.~\ref{fig:aslifetimes},
for a fairly low-luminosity $1 \Msun$ red giant.
Here we obtained $\bar \eta_0$ using the \citet{gough1977} 
and \citet{balmforth1992a} nonlocal treatment of the perturbations
to the convective properties, as described by \citet{houdek1999}.
We note that the acoustic-mode lifetimes are a factor 2 -- 3 larger
than those found by \citet{grosjean2014} and hence in substantially
better agreement with observations
\citep[][Handberg et al, submitted]{huber2010, corsaro2015}.
Comparison with Fig.~\ref{fig:coredamp} indicates that in this case
the near-surface damping dominates near $\nu = \nu_{\rm max}$,
even for the most g-dominated modes.

\subsubsection{Mode excitation}
\label{sec:modexcit}

Stochastically excited mode amplitudes were first estimated by
\citet{goldreich1977}.
Their results were used by \citet{jcd_sf1983stel} in a first estimate of
the expected mode amplitudes of stochastically excited oscillations
across the HR diagram.
This was later summarized by \citet{kjeldsen1995} in
a widely used relation according to which the mode amplitudes
roughly scale as $\CL/M$.
In particular, it was already then clear that evolved stars were expected
to have higher oscillation amplitudes, as has certainly been 
confirmed observationally.

The stochastic energy input from convection is driven by turbulent
Reynolds stresses and entropy fluctuations, their relative importance
depending on the detailed assumptions made in the calculation
\citep[e.g.,][]{balmforth1992b, goldreich1994, samadi2001, samadi2003,
chaplin2005}.
However, regardless of these details
the dominant contributions come from the near-surface layers where convection
is most vigorous.
Here the local properties of the oscillations are essentially independent
of the degree of the mode, at least for the low-degree modes that are relevant
in distant stars, and the rate of energy input is 
consequently a function of frequency but not degree.
It may be shown \citep[e.g.,][]{chaplin2005} that the resulting
mean square amplitude can be expressed as
\begin{equation}
\langle A^2\rangle \simeq {1 \over E \eta}
{\CF(\omega) \over E} \; ,
\label{eq:stochamp}
\end{equation}
where $\CF$ is the rate of energy input.
The details of this expression, including the precise form of 
$\CF(\omega)$, depend on the observed quantity represented by $A$
(see also Section~\ref{sec:introobs}).
However, since the ratio between the physical amplitudes of
different observable oscillation properties is generally a function of
frequency this does not change the form of the equation.
From Eq.~(\ref{eq:stochamp}) the rate of energy input can be determined 
from $\eta \langle A^2 \rangle$, with $\eta$ 
determined from the observed line width and the mode inertia obtained
from computed eigenfunctions.
In this manner \citet{stein2001} showed a remarkable agreement between 
the predictions from three-dimensional simulations of convection
and radial-velocity observations of solar oscillations.
Also, \citet{jacoutot2008} used such a comparison to constrain the detailed
properties of their simulations of solar convection.
A similar analysis based on the results from {\it Kepler}, including
for red giants, would clearly be very interesting.
\citet{samadi2007} compared the rate of energy input obtained from
hydrodynamical simulations with the results obtained for mixing-length
based models.
The analysis was extended to red giants by \citet{samadi2012}, 
comparing the results with CoRoT observations analysed by 
\citet{baudin2011}.
They noted the importance of the proper conversion of the predicted
velocity amplitudes to the photometric data, such as obtained by
CoRoT.

As discussed above, the properties of $\eta$ can be understood in terms
of the work integral, Eq.~(\ref{eq:workintg}).
Here (and in the expression for the contribution $\eta_{\rm t}$ from 
the turbulent pressure, Eq.~\ref{eq:workturbp})
the denominator essentially corresponds to the normalized mode inertia $E$,
apart from the normalization with the surface displacement
(cf. Eq.~\ref{eq:inertia}).
Near the surface the integrands in the numerators of 
Eqs~(\ref{eq:workintg}) and (\ref{eq:workturbp}) depend only on
frequency.
Thus if the contribution from the core to the work integral can be
neglected, $E \eta$ is just a function of frequency.
Consequently, $\langle A^2 \rangle \propto E^{-1}$,
and the amplitude of a general non-radial mode, relative to a radial mode
of the same frequency, scales as
\begin{equation}
\langle A^2 \rangle = Q^{-1} \langle \bar A_0^2 \rangle
\label{eq:qsclampl}
\end{equation}
(cf. Eq.~\ref{eq:sclinertia}), where $\langle \bar A_0^2 \rangle$ is
the radial-mode amplitude, interpolated to the frequency of the mode
considered.
However, in the mixed modes in red giants the contribution $\eta_{\rm core}$
from the damping in the core can be very significant and must be taken 
into account.
Using Eqs~(\ref{eq:sclinertia}) and (\ref{eq:combeta}) we obtain
\begin{eqnarray}
\langle A^2\rangle &\simeq& {1 \over Q \bar E_0(\omega)
[ Q^{-1} \bar \eta_0 + \zeta \eta_{\rm core}]}
{\CF(\omega) \over Q \bar E_0(\omega) } \nonumber \\
&\simeq& {1 \over 1 + Q \zeta \eta_{\rm core}/\bar \eta_0}
Q^{-1} {1 \over \bar E_0(\omega) \bar \eta_0}
{\CF(\omega) \over \bar E_0(\omega) } \nonumber \\
&\simeq& {1 \over 1 + (Q - 1) \eta_{\rm core}/\bar \eta_0}
Q^{-1} \langle \bar A_0^2\rangle \; ;
\label{eq:sclstochamp}
\end{eqnarray}
here, despite the {\it caveat} in footnote~\ref{fn:qzeta},
we used Eq.~(\ref{eq:qzeta}) to replace $Q \zeta$ by $Q-1$, with sufficient
accuracy for the present purpose.

In terms of the observed oscillation power spectrum $\langle A^2 \rangle$
corresponds to the area under a Lorentzian peak (cf. Eq.~\ref{eq:stochpow}).
Of more relevance to the interpretation of the observations is the
{\it peak height} $H$, related to $\langle A^2 \rangle$ by
\begin{equation}
H = {2 \over \eta} \langle A^2 \rangle 
\label{eq:hampl}
\end{equation}
\citep[see][and Section~\ref{sect:indf}]{baudin2005, chaplin2005}.
Relating as in Eq.~(\ref{eq:sclstochamp}) the peak height to the
peak height $\bar H_0$ of radial modes, interpolated to the relevant
frequency, we obtain
\begin{equation}
H \simeq {1 \over [1 + (Q - 1) \eta_{\rm core}/\bar \eta_0]^2} \bar H_0 \; .
\label{eq:hscale}
\end{equation}
In particular, in the lower part of the red-giant branch the
core damping is small compared with the envelope contribution
(cf. Fig.~\ref{fig:coredamp}) and hence can be neglected;
consequently the peak heights of all
mixed modes are predicted to be comparable to the peak height of the
adjacent radial modes, in clear contradiction to the observations.

In an important breakthrough in the interpretation of the observations of
red-giant oscillations it was noted by \citet{dupret2009} that the origin of
this discrepancy was the assumption that all peaks have a Lorentzian 
profile, leading to Eq.~(\ref{eq:hampl}). 
In fact, this is only strictly true if the oscillations are observed
for an infinite period.
For observations over a finite period $T_{\rm obs}$ the line profile is
a combination of a Lorentzian and a ${\rm sinc}^2$ function
(cf. footnote~\ref{fn:sinc}).
The broader peaks result in a reduction in $H$,
at the given $\langle A^2 \rangle$, and a behaviour that is
qualitatively consistent with the observations.
This analysis was extended, including a more detailed comparison with the
observations, by \citet{grosjean2014}.

\citet{fletcher2006} showed that the transition between unresolved
and fully resolved peaks can be approximated by replacing Eq.~(\ref{eq:hampl})
by
\begin{equation}
H = {2 \over \eta + 2/T_{\rm obs}} \langle A^2 \rangle 
\label{eq:thampl}
\end{equation}
As a result, Eq.~(\ref{eq:hscale}) is replaced by
\begin{equation}
H \simeq {1 + 2 \bar t_{{\rm damp},0}/T_{\rm obs}
\over [1 + (Q - 1) \eta_{\rm core}/\bar \eta_0]
[1 + (Q - 1) \eta_{\rm core}/\bar \eta_0+2 Q \bar t_{{\rm damp},0}/T_{\rm obs}]}
\bar H_0 \; ,
\label{eq:thscale}
\end{equation}
where we introduced the interpolated radial-mode damping time
$\bar t_{{\rm damp},0} = 1/\bar \eta_0$ and peak height $\bar H_0$.
For observations extending over 100 -- 1000 days, as was the case for
CoRoT and the nominal {\it Kepler} mission, typically
$\bar t_{{\rm damp},0} \ll T_{\rm obs}$ and the correction in the
numerator in Eq.~(\ref{eq:thscale}) can be neglected.
However, the same is not true of the corresponding term in
the denominator for the more g-dominated mixed modes, with $Q$ much
larger than 1 (cf. Fig.~\ref{fig:rginertia}), whose heights are 
therefore reduced.
A further reduction can clearly come from the core damping,
represented by $\eta_{\rm core}$.
It seems probably that this reduction of peak height is the reason that
\citet{frandsen2002} only identified the radial modes in 30-day observations
of $\xi$ Hydrae,
since even the acoustically dominated dipolar modes have a somewhat higher
inertia than the neighbouring radial modes.

\begin{figure}
\centering
\begin{minipage}{\linewidth}
\centering
\includegraphics[width=12cm]{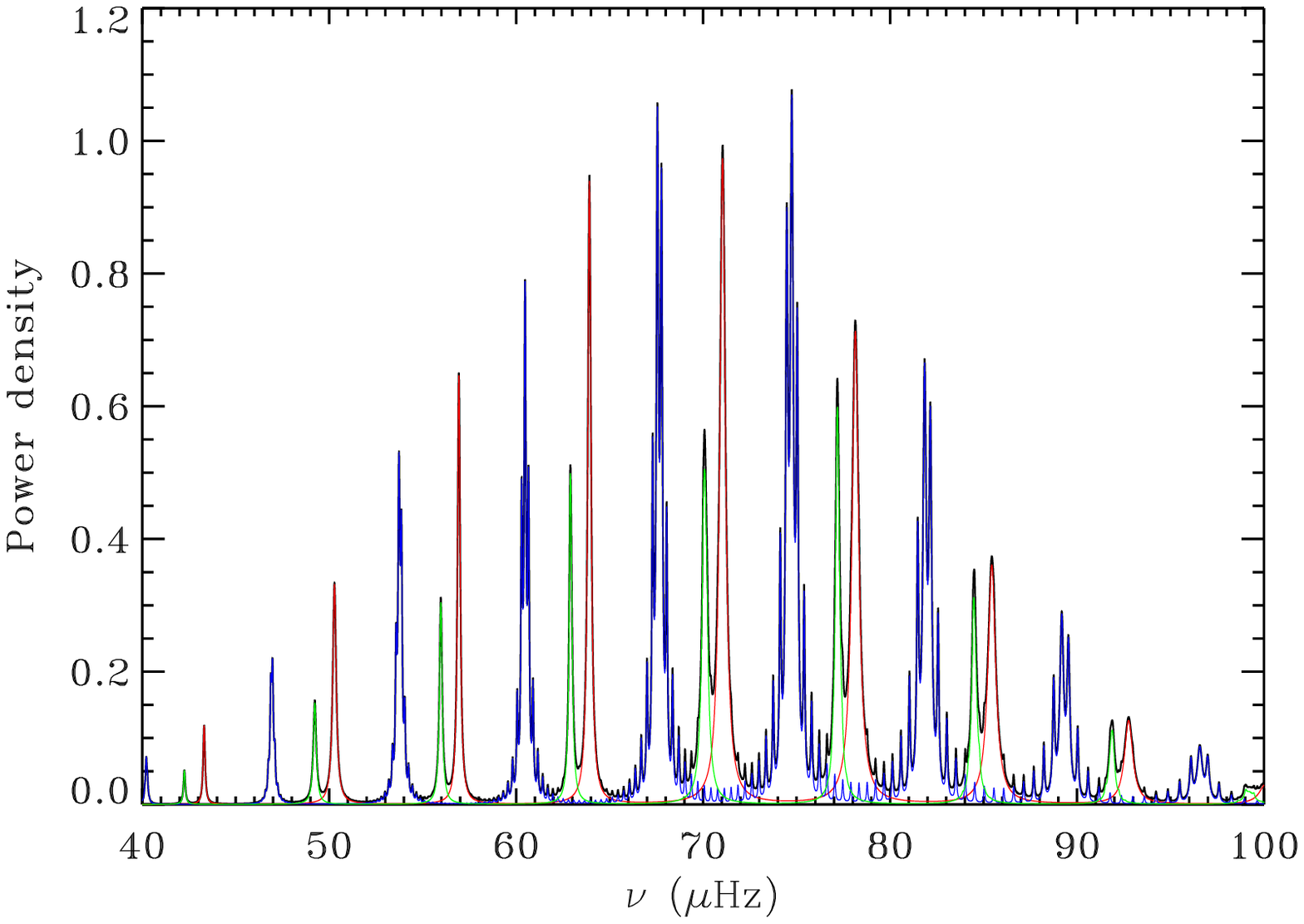}
\includegraphics[width=12cm]{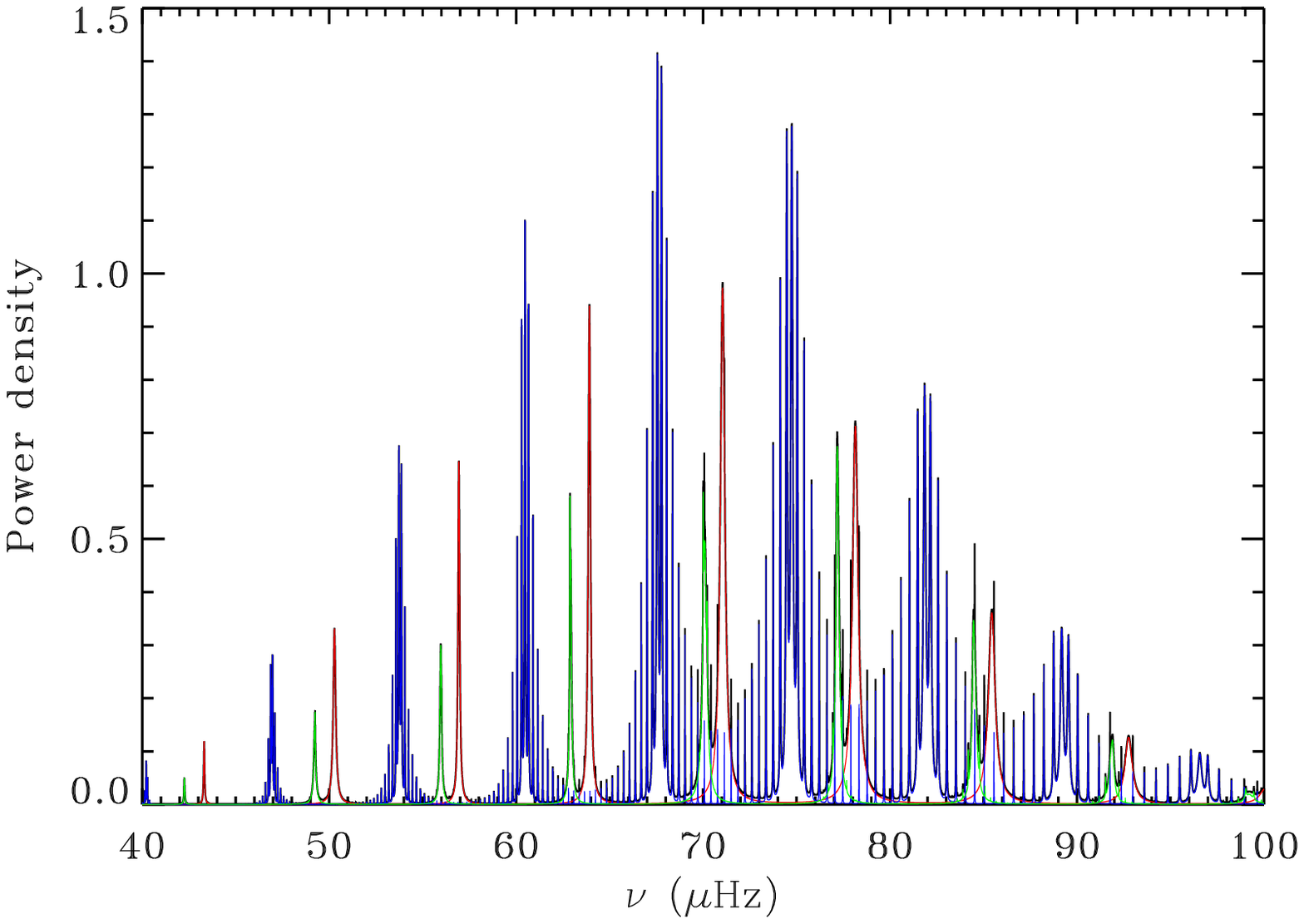}
\end{minipage}
\caption{Simulated power spectra for the model illustrated in
Fig.~\ref{fig:aslifetimes},
for observing times of 100 days (top) and 1000 days (bottom).
The combined power density is shown with the black curve, while the
red, blue and green curves show the contributions from modes with
degree $l = 0$, $1$ and $2$.
}
\label{fig:simpower}
\end{figure}

To illustrate these effects Fig.~\ref{fig:simpower} shows a simplified modelling
of power density spectra, for the $1 \Msun$, $7 \Rsun$ model illustrated
in Fig.~\ref{fig:aslifetimes}, based on the damping times shown there.
Here we have approximated $\bar H_0$ by a Gaussian centred on the estimated
$\nu_{\rm max} \simeq 68 \muHz$ and with a maximum of 1.
The total power is calculated as the sum over the modes of degree $l = 0 - 2$,
each mode represented by a Lorentzian with the width corrected for
the finite observing time, as implicit in Eq.~(\ref{eq:thampl}), and a height 
given by Eq.~(\ref{eq:thscale}).
In the top panel, for $T_{\rm obs} = 100 \, {\rm d}$, most mixed modes
have a strongly reduced height, owing to the relatively short observing time,
although some mixed dipolar modes may in principle be visible.
At high frequency the peaks are very substantially broadened by the
short mode lifetimes (cf. Fig.~\ref{fig:aslifetimes}).
With  $T_{\rm obs} = 1000 \, {\rm d}$, shown in the lower panel,
most dipolar mixed modes are excited to substantial heights, 
and hence one might expect to detect an almost complete dipolar spectrum,
at least at moderate and high frequency.
However, at low frequency the dipolar modes are strongly suppressed by
the increase in the damping rate 
(note that, according to Eq.~(\ref{eq:coredamp}),
$\eta_{\rm core} \propto \nu^{-2}$) and in particular the small damping
rate compared with $2/T_{\rm obs}$.
For $l = 2$ the suppression of the peak height, except for the most
acoustic modes, is far higher owing to the larger values of $Q$,
and very few if any mixed modes are predicted to be visible.

\begin{figure}
\centering
\begin{minipage}{\linewidth}
\centering
\includegraphics[width=9cm]{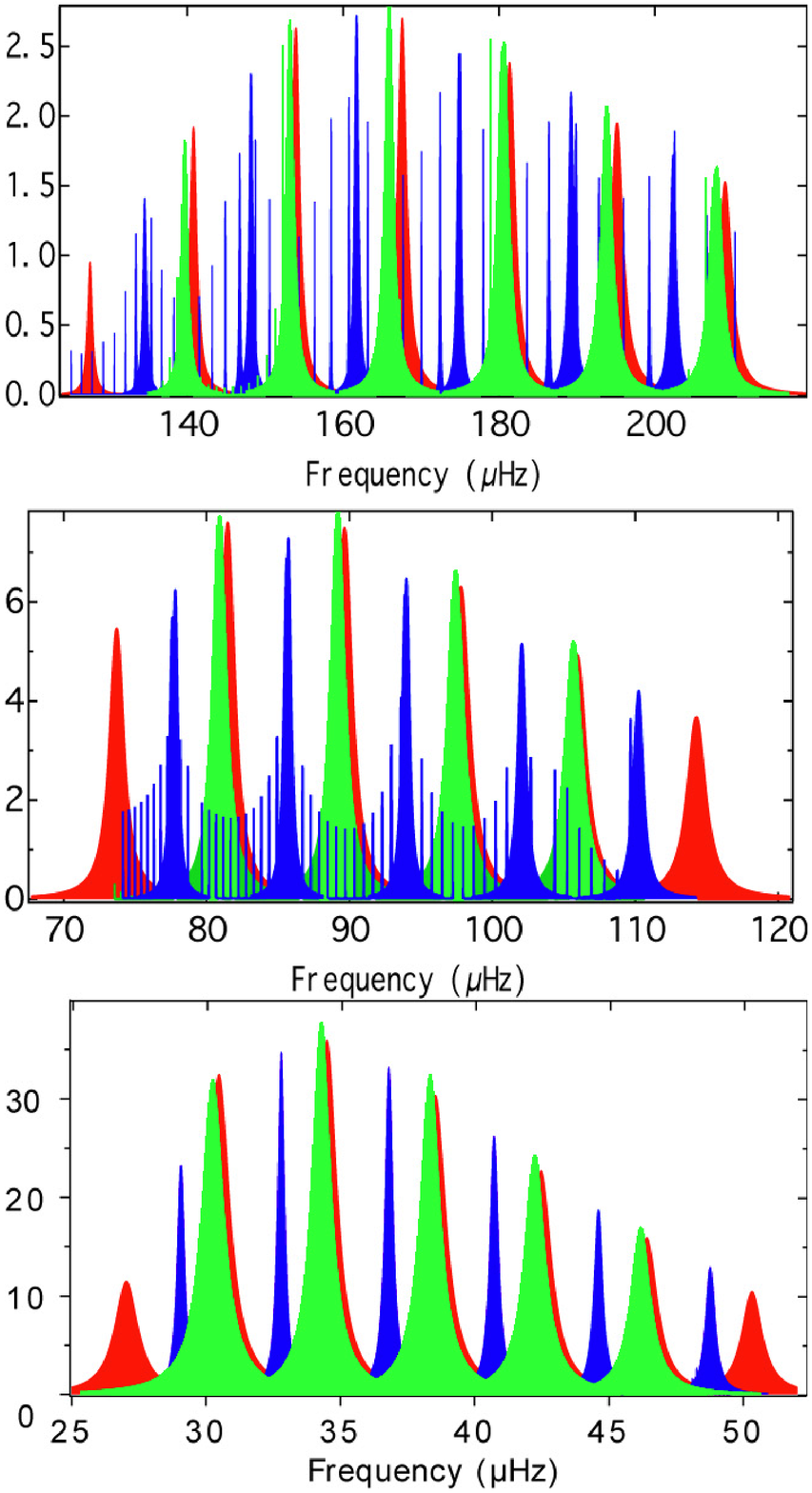}
\end{minipage}
\caption{Simulated power density, in ${\rm m^2 \, s^{-2} \, \muHz^{-1}}$, 
for models in a $1.5 \Msun$ evolution sequence with radii
$5.2 \Rsun$ (top), $7.3 \Rsun$ (middle) and $11.9 \Rsun$ (bottom).
{\orange  Image reproduced with permission from \citet{grosjean2014}, copyright by ESO.}
}
\label{fig:grosjean}
\end{figure}

The strong increase in the damping rates 
as the star moves up the red-giant branch (cf. Fig.~\ref{fig:coredamp}) 
has a dramatic effect on the predicted power density spectra;
this is illustrated in Fig.~\ref{fig:grosjean} taken
from \citet{grosjean2014}, for a $1.5 \Msun$ evolution sequence
here assuming $T_{\rm obs} = 360 \, {\rm d}$.%
\footnote{Note that, as already mentioned, the computed radial-mode
lifetime is substantially shorter than used in Fig.~\ref{fig:simpower}, 
leading to broader peaks of the acoustically-dominated modes.}
For the $5.2 \Rsun$ model all dipolar mixed modes are visible. 
When $R$ reaches $7.3 \Rsun$ the heights of the dipolar mixed modes are
substantially reduced, although many modes would still be expected to be
visible.
Finally, at a radius of $11.9 \Rsun$, just below the bump, the mixed modes
are no longer visible. 
Although further investigations along these lines, matching also the
observed linewidths, are needed, this clearly indicates a serious
limitation on the diagnostic possibilities for very evolved stars.

At even higher luminosity, the damping in the g-mode cavity becomes
so strong that gravity waves are damped before being reflected from the centre, essentially eliminating the g-dominated mixed modes.
In this case the remaining p-dominated oscillation spectrum 
can be computed for just the envelope model, applying boundary conditions
at the edge of the g-mode cavity which select those waves that propagate
towards the centre
\citep{dziembowski1977a, osaki1977, vanhoolst1998, dziembowski2001, 
dziembowski2012}.
It was noted by \citet{dziembowski2012} that towards the tip of the
red-giant branch the resulting loss of wave energy to the core becomes
essentially negligible.

As shown by Eqs~(\ref{eq:sclstochamp}) and (\ref{eq:thscale})
the mode amplitudes and peak heights of the
mixed modes depend strongly on the mode inertia, in units of the radial-mode
inertia.
It was pointed out by \citet{benomar2014} that this provides an opportunity
to use the observed amplitudes or peak heights as diagnostics of the stellar
interior, supplementing the information obtained from the frequencies.
In particular, given the sensitivity of $\zeta$, and equivalently $Q$, on the
coupling strength $q$ (cf. Eq.~\ref{eq:zeta_asymp})
this may provide information about the
evanescent region between the acoustic- and gravity-mode cavities.
These possibilities should be further explored, given the extended data
available from {\it Kepler}.

The original amplitude scaling relation with $\CL/M$ proposed by 
\citet{kjeldsen1995}, and the analysis carried out in this section,
strictly speaking only apply to the amplitudes observed in radial
velocity, which are directly related to the mode energy;
furthermore, for comparison with observations, 
the predicted amplitudes should be referred
to the effective height in the atmosphere where the radial-velocity
observations are carried out, requiring modelling of the oscillation
eigenfunctions in the atmosphere.
We also note that, as discussed briefly in the opening paragraph of
Section~\ref{sect:indf}, the observed amplitudes depend on the mode
visibility, determined by the geometric cancellation across the stellar disk,
which has not been taken into account here.
Predicting photometric amplitudes requires the ratio between the intensity
variations, in the appropriate wavelength band, and the velocity amplitudes
\citep[e.g.,][]{houdek2006, houdek2009, samadi2013, grosjean2014} 
which, as noted by \citet{kjeldsen1995}, depends on the effective temperature.
Analyses of observed amplitudes have typically started from the $\CL/M$
scaling but introduced different exponents and included the dependence on
$T_{\rm eff}$ \citep[e.g.][]{huber2011, corsaro2013}.
However, there is still a very substantial potential for more detailed 
comparisons between observed and modelled amplitudes and linewidths.

\subsection{Rotation}
\label{sec:rotation}

A striking result of the early analysis of red-giant observations from
{\it Kepler} was the detection by \citet{beck2012} of rotational
splitting of the observed frequencies, leading to a first inference of
the internal rotation of a red-giant star.
Extensive results are now available for both red-giant and clump stars
\citep[e.g.,][]{mosser2012rot, deheuvels2015, vrard2016}.
Strikingly, the inferred core rotation of
red giants is far slower than would be expected from models of 
angular-momentum evolution
\citep[e.g.,][see Section~\ref{sect:resrot}]
{eggenberger2012, marques2013, cantiello2014}.
Here we discuss the effects of rotation on the stellar oscillation 
frequencies.

For evolved stars it is generally assumed 
that rotation is so slow that second-order effects of rotation,
including the centrifugal acceleration, can be neglected.
Then in particular the hydrostatic structure of the star is not affected.
The oscillation frequencies are affected by advection of the pattern of
waves propagating in the azimuthal direction, 
such that the frequencies of
prograde waves (travelling in the direction of rotation) increase
and frequencies of retrograde waves decrease.
In addition, the oscillations are affected locally by the Coriolis force in 
a frame rotating with the star.
The result is the first-order rotational splitting, with the frequencies
given by
\begin{equation}
\omega_{nlm} = \omega_{nl0} + m \delta \omega_{nlm} \; 
\label{eq:rotsplit}
\end{equation}
(see also Section~\ref{sect:rot}).

We cannot assume that stars rotate as a solid body.
Hence the rotational splitting measures an average 
over the internal rotation rate,
determined by the properties of the mode.
For simplicity we only consider the case of the so-called
shellular rotation where the angular frequency
$\Omega = \Omega(r)$ only depends on the distance to the centre.
Then $\delta \omega_{nlm}$ does not depend on $m$, and we
can express it as
\begin{equation}
\delta \omega_{nl} = \int_0^R K_{nl}(r) \Omega(r) \dd r \; ,
\end{equation}
where the {\it rotational kernel} $K_{nl}$ is given by
\begin{equation}
K_{nl}(r) = {\left[ \xi_r^2 + l(l+1) \xi_{\rm h}^2 
- 2 \xi_r \xi_{\rm h} - \xi_{\rm h}^2 \right]
r^2 \rho \over \int_0^R \left[ \xi_r^2  + l(l+1) \xi_{\rm h}^2 \right] 
r^2 \rho \dd r} \; .
\label{eq:rotker}
\end{equation}
Here the first two terms provide a weighted average of $\Omega$ and
correspond to the advection, while the last two terms arise from the
Coriolis force.
It is common also to consider
\begin{equation}
\beta_{nl} = \int_0^R K_{nl}(r) \dd r \; ,
\end{equation}
such that for uniform rotation $\delta \omega_{nl} = \beta_{nl} \Omega$.
For acoustic modes with $|\xi_r| \gg |\xi_{\rm h}|$ we can neglect the last
two terms in the numerator of $K_{nl}$, so that $\beta_{nl} \simeq 1$.
For general rotation we obtain in this case
\begin{equation}
\delta \omega_{nl} \simeq \langle \Omega \rangle \; ,
\end{equation}
i.e., an average angular velocity weighted by the local contribution to
the inertia.
For high-order g modes, on the other hand, with $|\xi_{\rm h}| \gg |\xi_r|$
we can neglect the terms in $\xi_r$ to obtain
\begin{equation}
\beta_{nl} \simeq 1 - {1 \over l(l+1) } \; ;
\label{eq:beta_g}
\end{equation}
in particular, for dipolar modes, with $l = 1$, we obtain 
$\beta_{n1} \simeq 1/2$.

\begin{figure}
\centering
\begin{minipage}{\linewidth}
\centering
\includegraphics[width=9cm]{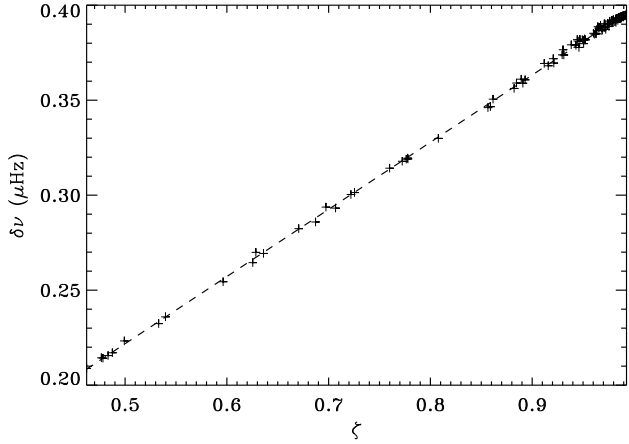}
\end{minipage}
\caption{Computed rotational splittings as functions of 
$\zeta$ for dipole modes, with $l = 1$, in the red-giant model $\modrg$ in Fig.~\ref{fig:chenhr}
$(1.3 \Msun \, , \; 6.2 \Rsun)$.
An angular velocity was imposed with
$\Omega = 4.98 \times 10^{-6} \,{\rm s}^{-1}$ in the core,
corresponding to a rotation period of 15 days,
$\Omega = 3.02 \times 10^{-7} \,{\rm s}^{-1}$ in the envelope
and a narrow transition at the hydrogen-burning shell, with $r = 0.09 R$.
All modes with frequency above $0.4 \nu_{\rm ac}$ were included.
We note the expected linear dependence (cf. Eq.~\ref{eq:zetasplit}).
The dashed line shows a uniformly weighted linear least-squares fit
to the results.
}
\label{fig:rotsplit}
\end{figure}

\def\nl{}
To analyse the rotational splitting for red giants, we follow 
\citet{goupil2013} and use the asymptotic description of the
mode inertia discussed in Section~\ref{sec:ascenrg}.
For simplicity, we suppress the subscript `$nl$' in the following.
Neglecting the term in $\xi_r \xi_{\rm h}$ in Eq.~(\ref{eq:rotker})
and using Eqs~(\ref{eq:as_zeta_I}) and (\ref{eq:beta_g}),
$\beta$ may be approximated by
\begin{equation}
\beta \simeq {[1 - (l(l+1))^{-1}] I_{\rm g} + I_{\rm p} \over
I_{\rm p} + I_{\rm g} } = [1 - (l(l+1))^{-1}] \zeta + (1 - \zeta)
= \beta_{\rm core} + \beta_{\rm env} \; ;
\end{equation}
here the contributions to $\beta$ from the core and the envelope are 
\begin{equation}
\beta_{\rm core} = \int_{\rm core} K_{\nl}(r) \dd r
\simeq [1 - (l(l+1))^{-1}] \zeta
\end{equation}
and
\begin{equation}
\beta_{\rm env} = \int_{\rm env} K_{\nl}(r) \dd r 
\simeq 1 - \zeta \; .
\end{equation}
We furthermore introduce the average
kernel-weighted core and envelope angular velocities:
\begin{equation}
\langle \Omega \rangle_{\rm core} =
{\int_{\rm core} \Omega(r) K_{\nl}(r) \dd r \over
\int_{\rm core} K_{\nl}(r) \dd r }
\end{equation}
and
\begin{equation}
\langle \Omega \rangle_{\rm env} =
{\int_{\rm env} \Omega(r) K_{\nl}(r) \dd r \over
\int_{\rm env} K_{\nl}(r) \dd r } \; .
\end{equation}
These averages formally depend on the mode;
however, the dependence on at least the mode order is weak,
given the asymptotic description of the modes and
assuming that $\Omega(r)$ varies smoothly with $r$ on the scale of the
eigenfunctions.
With these definitions we obtain
\begin{equation}
\delta\omega = \beta_{\rm core} \langle \Omega \rangle_{\rm core} +
\beta_{\rm env} \langle \Omega \rangle_{\rm env} \simeq
\langle \Omega \rangle_{\rm core} \{ [ 1 - (l(l+1))^{-1}] \zeta 
+ (1 - \zeta) \CR \} \; ,
\end{equation}
where
$\CR = \langle \Omega \rangle_{\rm env}/\langle \Omega \rangle_{\rm core}$.
Writing the splitting $\delta \nu = \delta \omega / 2 \pi$ 
in terms of cyclic frequency and specializing to $l = 1$ we finally obtain
\begin{equation}
\delta \nu = \delta \nu_{\rm max} [(1 - 2 \CR) \zeta + 2 \CR] 
\label{eq:zetasplit}
\end{equation}
\citep{goupil2013},
where $\delta \nu_{\rm max} = \langle \Omega \rangle_{\rm core}/4 \pi$,
assuming that the core is rotating substantially faster that the envelope,
so that the maximum splitting occurs for the g-dominated modes with
$\zeta \simeq 1$.
Thus we find that the splitting varies linearly with $\zeta$.
This is illustrated for a stellar model in Fig.~\ref{fig:rotsplit};
from the coefficients of the linear fit the core and envelope rotation
can be determined with very good precision:
the relative errors in the inferred  $\langle \Omega \rangle_{\rm core}$
and  $\langle \Omega \rangle_{\rm env}$ are 0.6 and 6\,\%, respectively.

In analyses of observed power spectra \citet{mosser2012core, mosser2012rot} 
assumed that the envelope contribution to the
splitting could be neglected, corresponding to taking $\CR = 0$, and
applied an empirically based approximation to the splitting, which
can be expressed as $\delta \nu = \zeta_{\rm Mosser} \delta \nu_{\rm max}$,
where%
\begin{equation}
\zeta_{\rm Mosser} = 1 - {\lambda \over 
\displaystyle 1 +
\left( {\nu - \nu_{n_{\rm p}\,l}^{\rm (p)} \over \gamma \Delta \nu} 
\right)^2} \; ,
\label{eq:zetamos}
\end{equation}
where $\lambda$ and $\gamma$ are empirically determined parameters.%
\footnote{\citet{mosser2012rot} denotes this $\CR_{n_{\rm p}}(\nu)$.}
An expression of this form can in fact be obtained from the full 
asymptotic expression for $\zeta$, Eq. (\ref{eq:zeta_asymp}),
by expanding the inverse,
assuming that $\zeta$ is close to 1, and expanding $\sin^2$
on the assumption that $\nu$ is close to $\nu_{n_{\rm p}\,l}^{\rm (p)}$.
A similar functional form, based on a Lorentzian departure from the asymptotic
period spacing, was used by \citet{stello2012} in the analysis of computed
period spacings.

\begin{figure}
\centering
\begin{minipage}{\linewidth}
\centering
\includegraphics[width=9cm]{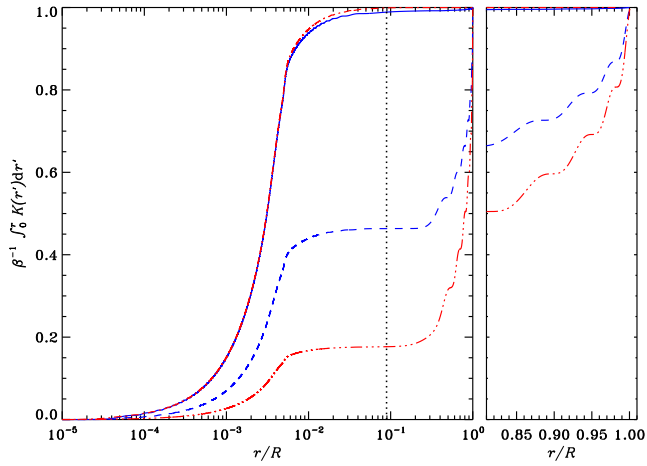}
\end{minipage}
\caption{Normalized partial kernel integrals for the model $\modrg$ in Fig.~\ref{fig:chenhr} $(1.3 \Msun \, , \; 6.2 \Rsun)$.
The modes illustrated are $l = 1, \nu = 79.09 \muHz$ (g-dominated; blue, solid),
$l = 1, \nu = 84.33 \muHz$ (p-dominated, blue, dashed),
$l = 2, \nu = 82.19 \muHz$ (g-dominated, red, dot-dashed),
$l = 2, \nu = 87.42 \muHz$ (p-dominated, red, triple-dot-dashed).
The right-hand panel shows details near the surface.
The vertical dotted line marks the base of the convective envelope.
}
\label{fig:introtker}
\end{figure}

The relative contribution to the rotational splittings
of the different parts of the star
can also be illustrated by the partial integrals of the kernels
\citep[see][supplementary material]{beck2012}.
These are shown in Fig.~\ref{fig:introtker} for selected 
p- and g-dominated modes with $l = 1$ and $2$
in a $1.3 \Msun$ red-giant model, normalized such that the surface values
are one.
This shows that even for the p-dominated modes a substantial contribution
to the splitting comes from the radiative core, particularly for $l = 1$.
This is related to the fact that $\zeta$ even for these modes
is substantially bigger than 0.
Given that generally $\CR \ll 1$ for red giants,
this makes it difficult to determine the envelope rotation rate;
observations of rotational splittings for modes with $l = 2$ or, even 
better, $l = 3$ would be very helpful.

\begin{figure}
\centering
\begin{minipage}{\linewidth}
\centering
\includegraphics[width=9cm]{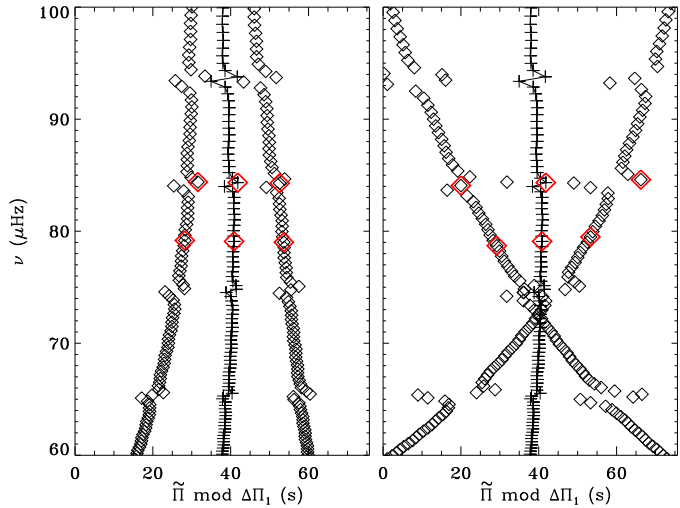}
\end{minipage}
\caption{Stretched period \'echelle diagrams for rotationally split
dipolar modes in the model $\modrg$ in Fig.~\ref{fig:chenhr} $(1.3 \Msun \, , \; 6.2 \Rsun)$.
In the left-hand panel slow rotation, with an angular velocity 5 times 
smaller than in Fig.~\ref{fig:rotsplit}, was assumed, while the right-hand
panel used the same angular velocity as in Fig.~\ref{fig:rotsplit}.
Plusses show modes with $m = 0$ and diamonds the rotationally split modes
with $m = \pm 1$.
The red diamonds mark the modes illustrated in Fig.~\ref{fig:rgamde}.
}
\label{fig:rotstretch}
\end{figure}

The analysis of rotationally split modes in red giants is complicated by
the possible overlap between neighbouring multiplets 
in the mixed-mode spectrum, when $\delta \nu_{\rm max} \sim \nu^2 \Delta \Pi$.
Even so, \citet{mosser2012rot} succeeded in determining the rotational
splitting in a large number of stars, including cases of overlap, by
identifying the pattern of rotationally split components using the
approximation in Eq.~(\ref{eq:zetamos}).
As pointed out by \citet{mosser2015}
the analysis is greatly simplified by using the stretching with the 
function $\CP$ introduced by Eq.~(\ref{eq:stretch}).
We first note that, according to Eq.~(\ref{eq:zetasplit}), the
frequencies of the rotationally split modes can be approximated by
\begin{equation}
\nu_{nlm} \simeq \nu_{nl0} + m \zeta \delta \nu_{\rm max} \; ,
\end{equation}
where for simplicity we neglected the envelope contribution and hence
assumed that $\CR \simeq 0$.
From this it follows,
using also that $\CP$ is defined as a function of frequency,
that the rotationally split modes for a given $m$ are approximately 
uniformly spaced in stretched period $\CP(\nu_{nlm})$,
with a period spacing given by
\begin{equation}
\Delta \Pi_{lm} \simeq \Delta \Pi_l\left(1 - 2 m \langle \zeta \rangle
{\delta \nu_{\rm max} \over \nu} \right) \; ,
\end{equation}
where $\langle \zeta \rangle$ is a suitable average of $\zeta$.
In a period \'echelle diagram based on the stretched periods
this corresponds of a sequence of modes on a line inclined relative to the
modes with $m = 0$.
Two examples of this are shown in Fig.~\ref{fig:rotstretch},
for $l = 1$ and two different rotation rates.
We note that in the right-hand panel 
the rotational splitting in terms of period is around 62\,s, i.e., comparable
with the period spacing of 75\,s and hence leading to a complex structure
of the power spectrum.
Even so, the structure of the rotationally split modes can be
unambiguously identified in the stretched period \'echelle diagram.

\citet{mosser2015} argued 
that $\langle \zeta \rangle$ 
can be represented by $\CN(\nu)/[\CN(\nu) + 1]$ where $\CN$
measures the number of gravity modes in a $\Delta \nu$-wide
interval around $\nu$ (see Eq.~\ref{eq:gnumber}).
Assuming also that the modes considered have frequencies close to
the frequency $\nu_{\rm max}$ of maximum oscillation power they 
obtain for the period spacings
\begin{equation}
\Delta \Pi_{lm} \simeq \Delta \Pi_l(1 - m x_{\rm rot}) \; ,
\end{equation}
where
\begin{equation}
x_{\rm rot} = 2 {\CN(\nu_{\rm max}) \over \CN(\nu_{\rm max}) +1}
{\delta \nu_{\rm max} \over \nu_{\rm max}} \; .
\end{equation}

As a complication in the treatment of red-giant 
rotational splitting it was noted by \citet{ouazzani2013} that
the core rotation may be so rapid that the perturbation analysis discussed
here is inadequate. 
In such cases Ouazzani et al demonstrated that a two-dimensional 
solution of the oscillation equations is required, involving coupling between
components of different degrees.
A complication in these cases is alsoß
that properties
of the eigenfunctions, in particular the mode inertia, can vary substantially
between the different components of the multiplet.

\section{Groundbreaking results}
\label{sect:results}
 Asteroseismology of red-giant stars has been very successful over the past decade with many publications and ground-breaking results. We anticipate that this will continue in the decade(s) to come (see next section). In this section we discuss results that we think have been seminal for the field. 

 Early observations and interpretations of oscillations in very large red giants such as Arcturus \citep{merline1995} and $\alpha$ UMa \citep{buzasi2000,dziembowski2001} have already been conducted during the last century and the beginning of this century. Following these discoveries, the study of solar-like oscillations in red-giant stars all along the red-giant and horizontal branch (Sections~\ref{sect:Hburn} $\&$ \ref{sect:Hecoreburn}) has effectively started with the spectroscopic campaigns in which $\xi$ Hydrae \citep{frandsen2002}, $\epsilon$ Ophiuchi \citep{deridder2006} and $\eta$ Serpentis \citep{barban2004} were observed. In these observations oscillation power excesses exhibiting a regular pattern of oscillation modes were observed unambiguously for the first time for early red giants. These discoveries led to many questions concerning e.g. whether non-radial oscillation modes would be observable or damped in the core or what the typical lifetime of the stochastically excited and damped modes would be. Data from the dedicated photometric space missions  MOST \citep[Microvariability and Oscillations of STars;][]{matthews2000}, CoRoT \citep[Convection Rotation and planetary Transits;][]{baglin2006} and \textit{Kepler} \citep{borucki2008} have been vital in answering these and subsequent questions.

\subsection{Non-radial oscillation modes}
\begin{figure}
\centering
\begin{minipage}{0.48\linewidth}
\centering
\includegraphics[width=\linewidth]{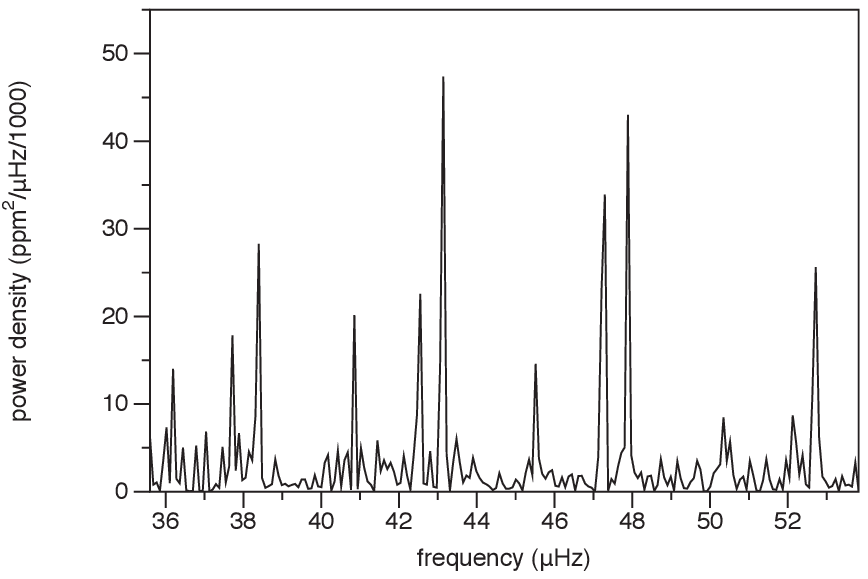}
\end{minipage}
\begin{minipage}{0.48\linewidth}
\centering
\includegraphics[width=\linewidth]{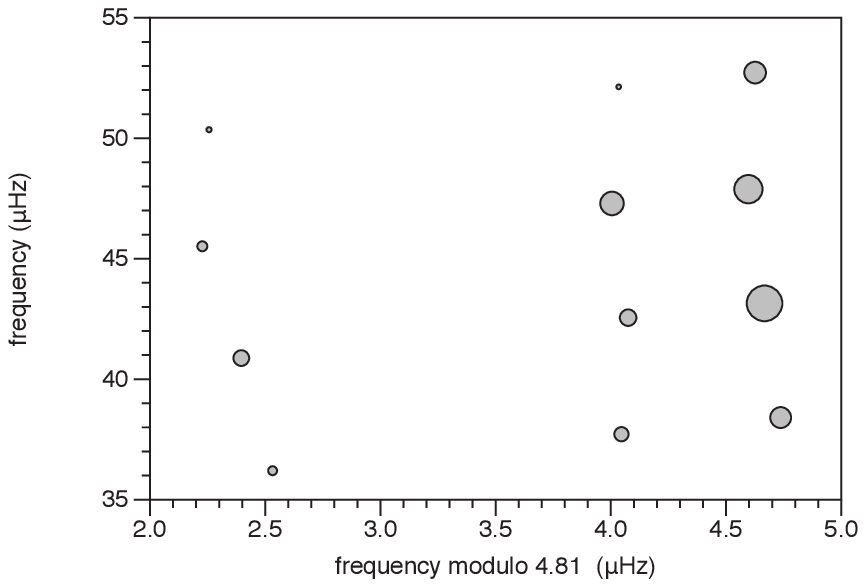}
\end{minipage}
\caption{Left: Power density spectrum of the red giant CoRoT-101034881 showing a frequency pattern  with a regular spacing. Right: \'{e}chelle diagram of the modes shown in the left panel showing 'ridges' related to radial and non-radial modes. {\orange Image reproduced with permission from \citet{deridder2009}, copyright by Macmillan.}}
\label{fig:nonradmodes}
\end{figure}

The theoretical work by \citet{dziembowski2001} for $\alpha$ UMa showed that high up on the red-giant branch non-radial modes are strongly damped in the core. Extrapolating from this result, it was initially thought that non-radial modes would be strongly damped in the cores of stars all along the red-giant branch \citep{jcd2004}. However, this was contradicted by observations. Firstly, \citet{hekker2006} claimed the detection of non-radial modes in red-giant stars. For this claim they analysed the moments of the cross-correlation functions (CCFs) of the spectra of $\xi$ Hydrae, $\epsilon$ Ophiuchi and $\eta$ Serpentis taken during the previously mentioned spectroscopic campaigns (see Section~\ref{sect:indf} for this diagnostic). The variations in the moments as a function of frequency were inconsistent with radial modes, and hence \citet{hekker2006} concluded that these oscillations are non-radial. This was subsequently followed by a similar claim based on MOST data of $\epsilon$ Ophiuchi \citep{kallinger2008}, where the authors extracted frequencies of non-radial modes from the power spectrum. These claims were finally unambiguously confirmed using CoRoT data \citep{deridder2009}. The CoRoT data showed that the oscillations in red giants form a regular pattern similar to that seen for the Sun. This includes oscillation modes with significant amplitudes at the expected locations of dipole ($l=1$) and quadrupole ($l=2$) modes. Hence, these observations confirmed that non-radial modes reach observable heights at the surface of red-giant stars \citep[see Fig.~\ref{fig:nonradmodes} taken from][]{deridder2009}. At the same time, the presence of non-radial modes with observable amplitudes at the stellar surface for stars along the red-giant and horizontal branch could be confirmed theoretically by \citet[][see also Section~\ref{sec:excitation}]{dupret2009}.

\subsection{Mode lifetimes}
The mode lifetimes of the individual oscillation frequencies (see Fig.~\ref{fig:aslifetimes}) are tied to the excitation and damping of the modes. From the first observations of $\xi$ Hydrae the lifetimes and amplitudes of the oscillation modes were examined by e.g. \citet{houdek2002} and \citet{stello2006}. \citet{houdek2002} reproduced the amplitudes tolerably well with their computations of a stochastic excitation model in which they used a non-local time-dependent generalisation of the mixing-length formulation of \citet{gough1977}.  Subsequently, \citet{stello2006} estimated the mode lifetime for $\xi$ Hydrae from the scatter of the measured frequencies about a regular pattern. With this method \citet{stello2006} found a substantially shorter mode lifetime than \citet{houdek2002}. Data from the CoRoT space mission were again seminal in showing that lifetimes of oscillation modes can vary from tens of days to of order one hundred days  \citep[see narrow peaks in the left panel of Fig.~\ref{fig:nonradmodes};][]{deridder2009}. 

\begin{figure}
\centering
\begin{minipage}{\linewidth}
\centering
\includegraphics[width=\linewidth]{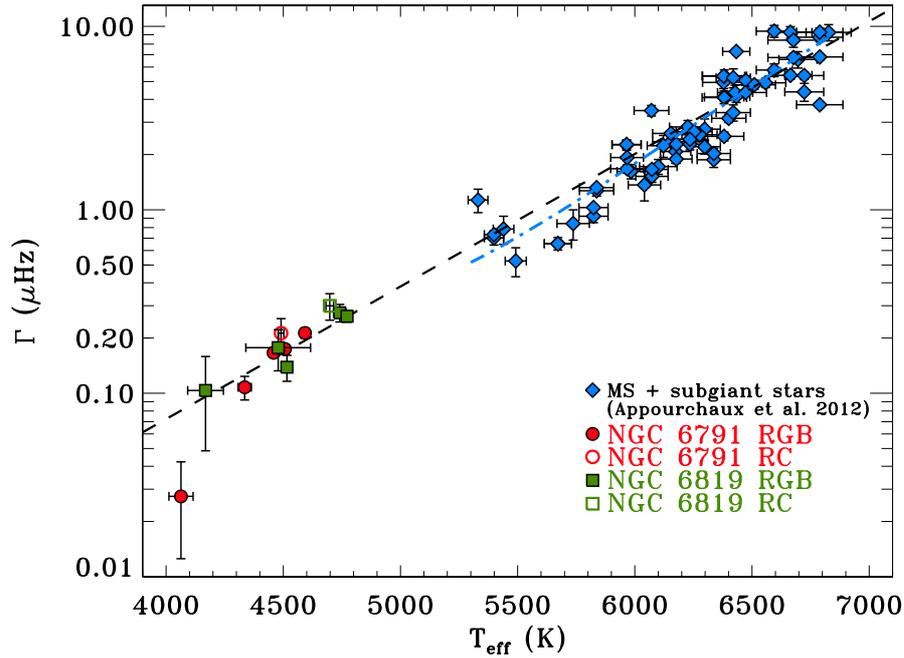}
\end{minipage}
\caption{Linewidths ($\Gamma$) of the $l=0$ ridge plotted against $T_{\rm eff}$ for the stars of NGC 6791 (red circles) and NGC 6819 (green squares). Red-giant-branch stars (RGB) are indicated with filled symbols and red-clump stars (RC) with open symbols (see legend). Also shown are measured linewidths for main-sequence and subgiant field stars (blue diamonds) from \citet{appourchaux2012a}. The fit to the main-sequence and subgiant stars taken from \citet{appourchaux2012a} is also shown (dot-dashed blue line). The dashed black line shows an exponential fit to all stars. {\orange Image reproduced with permission from \citet{corsaro2012}, copyright by AAS.}}
\label{fig:linewidthvstemp}
\end{figure}

Although the physics responsible for the damping mechanism is not yet fully understood (see also Section~\ref{sec:excitation}), the mode lifetime ($\tau\propto1/\Gamma$, with $\Gamma$ indicating the FWHM mode linewidth) follows a trend with temperature \citep[see Fig.~\ref{fig:linewidthvstemp} and][and references therein]{chaplin2009,baudin2011,corsaro2012,corsaro2015}. Interestingly, there seems to be a steeper temperature gradient for the hotter main-sequence stars than for the cooler red-giant stars, whose origin is yet to be explained.

\begin{figure}
\centering
\begin{minipage}{\linewidth}
\centering
\includegraphics[width=\linewidth]{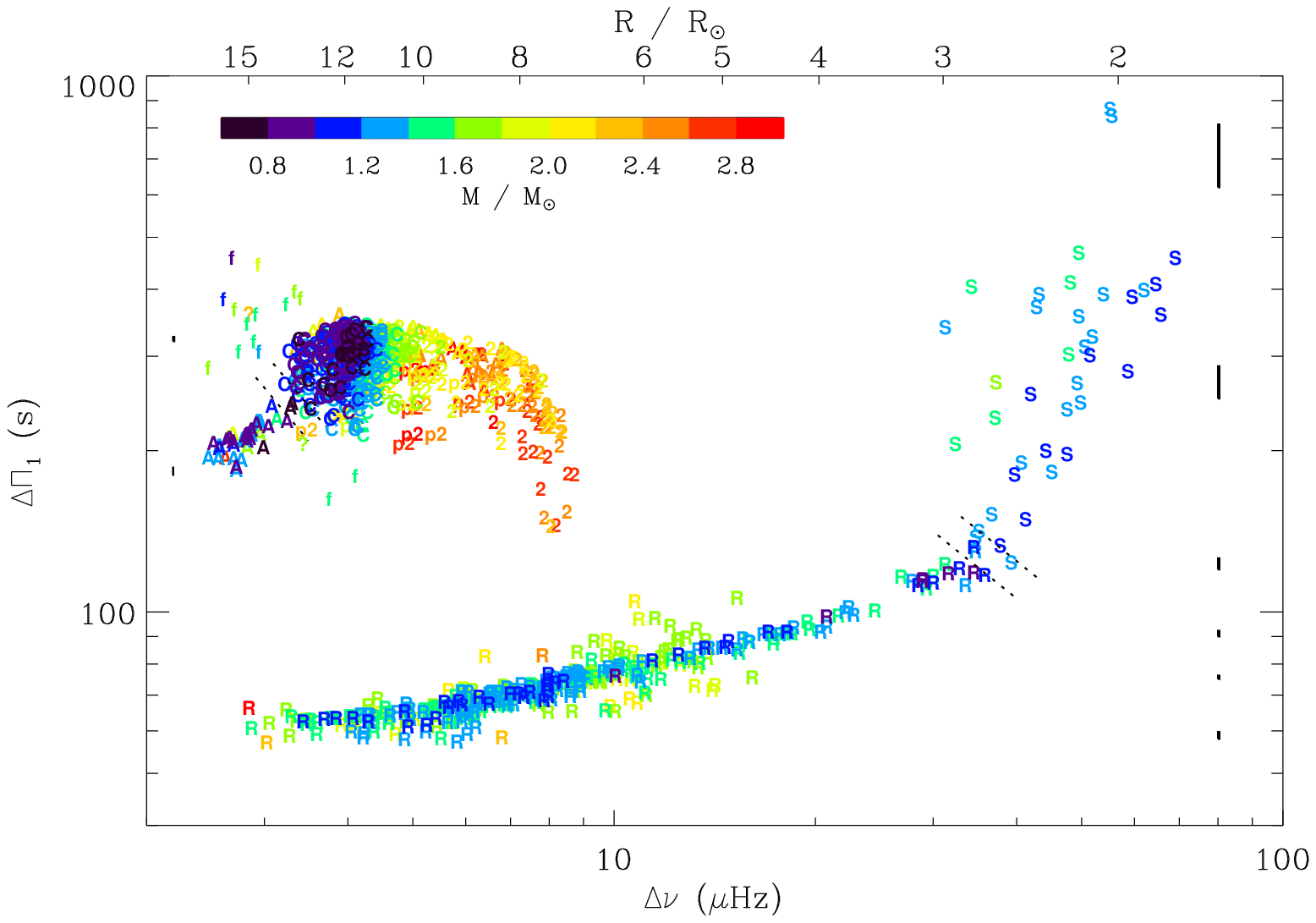}
\end{minipage}
\begin{minipage}{\linewidth}
\centering
\includegraphics[width=\linewidth]{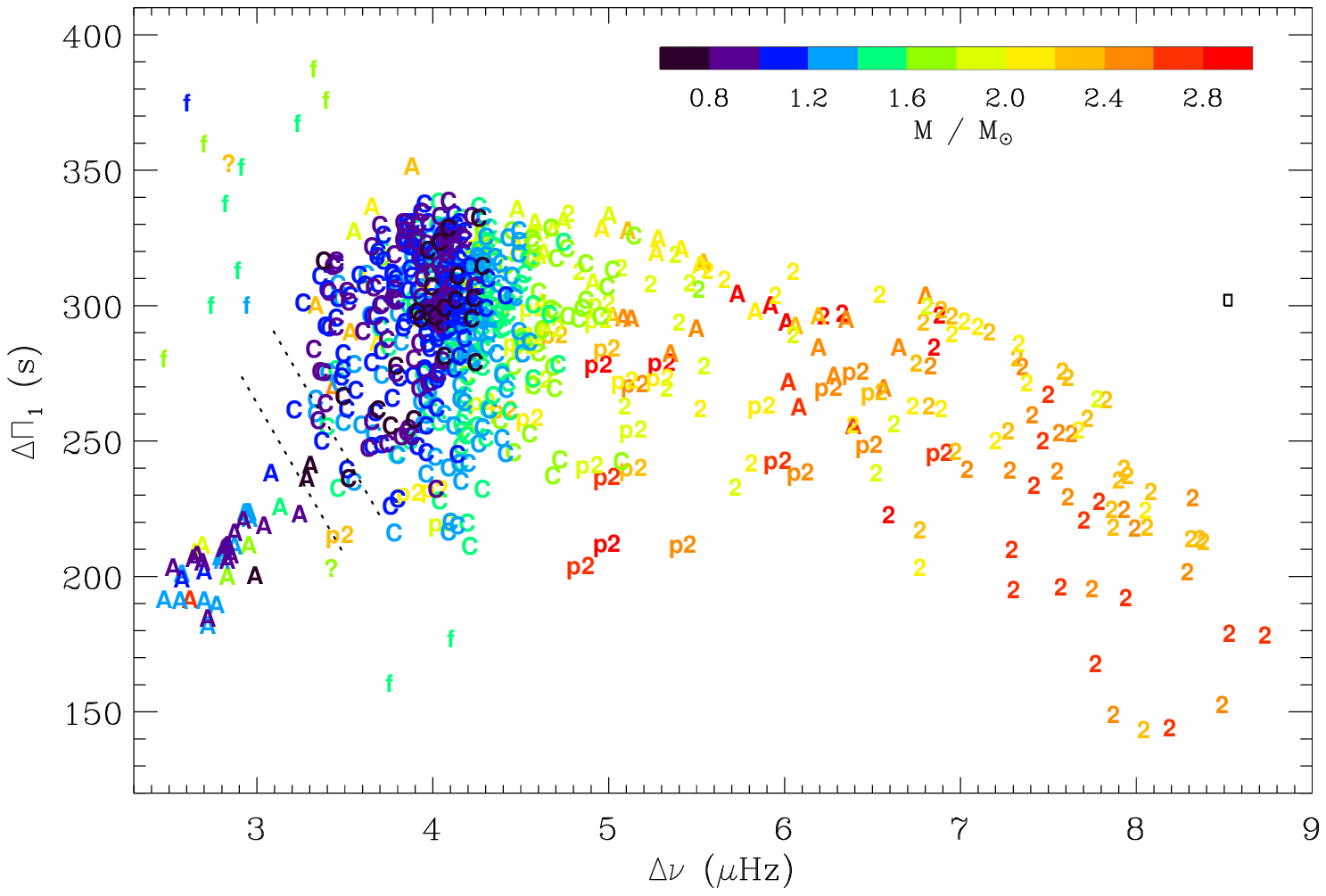}
\end{minipage}
\caption{Dipolar period spacing $\Delta\Pi_{1}$ as a function of the frequency spacing $\Delta\nu$. Top: the seismic proxy for the stellar mass is indicated by the colour code. The evolutionary states are indicated by S (subgiants), R (RGB), f (helium subflash stage), C (red clump), p2 (pre secondary clump), 2 (secondary clump), and A (stars leaving the red clump moving toward the AGB). The error boxes on the right side indicate the mean uncertainties, as a function of $\Delta\Pi_{1}$, for stars on the RGB; for clump stars, uncertainties are indicated on the left side. Dotted lines indicate the boundaries between evolutionary stages. We note that only \citet{mosser2014} have so far claimed evidence of stars with sub-flashes (f); these assignments have not been assessed nor dismissed by other work. Bottom: zoom in the red-clump region. {\orange Image reproduced with permission from \citet{mosser2014}, copyright by ESO.}}
\label{fig:dPvsdnu}
\end{figure}

\subsection{Glitches}
\label{sect:resglitches}
The first evidence for the helium glitch in a red-giant star was
found using CoRoT data by \citet{miglio2010},
who pointed out that the properties of the glitch could be used as
additional seismic diagnostics in the global characterization of the star.
\citet{broomhall2014} made an extensive analysis of the effects of the
helium glitch in red-giant models.
They noted the difficulty in obtaining meaningful inferences of the
helium abundance, given the limited number of modes available for the
analysis.
Encouraging results were obtained by \citet{corsaro2015he} for 18
low-luminosity red giants observed for the full {\it Kepler} mission;
they showed that the formulation derived by \citet{houdek2007} gave
an excellent fit to the observed glitch signatures and obtained relatively
precise determinations of the amplitude and acoustic depth of the
glitch signal.
\citet{vrard2015a} analysed a large sample of {\it Kepler} red giants
and found interesting differences in the glitch properties of 
red-giant-branch and clump stars. 
This may be related to the difference found by \citet{kallinger2012} 
between these two evolutionary stages in the phase term $\epsilon$
(cf. Section \ref{sect:epsilon}),
which appears to be caused by differences in the ionization behaviour
of helium \citep{jcd2014}.

\subsection{Mixed modes}
\label{sect:resmixedmodes}
A first mention of dense and/or irregular frequency patterns in the solar-like oscillations of red giants was made by \citet{hekker2009}. These authors already indicated that this could be explained by the fact that the observed oscillations are influenced by their behaviour in both the p-mode and g-mode cavity. \textit{Kepler} observations were needed to resolve these modes and identify that non-radial modes in red-giant stars are mixed modes \citep{bedding2010}, i.e. they propagate in both the outer acoustic cavity as well as in the inner buoyancy cavity. Results of mixed modes for stars observed by both CoRoT and \textit{Kepler} have been presented by e.g. \citet{beck2011,bedding2011,mosser2011mm,mosser2014} The detection of mixed modes in timeseries data from the \textit{Kepler} mission for a few hundred stars showed that the period spacing between the mixed modes is a direct measure of whether a star is in the hydrogen shell burning phase or also burning helium in the core \citep{bedding2011,mosser2011mm}. Subsequently, \citet{mosser2014} used the combined period spacing and large frequency separation to also identify stars in short evolutionary phases such as the helium subflash stage, pre-secondary clump stars and stars leaving the red clump moving towards the AGB (see Fig.~\ref{fig:dPvsdnu}). The differences in the period spacings between stars with an inert helium core and stars with core-helium fusion are in part attributed to the presence of a convective core in stars with helium-core fusion \citep[][see also Section~\ref{sec:redclump}]{jcd2014}.

\begin{figure}
\centering
\begin{minipage}{\linewidth}
\centering
\includegraphics[width=\linewidth]{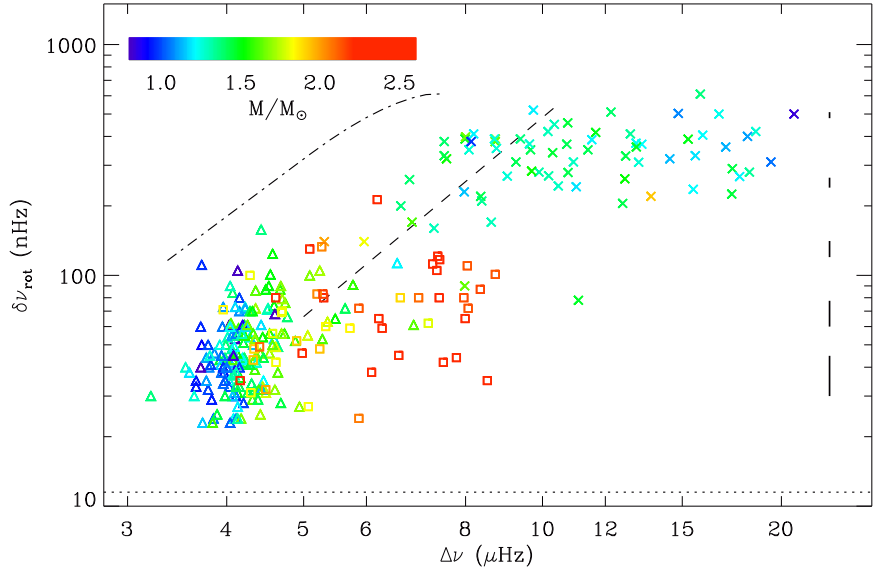}
\end{minipage}
\caption{Rotational splitting $\delta\nu_{\rm rot}$ as a function of large frequency separation $\Delta\nu$ in log-log scale. RGB stars are indicated with crosses, red-clump stars are indicated with triangles and secondary-clump stars are indicated with squares. All symbols are colour coded by the mass estimate from asteroseismic parameters (see colour bar). The mean uncertainties in $\delta\nu_{\rm rot}$ are indicated by the vertical bars on the right. The horizontal dotted line indicates the frequency resolution. The dashed (dotted-dashed) line shows the confusion limit with mixed modes in RGB (RC) stars. {\orange Image reproduced with permission from \citet{mosser2012rot}, copyright by ESO.}}
\label{fig:mosserrot}
\end{figure}

\begin{figure}
\centering
\begin{minipage}{\linewidth}
\centering
\includegraphics[width=\linewidth]{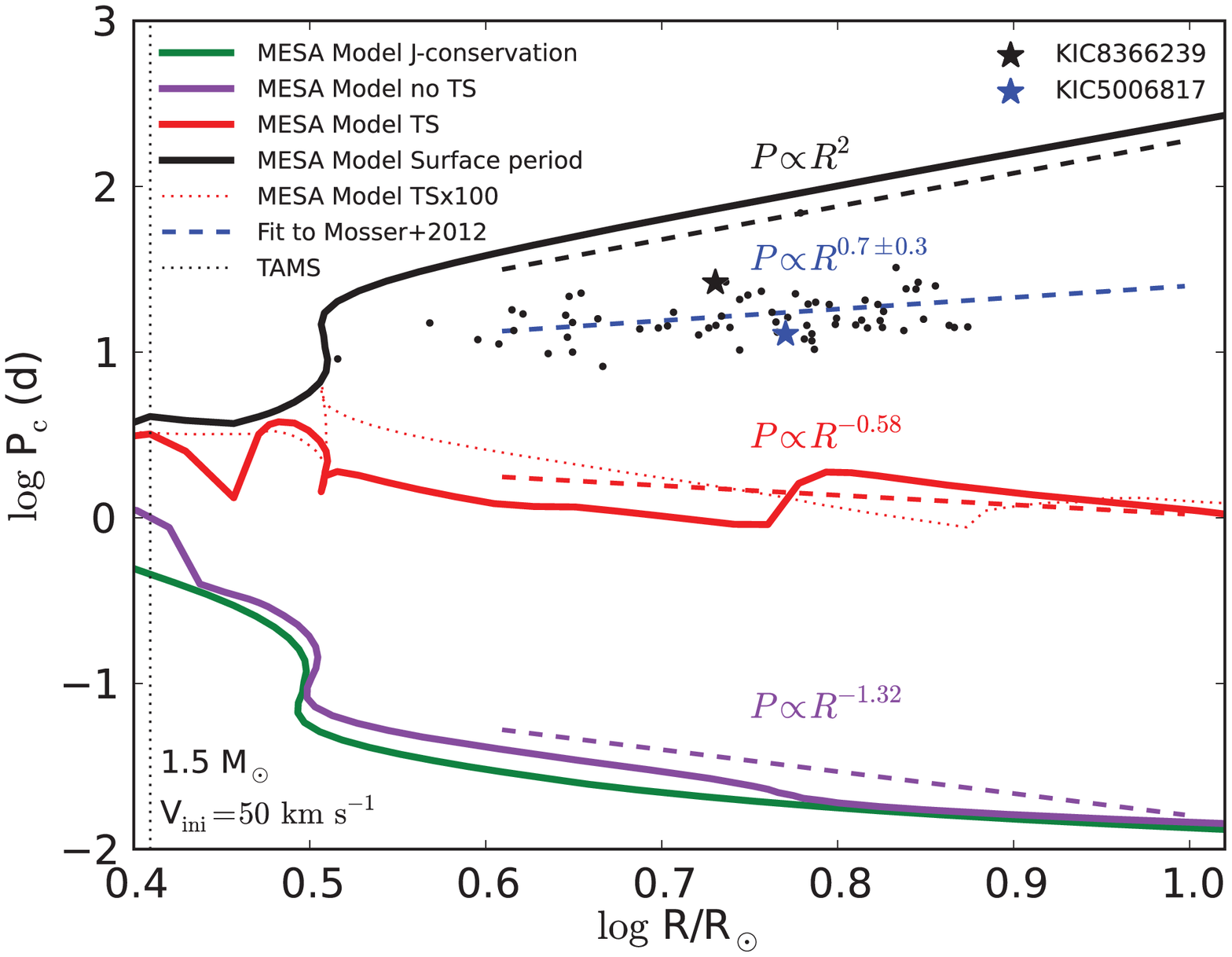}
\end{minipage}
\caption{Evolution of the average core rotation period as a function of stellar radius for different assumptions of angular momentum transport in a 1.5~M$_{\odot}$ model initially rotating at 50~km\,s$^{-1}$. Models without angular momentum transport (green), including transport of angular momentum due to rotational instabilities (purple) and accounting for magnetic torques in radiative regions (red, Tayler-Spruit magnetic fields) are shown. The star symbols indicate the locations of KIC8366239 and KIC5006817 as derived using the maximum observed splitting of their mixed modes \citep{beck2012,beck2014}. Dashed lines indicate a linear fit to the different curves during the early RGB. The vertical dotted line shows the location of H-core exhaustion {\orange (TAMS, terminal-age main sequence)}. The red dotted line shows the evolution of core rotational period for a model where the resulting Tayler-Spruit diffusion coefficient has been multiplied by a factor of 100. Stars in the red giant sample of \citet{mosser2012rot} with $R < 7.5$~R$_{\odot}$ are shown as black dots. The best fit to the core rotation of the sample \citet{mosser2012rot} is also shown as a dashed blue line. {\orange Image reproduced with permission from \citet{cantiello2014}, copyright by AAS.}}
\label{fig:cantiellorot}
\end{figure}

\subsection{Radial differential rotation}
\label{sect:resrot}
Using the fact that different mixed modes probe different radial regions in stars, \citet{beck2012} investigated the mixed modes for rotational splittings. They found that rotational splittings show different behaviour depending on whether the rotationally-split modes are pressure dominated or gravity dominated. Qualitative comparison with models revealed that subgiants and red-giant stars ascending the red-giant branch exhibit radial differential rotation with the core rotating faster than the surface \citep{beck2012,deheuvels2012}. \citet{deheuvels2014} showed that the core of stars in the subgiant phase spins up before reaching the base of the RGB and subsequently {\orange spins down} on the RGB \citep[][(see their Fig.~9)]{mosser2012rot}  due to efficient transport  of angular momentum (AM) from the core to the envelope the origin of which is still unknown. At the same time \citet{mosser2012} also found an important slow down for red-clump stars compared with the red-giant-branch stars (see Fig.~\ref{fig:mosserrot}). \citet{deheuvels2012,deheuvels2014,dimauro2016} subsequently performed rotational inversions to investigate the radial rotation profiles. They showed that the core rotation rate can be determined with only a weak model dependence. Additionally, an upper limit for the surface rotation could be obtained.

Interestingly, from a theoretical point of view the observed core rotation cannot be explained. Current models include transport of angular momentum due to rotationally induced instabilities and circulations as well as magnetic fields in radial zones  \citep[generated by the Tayler-Spruit dynamo;][]{tayler1973, spruit1999} and internal gravity waves. However, these models over-predict the core rotation rate by about one order of magnitude \citep[e.g.][]{eggenberger2012,marques2013,fuller2014,cantiello2014}, as illustrated in Fig.~\ref{fig:cantiellorot}. This leads to the conclusion that an additional angular momentum transport process must be operating that is currently not included in the models. \citet{belkacem2015I,belkacem2015II} investigated the efficiency of mixed modes in extracting angular momentum from the innermost regions of subgiants and red giants. They concluded that for evolved red giants, mixed modes are sufficiently efficient to balance and exceed the effect of the core contraction, in particular in the hydrogen-burning shell. However, this is not the case for subgiants and early red giants.

\begin{figure}
\centering
\begin{minipage}{0.48 \linewidth}
\centering
\includegraphics[width=\linewidth]{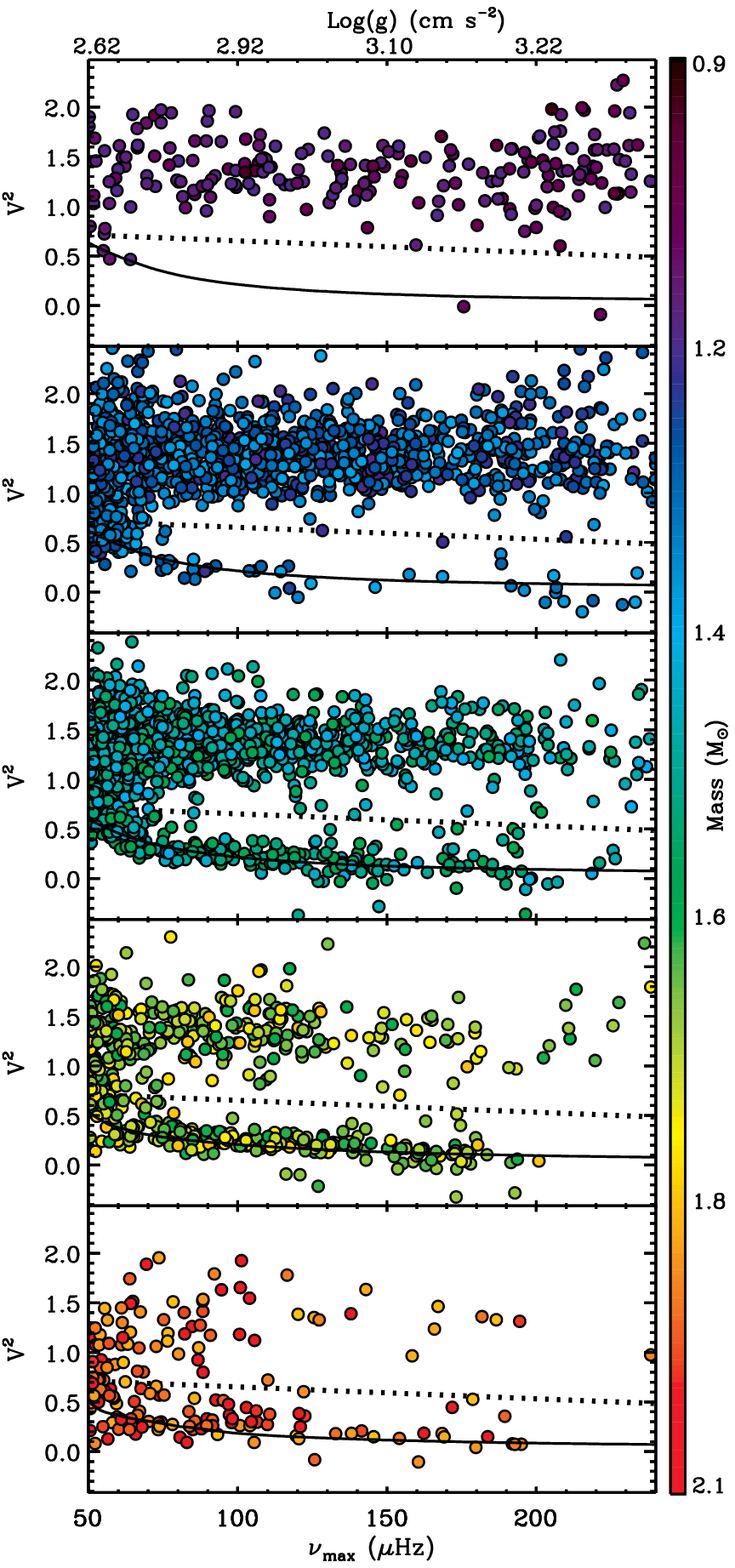}
\end{minipage}
\centering
\begin{minipage}{0.45 \linewidth}
\centering
\includegraphics[width=\linewidth]{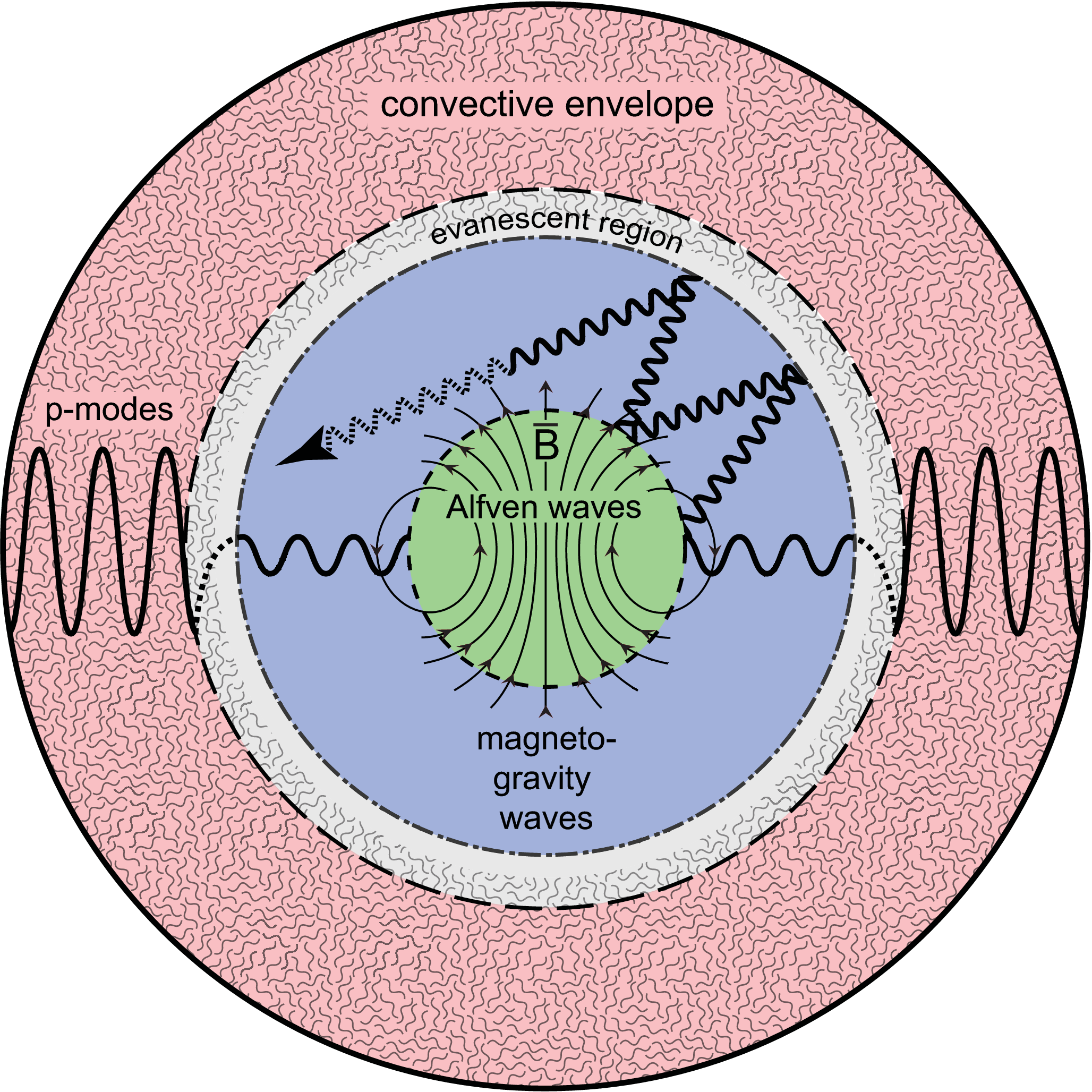}
\end{minipage}
\caption{Left: Average visibility of each star vs. $\nu_{\rm max}$, which correlates closely with surface gravity (shown on the top axis) for stars with masses ranging from 0.9 to 2.1 (top to bottom and colour scale). Stellar mass has a formal $1\sigma$ uncertainty of 10\%. Stars evolve from right to left in the diagrams, corresponding roughly with the beginning of the red giant phase to the red giant luminosity bump.  The solid black line shows the theoretical predicted dipole-mode suppression for $1.1$M$_{\odot}$, $1.3$M$_{\odot}$, 1.5M$_{\odot}$, 1.7M$_{\odot}$ and 1.9M$_{\odot}$ (top to bottom panels) and a radial-mode lifetime of 20 days. The fiducial dotted lines separate normal and dipole-suppressed stars. {\orange Image reproduced with permission from \citet{stello2016}, copyright by AAAS.}
Right: Schematic representation of the magnetic greenhouse effect. Acoustic waves excited in the envelope couple to gravity waves in the radiative core. In the presence of a magnetic field in the core, the gravity waves are scattered at regions of high field strength. Because the field cannot be spherically symmetric, the waves are scattered to high angular degree and become trapped within the core, where they eventually dissipate (dashed wave with arrow). {\orange Image reproduced with permission from \citet{fuller2015}, copyright by Macmillan.}}
\label{fig:supdipmod}
\end{figure}

\subsection{Suppressed dipole modes}
\label{sect:ressupmodes}
For a small subset of stars the dipole  ($l=1$) modes seem to be suppressed, i.e. they have a low height or visibility ($V^2${\orange , i.e. amount of integrated power compared with that of the radial modes}) in the power spectrum. This was first flagged by \citet{mosser2012} and followed by a detailed investigation for a single star by \citet{garcia2014}. These authors concluded that the low visibility cannot be explained by damping effects nor by a fast-rotating core. Recently, \citet{fuller2015} showed in a theoretical study that a high magnetic field in the stellar core, i.e. a magnetic greenhouse effect, would induce low mode visibilities.
In parallel, \citet{stello2016} showed that there is agreement between the observed and simulated visibilities of the suppressed oscillation modes and that this lends evidence for the presence of a high magnetic field in the cores of the red giants with suppressed dipole modes.
Further support for this model follows from the fact that the suppressed modes
are only found in stars with masses above ${\sim} 1.1 \Msun$, i.e., stars that
in the main-sequence phase would have had a convective core which could have
given rise to a dynamo-generated field in the core.
Fig.~\ref{fig:supdipmod} shows a schematic diagram of the magnetic greenhouse scenario as well as a comparison of the predicted visibilities from this scenario with observed visibilities of dipole modes as a function of stellar mass.
Subsequently, \citet{stello2016qo} also investigated the suppression of quadrupole and octupole modes. They found that mode suppression weakens for higher-degree modes with a reduction in the quadrupole mode visibility of up to 49\% and no detectable suppression in octuple modes. This is consistent with predictions based on the theory of the magnetic greenhouse effect \citep{fuller2015} applied to these higher-degree modes.
Recently, \citet{mosser2017} analysed a sub-sample of red giants
with suppressed modes, 
characterized by having high signal-to-noise ratio, and showed that 
in these cases the suppressed dipole modes are mixed modes, 
with a character similar to that of modes in stars with no suppression.
Therefore, in this case a mechanism that only partially damps 
the oscillations without affecting the basic frequency structure, is implied.
On this basis \citet{mosser2017} questioned the scenario of \citet{fuller2015}.
However, it remains to be seen whether the magnetic mechanism,
perhaps with suitable modifications, is consistent with these observations.

\subsection{Galactic Archaeology}
The large number of dwarfs and intrinsically bright red-giant stars for which accurate stellar parameters are now becoming available from asteroseismology has added a new dimension to the field of Galactic Archaeology, i.e. the study of the formation and evolution of the Milky Way by reconstructing its past from its current constituents. The first study using asteroseismic data to probe the galactic disk was performed on a set of about 800 CoRoT stars. This study showed qualitative agreement between the models and observations, but also flagged differences in $\nu_{\rm max}$ distributions \citep{miglio2009}. This was subsequently followed by a population synthesis study using asteroseismic data of about 500 dwarfs observed with \textit{Kepler}. Again the models and observations were in qualitative agreement, although there were differences present in the mass distributions \citep{chaplin2011}.

These early investigations have initiated major collaborations between asteroseismic experts and experts in galactic astronomy to combine the efforts of large spectroscopic surveys \citep[see][for an overview of the ongoing spectroscopic surveys and the requirements for them]{feltzing2015} with the asteroseismic measurements. These collaborations have among others led to improvements in  red giant spectroscopy \citep{pinsonneault2014,ren2016,valentini2016}; to improved estimates of distances and extinctions \citep{rodrigues2014}; to a direct measurement of a vertical age gradient in the Milky Way \citep{casagrande2016}; as well as to the detection of $\alpha$-rich young stars, i.e. stars that are young according to their asteroseismic measures and old based on their chemical abundance of $\alpha$ elements \citep{chiappini2015,martig2015}. First studies investigating these young $\alpha$-rich stars hint towards these stars being blue stragglers \citep{yong2016,jofre2016}.

\subsection{Extra-solar planets}
The photometric timeseries obtained by the CoRoT and \textit{Kepler} space instruments are suitable for both exoplanet studies and asteroseismology. The combination of both has been the basis of many ground-breaking planet discoveries \citep[e.g.][]{carter2012,barclay2013,campante2015}. Transiting exoplanets leave larger signatures in the timeseries data when orbiting a smaller star, hence one may expect an observational bias against stars with larger radii. Nevertheless, it is possible to detect transiting planets around early red giants. \citet{huber2013} detected a planetary system consisting of two transiting planets around a low-luminosity red giant. Using the rotationally split (mixed) oscillation modes (Section~\ref{sect:rot}) they also found that the spin axis of the star is not aligned with the orbital axis of the two planets. This could only be explained by a third wide compagnion, which has indeed been detected \citep{otor2016}. Studies like the one by Huber et al. are essential to unravel planet formation scenarios.

\section{Future}
\label{sect:future}
The wealth of data obtained by {\orange the} photometric space missions CoRoT \citep{baglin2006}, \textit{Kepler} \citep{borucki2008} and K2 \citep{howell2014} are currently being explored. Exploiting these as well as data from future complementary (space) telescopes such as TESS \citep{ricker2014}, PLATO \citep{rauer2014} and the ground-based SONG network \citep{grundahl2014} will be essential to study many of the questions that are still open in stellar structure and evolution of giant stars exhibiting solar-like oscillations. These questions include, but are not limited to convection, rotation, (core) overshoot, additional mixing and stellar ages. We discuss these in some detail here.

\paragraph{Convection and surface effects}
Deep in stars convection leads to a temperature stratification that is essentially adiabatic. Near the surface, where the convective transport is less efficient, a substantially superadiabatic region is present. In stellar models this is typically described with some form of a mixing-length approach \citep[e.g.,][]{bohmvitense1958,canuto1991}. To take the next step forward it is important to improve the models to take convection properly into account. Very promising efforts are in progress to perform 3D hydrodynamical simulations of the atmosphere and near-surface part of the interior in which convection is realistically modelled \citep[see for instance][]{trampedach2014I,trampedach2014II,magic2016}. The results of the simulations can be used in stellar modelling through a calibration of parameters of simpler formulations, such as the mixing length \citep{salaris2015}. Alternatively, the outer layers of the model can be replaced by suitably averaged versions of the simulations, interpolated to the parameters of the star.

The inadequate modelling of the structure of the outermost layers is an important contribution to the frequency dependent offset between the model frequencies and observed frequencies, i.e. the surface effect. As discussed in Section~\ref{sect:indf} several approaches exist to mitigate this, although these all require adjustments on a star by star basis. Thus a better theoretical understanding of these effects is highly desirable. Replacing the outer layers of the model by an averaged 3D simulation goes some way towards reducing these effects \citep{rosenthal1999,robinson2003,sonoi2015,ball2016}. An additional contribution comes from non-adiabatic effects and the influence of turbulent pressure which are typically ignored in the modelling. Interestingly, combining 3D simulations with a non-local treatment of time-dependent convection and non-adiabaticity very substantially improves the agreement between observed and modelled solar oscillation frequencies \citep{houdek2017}.

\paragraph{Rotation}
The current stellar models cannot reproduce the core rotation rates observed using rotationally split mixed modes (see Section~\ref{sect:resrot}). In these models transport of angular momentum due to rotationally induced instability and circulation, as well as magnetic fields in radiative zones generated by the Tayler-Spruit dynamo are included \citep{cantiello2014}. The fact that these models can still not match the observations is most likely due to missing physics in the models. 
Observations of (radially differential) rotation in stars across the HR diagram may provide indications of what additional physical effects play a role.
Additionally, the impact on red-giant models of the recent results by \citet{vansaders2016} and \citet{metcalfe2016} regarding rotation in main-sequence stars and subgiants may need to be investigated. \citet{vansaders2016} showed that the effective loss of angular momentum ceases above a critical Rossby number (Ro):
\begin{equation}
\rm{Ro} = \frac{\text{rotation\,\,period}}{\text{convective\,\,turn-over\,\,time}} \approx 2.1
\end{equation}
Subsequently, \citet{metcalfe2016} aimed to explain the underlying reasons for this by proposing a scenario implying a change in the character of differential rotation that ultimately disrupts the large-scale organisation of the magnetic field in solar-type stars. This process begins at Ro $\approx$ 1, where the rotation period becomes comparable to the convective turn-over time. \citet{metcalfe2016} speculated that this may lead to an accelerated decrease in the surface area of spots while the star goes through a rapid phase of magnetic evolution while crossing the Vaughan-Preston gap \citep{vaughan1980}. Due to the changes in magnetic topology during this fast phase of magnetic evolution stars reach Ro $\approx$ 2 where magnetic breaking operates with a dramatically reduced efficiency \citep{vansaders2016}. Although this scenario is still speculative the proposed changes in the rotation of main-sequence stars may impact the rotation profiles of red giants.

Finally, in terms of observations of the envelope rotation rate it will be of importance to measure rotational splittings of p-dominated $l=2$ and $l=3$ modes (see Sections~\ref{sect:rot} and \ref{sec:rotation}) as these have larger sensitivity to the surface layers (see red triple-dot-dashed line in Fig.~\ref{fig:introtker}). Radial-velocity observations are more sensitive to these higher-degree modes and hence SONG could play a major role in this.

\paragraph{Additional mixing}
The standard theory of stellar evolution fails to explain abundance anomalies observed in stars ascending the red-giant branch. Spectroscopic studies showed that when stars reach the bump (see Section~\ref{par:bump}) a drop in the surface carbon isotopic ratio, lithium and carbon abundances is present, while nitrogen increases slightly \citep[e.g.][]{gilroy1991,tautvaisiene2013}. This is not seen in standard stellar models and provides evidence that an extra-mixing process should occur when low-mass stars reach the bump \citep[e.g.][for a study into the effects of rotation-induced mixing using both asteroseismic and spectroscopic constraints]{lagarde2015}.

Asteroseismic measurements of the strength and locations of boundaries between radiative and convective regions that are observable as glitches (see Sections~\ref{sect:glitches} and \ref{sect:glitchestheory}) could be very important to determine the strength, location and efficiency of the additional mixing that needs to be added to the models to match both the observed stellar internal structure and the observed chemical yields.

\paragraph{Core overshoot}
In models the edge of a convective core can be described by different criteria (Section~\ref{sect:cc}). Whether semi-convection exists or another form of slow mixing is present can currently not be directly constrained from observations \citep{constantino2015,constantino2016}.
{\orange With the mixed modes it is possible to probe the stellar core 
\citep[e.g.][]{bedding2011,mosser2014};
model investigations show that there are prospects to investigate
sharp features around the core \citep{cunha2015},
and this will be essential in constraining the properties of the edge 
of a convective core.
Such glitch analysis 
(Sections~\ref{sect:glitches} and \ref{sec:glitch}) based on} mixed modes may be the only source of direct observational evidence that can be obtained for the region where core overshoot may take place. A full exploitation of the archival data from \textit{Kepler} may provide these observational constraints on the core overshoot processes.

\paragraph{Suppressed dipole modes}
As described in Section~\ref{sect:ressupmodes} for a fraction of the observed red-giant stars the dipole modes are suppressed. Currently there exists one scenario \citep{fuller2015} to explain the presence of these suppressed dipole modes. For this scenario, the predictions are consistent with observations \citep[][see also Fig.~\ref{fig:supdipmod}]{stello2016}, although it was questioned by \citet{mosser2017}. However, the presence of large magnetic fields in the core can currently not be tested directly, while further theoretical elaboration of the magnetic model is required \citep[e.g.][]{cantiello2016}. Hence, it is hard to confirm or reject this scenario at the moment. In this respect, it may be interesting if other scenarios that can be confirmed/rejected or correlations with other stellar parameters can be established.

\paragraph{Ages}
There is no observable that is sensitive to age and age only \citep{soderblom2010}, so all measures of stellar ages are either empirical (with a model dependent calibration) or model dependent. Hence, any improvement in the models and/or matching the data with models by mitigating the surface effect will influence the age determinations of stars. This is particularly relevant for evolved stars as uncertainties in earlier evolution phases accumulate. Despite these limitations, it may be possible to find combinations of (asteroseismic) observables that are more sensitive to age and to improve age estimates using asteroseismology. 

\paragraph{Red-clump stars}
Red-clump stars are both from an observational as well as from a modelling point of view not well understood. Red-clump models carry information from the main-sequence stellar structure. Even a small change in the stellar structure at the main sequence can lead to significant differences in terms of effective temperature, luminosity and internal structure after the onset of helium-core burning. Additionally, the subflashes that are predicted to occur at the onset of helium-core burning for low-mass stars with degenerate cores (see Section~\ref{sect:He-flash}) leave sharp features in the model internal structure. This results in a very irregular pattern of frequencies and period spacings for stars entering the red clump \citep[][see also Section~\ref{sec:redclump}]{constantino2015}. Information on the reality and nature of the subflashes from observations will be essential to better understand the red-clump models. This may be obtained from stars that are descending the red-giant branch after helium ignition. However, this is challenging as this is a very short phase in evolution, reducing the probability that any observed star is in that phase.  Additionally, stars already in the red clump may provide the observational evidence for the presence of the flashes. However, from an observational point of view, the combination of period spacings, rotational splittings and noise that seems to be present in the observed power density spectra make it very difficult to disentangle all individual oscillation modes. Hence, observational evidence of regularity or irregularity is still awaiting for many of these stars. We anticipate that with the long timeseries of \textit{Kepler} data it will be possible to disentangle these observational features, but a detailed mining of the data with tailor-made tools will be required. \newline
\newline
The future for asteroseismology of giant stars with solar-like oscillations looks very promising. In addition to the wealth of archival data there are two new space missions planned which will again provide complementary data in terms of nearly all-sky coverage with TESS and brighter stars with PLATO. Additionally, the Stellar Oscillations Network Group (SONG) has one node fully operational and one node in development. SONG provides spectroscopic data from the ground, which will be valuable to study higher degree modes, as well as stars with long periods. These observational data together with improved 1-D and 3-D models are bound to improve significantly our knowledge of stellar structure and evolution of giant stars over the next decade(s).

\begin{acknowledgements}
We thank George Angelou, Achim Weiss, Yvonne Elsworth, Maarten Mooij, Douglas Gough, Margarida Cunha, Benoit Mosser, Dennis Stello, Marc-Antoine Dupret and Chen Jiang for discussions and constructive comments on earlier versions of the draft. Their input improved the manuscript considerably.
G\"unter Houdek and Marcelo Miguel Miller Bertolami are thanked for providing data for figures and discussions of these results.
We would like to thank the anonymous referee and Benoit Mosser for their comments on the submitted version of the manuscript, {\orange which improved the manuscript substantially, as did comments on the penultimate version by
Andreas Quirrenbach}.
We acknowledge funding from the
European Research Council under the European Community's Seventh Framework
Programme (FP7/2007-2013) / ERC grant agreements no 338251 (StellarAges) and 267864 (ASTERISK; ASTERoseismic Investigations with SONG and Kepler).
Funding for the Stellar Astrophysics Centre is provided by The Danish National Research Foundation (Grant DNRF106). 
\end{acknowledgements}

\bibliographystyle{spbasic}      
\bibliography{AARv15}   

\end{document}